\documentclass[11pt,a4paper]{article}
\usepackage{jheppub}

\usepackage{amsmath}
\usepackage{epsfig,multicol,bbm}
\usepackage{graphicx}
\usepackage{bm}
\usepackage{multirow}
\usepackage{hyperref}

\newcommand{\nn}{\nonumber} 
\newcommand{\bn}{{\bar n}}

\newcommand{\be}{\begin{equation}}
\newcommand{\ee}{\end{equation}}

\newcommand{\abs}[1]{\left\lvert #1\right\rvert}
\newcommand{\bra}[1]{\left\langle #1\right\rvert}
\newcommand{\ket}[1]{\left\lvert #1\right\rangle}
\newcommand{\bbra}[1]{\big\langle #1\big\rvert}
\newcommand{\bket}[1]{\big\lvert #1\big\rangle}
\newcommand{\bbbra}[1]{\Big\langle #1\Big\rvert}
\newcommand{\bbket}[1]{\Big\lvert #1\Big\rangle}
 
\newcommand{\minus}{\!-\!}
\newcommand{\plus}{\!+\!}

\newcommand{\as}{\alpha_s}

\newcommand{\cO}{\mathcal{O}}
\newcommand{\cM}{\mathcal{M}}
\newcommand{\cC}{\mathcal{C}}
\newcommand{\cI}{\mathcal{I}}
\newcommand{\cT}{\mathcal{T}}
\newcommand{\cG}{\mathcal{G}}
\newcommand{\cQ}{\mathcal{Q}}
\newcommand{\cH}{\mathcal{H}}
\newcommand{\cF}{\mathcal{F}}
\newcommand{\cL}{\mathcal{L}}
\newcommand{\cA}{\mathcal{A}}
\newcommand{\cS}{\mathcal{S}}
\newcommand{\la}{\ell_1}
\newcommand{\lb}{\ell_2}
\newcommand{\lac}{\la^{\,c}}
\newcommand{\lbc}{\lb^{\,c}}

\newcommand{\e}{\epsilon}
\newcommand{\kap}{k_1^+}
\newcommand{\kam}{k_1^-}
\newcommand{\kbp}{k_2^+}
\newcommand{\kbm}{k_2^-}
\newcommand{\meas}[1]{\frac{d^d #1}{(2\pi)^d}}
\newcommand{\F}{{}_2F_{1}}
\newcommand{\opp}{{\rm opp}}
\newcommand{\Sopp}{S^{\opp}}

\newcommand{\SigmaNG}{\Sigma^{\rm NG}}
\newcommand{\Scum}{{{\cal S}_c}}

\newcommand{\Lac}{L_1}
\newcommand{\Lbc}{L_2}
\newcommand{\Lax}{\widetilde L_1}
\newcommand{\Lbx}{\widetilde L_2}

\newcommand{\eq}[1]{Eq.~\eqref{#1}}
\newcommand{\eqs}[2]{Eqs.~\eqref{#1} and \eqref{#2}}
\newcommand{\eqss}[3]{Eqs.~\eqref{#1}, \eqref{#2}, and \eqref{#3}}
\renewcommand{\sec}[1]{Sec.~\ref{#1}}
\newcommand{\ssec}[1]{Sec.~\ref{ssec:#1}}
\newcommand{\fig}[1]{Fig.~\ref{#1}}
\newcommand{\figs}[2]{Figs.~\ref{#1} and \ref{#2}}

\allowdisplaybreaks[3]

\DeclareMathOperator{\Tr}{tr}
\DeclareMathOperator{\Real}{Re}

\DeclareMathOperator{\Li}{Li}

\begin{document}

%\preprint{MIT-CTP 4268}

\title{Non-global Structure of the ${\cal O}(\alpha_s^2)$ Dijet Soft Function}

 \author[a]{Andrew Hornig,}
 \author[b]{Christopher Lee,}
 \author[b,c]{Iain W. Stewart,}
 \author[d]{Jonathan R. Walsh,}
 \author[e]{Saba Zuberi}
 
\affiliation[a]{Department of Physics, University of Washington, Box 351560, Seattle, WA  98195, USA}
\affiliation[b]{Center for Theoretical Physics, Massachusetts Institute of Technology, 77 Massachusetts Ave., 
Cambridge, MA 02139, USA}
\affiliation[c]{Center for the Fundamental Laws of Nature, Harvard University, 17 Oxford St.,
Cambridge, MA 02138, USA}
\affiliation[d]{Theoretical Physics Group, Ernest Orlando Lawrence Berkeley National Laboratory, 1 Cyclotron Rd., 
and Center for Theoretical Physics, University of California,
366 LeConte Hall \#7300, Berkeley, CA 94720, USA}

\emailAdd{ahornig@uw.edu}
\emailAdd{clee137@mit.edu}
\emailAdd{iains@mit.edu}
\emailAdd{jwalsh@lbl.gov}
\emailAdd{szuberi@lbl.gov}

 %\received{\today}             %%
%\revised{}
%\accepted{\today}              %% These are for published papers.

\abstract{
  High energy scattering processes involving jets generically involve matrix
  elements of light-like Wilson lines, known as soft functions. These describe
  the structure of soft contributions to observables and encode color and
  kinematic correlations between jets. We compute the dijet soft function to
  ${\cal O}(\alpha_s^2)$ as a function of the two jet invariant masses, focusing
  on terms that have a non-separable dependence on these masses and are not
  determined by the renormalization group evolution of the soft function. Our
  results include non-global single and double logarithms, and analytic results
  for the full set of non-logarithmic contributions as well.  Using a recent
  result for the thrust constant, we present the complete ${\cal O}(\alpha_s^2)$
  soft function for dijet production in both position and momentum space.
}


\maketitle

%Main body of the paper

%%%%%%%%%%%%%%%%%%%%%%%%%%%%%%%%%%%%%%%%%%%%%%%%%%%%%%%%%%%%%%%%%%%%%%%%%%%%%%%%%%%%%%%%%%%%%%%%%%%%%%%%%
\section{Introduction}
\label{sec:intro}
%%%%%%%%%%%%%%%%%%%%%%%%%%%%%%%%%%%%%%%%%%%%%%%%%%%%%%%%%%%%%%%%%%%%%%%%%%%%%%%%%%%%%%%%%%%%%%%%%%%%%%%%%

Jets play a ubiquitous role at all high-energy colliders, both as signals for
new particles which interact strongly, and as backgrounds for such signals
through Standard Model processes involving the strong interactions.
Thus it is crucial to achieve reliable and precise theoretical predictions for
many types of jet cross sections.

Two obstacles to achieving precision predictions for jet observables are poor
behavior of perturbative expansions and uncontrolled nonperturbative
corrections. Two key tools for overcoming these challenges are factorization and
resummation of large logarithms~\cite{Sterman:1995fz}. For a jet cross section
which factorizes, the separation of perturbative and nonperturbative corrections
increases the predictive power thanks to universality of the nonperturbative
contributions, and the perturbative contributions can be organized according to
the different dynamical scales (e.g.  hard, jet, soft) contributing to the cross
section. The presence of these disparate scales allows factorization but is also
the culprit in producing large logarithms in perturbative results for jet cross
sections. Techniques to resum these logarithms to all orders in perturbation
theory are often critical to obtaining well-behaved predictions.

The most powerful methods to resum such logarithms rely on the renormalization
group evolution of the different factors that appear in a factorized jet cross
section, which are separated at an arbitrary scale $\mu$.  Such methods have
been developed directly within the language of perturbative QCD itself
\cite{Catani:1991kz,Catani:1992ua,Contopanagos:1996nh}. An alternative approach
has been formulated in the powerful language of the soft-collinear effective
field theory
(SCET)~\cite{Bauer:2000ew,Bauer:2000yr,Bauer:2001ct,Bauer:2001yt,Bauer:2002nz}.
For $e^+e^-$ annihilation cross sections in which the jet-like structure of an
entire event is probed with a single (``global'') variable, such as thrust
\cite{Farhi:1977sg}, heavy jet mass \cite{Clavelli:1979md}, angularities
\cite{Berger:2003iw}, etc., these methods succeed in resumming all logarithms of
the ``event shape'' that become large in the two-jet limit to a well defined
order in resummed perturbation theory
\cite{Catani:1991kz,Catani:1992ua,Berger:2003iw,Fleming:2007qr,Schwartz:2007ib,Fleming:2007xt,Becher:2008cf,Hornig:2009vb,Abbate:2010xh}.
For such a global two-jet event shape, $e$, the cross section takes the form
$d\sigma/de = d\sigma/de\bigr\rvert_{\text{dijet}}[1+O(e)]$, where the leading
order piece factorizes schematically as~\cite{Berger:2003iw,Bauer:2008dt},
\be
 \label{globalCS}
 \frac{d\sigma}{de}\Big|_{\rm dijet} = \sigma_0 H(Q;\mu) \bigl[J_n(Qe^{1/j};\mu) \otimes J_{\bn}(Qe^{1/j};\mu)\otimes S(Qe;\mu)\bigr]\,,
\ee
in terms of a hard function $H$, jet functions $J_{n,\bn}$, and soft function
$S$. The $\otimes$ denotes convolutions in the $e$-dependent arguments of $J_{n,\bn}$ and $S$.  For  event shapes like thrust and jet mass, the exponent $j=2$, but for angularities, $j$ takes a range of values greater than 1.  $\sigma_0$ is the total Born cross section. Each function depends
on logs only of a single ratio of scales, $\mu/Q$ for $H$, $\mu/(Qe^{1/j})$ for
$J$, and $\mu/(Qe)$ for $S$.  Solving the renormalization group
evolution equations in $\mu$ for each of these functions produces a form for
$d\sigma/de$ in which all large logarithms of $e$ are resummed to a given order
in resummed perturbation theory. The same technology can also be used to
factorize and resum cumulative distributions or ``cumulant'' observables, \be
  \label{cumulant}
  \Sigma(e_c) = \int_0^{e_c} de \: \frac{d\sigma}{de} \,.
\ee

As a step towards more exclusive probes of jets than \eq{globalCS}, we can
consider cross sections differential in more than one measure of the
``jettiness'' of a final state. We will focus here on the particular example of
dijet cross sections in the context of $e^+e^-$ annihilation at center-of-mass
energy $Q$, in particular, the dijet invariant mass distribution to hadronic
final states $X$ defined by
\be
\label{doublemass}
\begin{split}
 \frac{d\sigma}{dm_1^2 dm_2^2} = \frac{1}{2Q^2}\sum_X & \abs{\bra{X} j^\mu L_\mu \ket{e^+ e^-}}^2 (2\pi)^4\delta^4(Q - p_X) \\ 
&\times\delta\Biggl(m_1^2 - \biggl(\sum_{i\in L} p_i \biggr)^2\Biggr) \delta\Biggl(m_2^2 - \biggl(\sum_{i\in R} p_i \biggr)^2\Biggr)\,,
\end{split}
\ee where $j^\mu$ and $L_\mu$ are sums of the vector and axial currents in QCD
and QED, respectively. $L,R$ are the two hemispheres defined with respect to the
thrust axis of the final state $X$, and $p_i$ is the four-momentum of the $i$th
particle in $L,R$. Using the formalism of SCET
\cite{Bauer:2000ew,Bauer:2000yr,Bauer:2001ct,Bauer:2001yt,Bauer:2002nz}, it has been shown
that this cross section factorizes into the form
\cite{Fleming:2007qr,Fleming:2007xt}, 
\be 
\label{dijetcsfactorized}
\frac{d\sigma}{dm_1^2dm_2^2} = \sigma_0 H(Q,\mu) \int d\la d\lb
J_\bn(m_1^2 - Q\la,\mu) J_n (m_2^2 - Q\lb,\mu) S(\la,\lb,\mu) + \cdots \,,
\ee
%\cite{}
to all orders in $\as$. The soft function $S(\la,\lb,\mu)$ is given by a matrix element of Wilson lines and is defined below in \eq{hemisoftdef}.
 The ellipses denote that the result is at leading order
in the power expansion in $m_{1,2}^2/Q^2\ll 1$. The corresponding factorization
theorem for the double cumulant 
$\Sigma(m_1^{c2},m_2^{c2}) = \int dm_1^2  dm_2^2\: \theta(m_1^{c2}-m_1^2)
\theta(m_2^{c2}-m_2^2) \:
d^2\sigma/(dm_1^2 dm_2^2)$ is
\begin{equation}
\label{dijetcumfactorized}
\Sigma(m_1^{c2},m_2^{c2}) 
= \sigma_0 H(Q,\mu)Q^2\!\! \int\!\! d\la^c\, d\lb^c\:
J_\bn(m_1^{c2} \minus Q\la^c,\mu) J_n (m_2^{c2} \minus Q\lb^c,\mu) 
\Scum(\la^c,\lb^c,\mu) + \cdots ,
\end{equation}
where $\Scum$ is the double cumulant soft function
\begin{align} \label{Scumdefn}
  \Scum(\la^c,\lb^c,\mu) = \int^{\la^c} d\la \int^{\lb^c} d\lb\ S(\la,\lb,\mu)\,.
\end{align}

\eqs{dijetcsfactorized}{dijetcumfactorized} exhibit a richer structure than \eq{globalCS},
containing more information about the two-jet-like final state, and presenting
additional challenges to resumming all potentially large logarithms. Each of the
hard and jet functions depend only on ratios of $\mu$ to a single scale, $Q$ or
$m_i$, but in contrast to \eq{globalCS} the soft function now depends on the
ratios $\mu/\la$, $\mu/\lb$, and $\la/\lb$.  When $m_1\sim m_2$ the
factorization theorem \eqs{dijetcsfactorized}{dijetcumfactorized} allow resummation of logs of
$m_1/Q$ and $m_2/Q$ to arbitrarily high accuracy. Since it leaves logs of
$m_1/m_2$ in fixed-order perturbation theory without resummation, it does not
handle log resummation for $m_1\gg m_2$. In the latter situation hierarchies
appear in the ratios appearing in the hemisphere soft function.  These latter
logs are examples of what have been dubbed ``non-global logarithms'' (NGLs)
\cite{Dasgupta:2001sh,Dasgupta:2002dc}.

At present, all of the ingredients in \eq{dijetcsfactorized} are known
analytically at $\mathcal{O}(\as^2)$ except for the soft function
$S(\la,\lb,\mu)$.  What is known so far about $S$ at this order are all the logs
of $\mu/\la$ and $\mu/\lb$ thanks to knowledge of the anomalous dimensions to
two-loop order (the cusp anomalous dimension is known to three-loop order).
These terms are constrained by renormalization group invariance of the physical
cross section \eq{dijetcsfactorized}, requiring the sum of the anomalous
dimensions of $H, J_n, J_\bn$, and $S$ to add to zero. What is not known are the
functions of $l_1/l_2$ that can arise in $S$ at $\mathcal{O}(\as^2)$.
Conjectures have been made about what types of functions can arise
\cite{Hoang:2008fs}, but these have never been validated nor their coefficients
calculated analytically.

Knowledge of the soft function $S(l_1,l_2,\mu)$ is applicable not only in the
dijet invariant mass distribution \eq{dijetcsfactorized} itself, but also to a
wide class of event shapes in $e^+ e^-$ annihilation, including thrust, heavy
jet mass, and the ``asymmetric'' thrust and jet mass variables \cite{Hoang:2008fs}.
Knowledge of $S$ to $\mathcal{O}(\as^2)$ is required to achieve NNLL (and
higher) accuracy in resummed predictions for distributions in these observables,
and is important input for recent extractions of the strong coupling $\as$ at
N$^3$LL from thrust~\cite{Becher:2008cf,Abbate:2010xh} and the heavy jet
mass~\cite{Chien:2010kc}.  These analyses currently depend on numerical
extraction of ${\cal O}(\alpha_s^2)$ constants from the Monte Carlo generator 
EVENT2~\cite{Catani:1996jh,Catani:1996vz}. The same $S$ also enters event shape
distributions for massive jets~\cite{Fleming:2007qr,Fleming:2007xt,Jain:2008gb}, applicable
for example for extracting the top quark mass from jet mass distributions.
Although \eq{dijetcsfactorized} is formulated for $e^+e^-$ collisions, the
hemisphere dijet soft function is actually closely related to an incoming dijet
soft function that appears in event shapes for hadron-hadron collisions, such as
the ``beam thrust'' or ``0-jettiness'' cross sections introduced in
\cite{Stewart:2009yx,Stewart:2010tn}. In that case, the two masses are those of
the measured radiation in hemispheres determined by the beam directions.

Observables like the dijet invariant mass distribution can probe the jet-like
structure of an event in a ``non-global'' way, meaning that they are sensitive
to soft radiation at different scales in sharply divided regions of phase space.
The remaining sensitivity to these scales in the soft functions produces NGLs.
Other examples of such non-global observables are exclusive jet cross sections
\cite{Dasgupta:2003mk} and jet shape distributions
\cite{Ellis:2010rw,Banfi:2010pa} using particular jet algorithms, as well as
distributions from jet substructure algorithms \cite{Rubin:2010fc}.  Accounting
for and resumming these logarithms will be important for achieving precision jet
phenomenology in this era where we probe jets with ever more exclusive measures.

NGLs thus far have not been resummed using the renormalization group-based
techniques mentioned above. Their presence was first pointed out by Dasgupta and
Salam in \cite{Dasgupta:2001sh,Dasgupta:2002dc}. They arise in observables
probing soft radiation only in a part of phase space (type 1) or in sharply
divided parts of phase space with different scales (type 2).  An example of the
first type was given in \cite{Dasgupta:2001sh}, which considered the cumulative
\emph{single}-hemisphere invariant mass distribution $\Sigma(\rho_R)$, related
to our dijet invariant mass distribution \eq{dijetcsfactorized} by
\be
\label{singlemass}
\Sigma(\rho_R) = 
\int_0^\infty dm_1^2
\int_0^{Q^2\rho_R} dm_2^2\: \frac{d\sigma}{dm_1^2 dm_2^2}\,.
\ee
Dasgupta and Salam noted (in our language\footnote{In the original language of
  \cite{Dasgupta:2001sh}, the resummed prediction was based on the
  Catani-Trentadue quark and gluon ``jet'' functions $\Sigma_{q,g}$ defined by
  \cite{Catani:1988tn,Catani:1989ne} for use in resumming logs in global two-jet
  event shape distributions.}) that a prediction for $\Sigma(\rho_R)$ based on
inserting the dijet factorization theorem \eq{dijetcsfactorized} in the relation
\eq{singlemass} would not resum all logs of $\rho_R$ from renormalization
group evolution. Resummed results therefore need to be
supplemented by a factor containing the NGLs,
\begin{align}
\label{DSNGL}
 \Sigma(\rho_R) = \Sigma^{\rm dijet}_{\text{resum}}(\rho_R) \bigg[ 1 -
 \frac{\alpha_s^2 C_F C_A}{(2\pi)^2} \frac{\pi^2}{3}
 \ln^2\rho_R + \cdots \bigg] + \Sigma^{\text{p.c.}}(\rho_R)\,.
\end{align}
They identified the physical source of the additional logarithms in brackets as
soft gluons being emitted into opposite hemispheres, with only those in one
hemisphere being probed with the event shape $\rho_R$. The additional term
$\Sigma^{\text{p.c.}}(\rho_R)$ is generated by the terms in the ellipses in
\eq{dijetcsfactorized} that are power suppressed when $m_{1,2}^2/Q^2\ll 1$, but
contribute at leading order in \eq{singlemass} because $m_1^2$ is integrated
over all values. The leading double log comes from a light gluon jet in the
$m_2$ hemisphere recoiling against a hard $q\bar q$ pair in the $m_1$
hemisphere. Explicitly \cite{Dasgupta:2001sh,Burby:2001uz},
\be
\label{eq:hardlogs}
\Sigma^{\text{p.c.}}(\rho_R) =- \Big(\frac{\alpha_s}{2\pi}\Bigr)^2 
 \frac{C_F C_A}{2} \bigg[2\ln^2 2 - \frac{5}{4}\ln 3 + 4\Li_2\Big(-\frac{1}{2}\Bigr) + \frac{\pi^2}{3} - \frac{1}{6}\bigg] \ln^2\rho_R + \cdots\,,
\ee
where the ellipses denote subleading logarithms and higher-order terms in $\as$.
The logarithm shown in \eq{eq:hardlogs} is not non-global in origin and begins a
series of logs that can be resummed by standard techniques at the leading-log
level. Additional soft gluon emissions from the hard partons into opposite
hemispheres can generate another series of non-global logarithms that should be
included in the last term of \eq{DSNGL}, but they begin to contribute at yet
higher order in $\as$, beyond the scope of this paper.

Our examination of NGLs will focus on probing soft radiation in separate regions
with different scales, by examining hemisphere masses with $m_1^2, m_2^2 \ll
Q^2$ using the dijet invariant mass distribution in \eq{doublemass}. In this
regime we need only consider soft emissions from hard $q$ and $\bar q$ partons
in opposite hemispheres.  By simultaneously considering $m_1^2$ and $m_2^2$ the
two sources of large log contributions in \eq{DSNGL} can be cleanly
distinguished at $\cO(\as^2)$, since they arise from different parametric
regimes, namely $m_2^2 \ll m_1^2\ll Q^2$ and $m_2^2 \ll m_1^2\sim Q^2$
respectively.  With $m_2^2 \ll m_1^2\ll Q^2$ we have an NGL of type 2, from
probing soft radiation in separate regions of phase space with different
scales~\cite{Dasgupta:2002dc}.  \eq{singlemass} suggests that the NGLs of type 1
and type 2 are in fact related, with the double log of $\rho_R$ in \eq{DSNGL}
found in \cite{Dasgupta:2001sh} arising from a double non-global log of
$l_1/l_2$ in the fixed order $\mathcal{O}(\as^2)$ part of the soft function
$S(l_1,l_2,\mu)$ in \eq{dijetcsfactorized}. We will use our results to verify
this statement.

In this paper, we calculate the full structure of the dijet soft function
$S(l_1,l_2,\mu)$ at $\mathcal{O}(\as^2)$ analytically, and also explore the
relation to NGLs, which first arise at this order. We will learn that NGLs of a
given observable can be understood as logs of ratios of multiple soft scales
that are left over in the perturbative expansion of the relevant soft function
even after a standard factorization theorem for the observable has been established.
This perspective allows one to calculate systematically not only the leading
non-global logarithms that have been identified in \cite{Dasgupta:2001sh,Dasgupta:2002dc}, but
also to calculate straightforwardly the full set of other non-global structures
that actually arise.

We will find it simpler for many purposes to work with the Fourier transform of
$S(\la,\lb,\mu)$ to position space.  The cross section transforms as
\be
\label{positionsigma}
\widetilde \sigma(y_1,y_2) = \int_{-\infty}^\infty dm_1^2 dm_2^2\:
  e^{-i m_1^2 y_1} e^{-i m_2^2 y_2}\,\frac{d\sigma}{dm_1^2 dm_2^2}\,.
\ee
In position space, the convolution in \eq{dijetcsfactorized} becomes a product,
\be
\label{positionfactorization}
\widetilde \sigma(y_1,y_2) = \sigma_0 H_2(Q;\mu)\, \widetilde
J_\bn(y_1,\mu)\, \widetilde J_n(y_2,\mu)\, \widetilde S(Q y_1,Q y_2,\mu)\,.
\ee
In position space, the renormalization group evolution of each factor becomes
particularly simple as they each renormalize multiplicatively, as opposed to
convolutions in momentum space. In fact, RGEs in momentum space are most easily
solved by first going through position space
\cite{Korchemsky:1993uz,Balzereit:1998yf}, and there are several formalisms in use  to do this to arbitrarily high orders as long as the anomalous
dimensions are known (using derivatives on the evolution
kernels~\cite{Becher:2006nr}, converting analytically back to momentum
space~\cite{Ligeti:2008ac}, or even transforming numerically). For example, the
 formalism of \cite{Becher:2006nr} for summing logarithms in momentum
space relies on the form of the position (or Laplace)-space soft function using derivatives as arguments (the
``associated jet/soft functions'' in \cite{Becher:2006nr}).  Thus the position-space soft function $\widetilde S(x_1,x_2,\mu)$ is a crucial ingredient in
resummation of logs in practically all dijet observables.

We will present results for the dijet soft function to $\mathcal{O}(\as^2)$ both
in position space, $\widetilde S(x_1,x_2,\mu)$, and for the double cumulant in
momentum space, $\Scum(\lac,\lbc,\mu)$. We determine analytically for the first
time the full set of functions of $x_1/x_2$ or $\lac/\lbc$ that appear in
$\widetilde S$ and $\Scum$ at this order.\footnote{There is a constant term in
  $\widetilde S(x_1,x_2,\mu)$, or coefficient of $\delta(l_1)\delta(l_2)$ in
  $S(l_1,l_2,\mu)$, that we do not calculate, but which can be deduced from the
  recent result for the thrust soft function by \cite{Monni}. See also the note added to the Conclusions.}

From our results, we learn:
\begin{itemize}

\item There is a non-global double logarithm with $C_F C_A$ color structure in
  $\Scum(\lac,\lbc,\mu)$ and $\widetilde S(x_1,x_2,\mu)$, which corresponds exactly to the
  non-global log first identified by Dasgupta and Salam in
  \cite{Dasgupta:2001sh} from the single-hemisphere mass
  distribution.

\item There is a single non-global logarithm in $\Scum(\lac,\lbc,\mu)$ and
  $\widetilde S(x_1,x_2,\mu)$ that appears when the ratio $\lac/\lbc$ or
  $\lbc/\lac$ becomes large, or when $x_1/x_2$ or $x_2/x_1$ becomes large. Both the
  color structures $C_F C_A$ and $C_F T_R n_f$ have this single log.

\item In addition, other non-logarithmic non-global structures arise in $\Scum$ and
  $\widetilde S$ at $\mathcal{O}(\as^2)$. These structures and the single log
  were not accounted for in previous conjectures about their form
  \cite{Hoang:2008fs}.
  
\end{itemize}
These results not only complete our knowledge of the soft function at $\cO(\as^2)$, together with known results for the hard and jet functions in \eqs{dijetcsfactorized}{positionfactorization}, they make possible resummation to N$^3$LL accuracy in  these doubly-differential cross sections when $\la/\lb\sim 1$ or $x_1/x_2\sim 1$ (when these ratios are large/small the NGLs must be resummed as well). In position space the complete analytic N$^3$LL result can immediately be obtained from our results through multiplicative RG evolution, while in momentum space convolutions between the evolution factors and fixed order functions must still be performed. This becomes  a more nontrivial exercise to perform in the presence of the non-logarithmic non-global structures found here.

We arrive at the above results as follows: 

In~\sec{sec:hemisphere}, we review
properties of the dijet soft function that are already known, and an existing
conjecture for the ${\cal O}(\alpha_s^2)$ parts of it that are (so far) unknown.
We also explore the relation to NGLs in detail. 
 
 In~\sec{sec:calculation}, we
give our final results for the calculation of the dijet soft functions in
position space, $\widetilde S(x_1,x_2,\mu)$, and the double cumulant projection
of the momentum space soft function, $\Scum (\la^c,\lb^c,\mu)$, to
$\mathcal{O}(\as^2)$. We organize contributions to the two-loop soft function
according to whether one or both hemispheres are populated by final-state
particles. Thanks to renormalization group invariance and known anomalous
dimensions of the two-loop soft function, we obtain all the new information we
need from contributions with two particles (gluons or quark-antiquark) in the
final state going into opposite hemispheres.  These contributions encode all the
non-global structure in the dijet soft function. In addition, we will give a
remarkably simple function that approximates very closely the total
non-logarithmic non-global terms in the double cumulant $\Scum$.
 
In~\sec{sec:projections}, we discuss several projections of our soft functions
$S(\la,\lb,\mu)$ and $\widetilde S(x_1,x_2,\mu)$, including how we obtain
analytic results for the double cumulant, $\Scum(\la^c,\lb^c,\mu)$, and the
asymmetric thrust and heavy jet-mass event shapes of Ref.~\cite{Hoang:2008fs}.
The latter include the standard thrust and heavy jet-mass projections as special
cases, and we compare our analytic results to EVENT2, providing a strong
consistency check.  

In~\sec{sec:event2} we carry out another consistency check
by comparing the dijet factorization theorem including $\Scum(\la^c,\lb^c,\mu)$,
expanded to ${\cal O}(\alpha_s^2)$, to the two-dimensional double cumulant
distribution from EVENT2. We demonstrate that when our results are included in
the factorization theorem in \eq{dijetcsfactorized} at ${\cal O}(\alpha_s^2)$
that the remaining terms are truly power suppressed. In this section we also
study the region $m_2^{c2}\ll m_1^{c2}\sim Q^2$, and calculate the additional
global double log of $m_2^{c2}/Q^2$ which arises in the double cumulant
\eq{dijetcumfactorized} when $m_1^{c2}$ enters the hard regime. These additional
logs have the same origin as those computed in Ref.~\cite{Burby:2001uz}, and are
the origin of the logs of $\rho_R$ in the $\Sigma^{\rm p.c.}$ term of
\eq{DSNGL}. By projecting our double cumulant onto the single cumulant in $\rho_R$,
we reproduce the value of the double log given in \eq{eq:hardlogs}.  

In Section~\ref{sec:conclusions}, we conclude.

In Appendix~\ref{appx:anomdims}, we record the known anomalous dimensions of the
dijet soft function to two-loop order, which we need to assemble our result for
the full soft function.  

In Appendix~\ref{appx:twoloopcalc}, we provide details
of the calculation of the various parts of the soft function given in
Sec.~\ref{sec:calculation}, organized by color factors and Feynman diagram
topologies. We give results in both momentum and position space and provide
details of the translation between the two.

%%%%%%%%%%%%%%%%%%%%%%%%%%%%%%%%%%%%%%%%%%%%%%%%%%%%%%%%%%%%%%%%%%%%%%%%%%%%%%%%%%%%%%%%%%%%%%%%%%%%%%%%%
\section{The Dijet Hemisphere Soft Function}
\label{sec:hemisphere}
%%%%%%%%%%%%%%%%%%%%%%%%%%%%%%%%%%%%%%%%%%%%%%%%%%%%%%%%%%%%%%%%%%%%%%%%%%%%%%%%%%%%%%%%%%%%%%%%%%%%%%%%%

\subsection{Definitions}

The dijet hemisphere soft function is defined as
\be
\label{hemisoftdef}
S(\la,\lb) = \frac{1}{N_c}\Tr\sum_{X_S} \abs{\bbra{X_S} T[ Y_n^\dag Y_\bn]
  \bket{0}} ^2 
  \delta\bigg(\la - \sum_{i\in L} \bn \cdot k_i\bigg) 
  \delta\bigg(\lb - \sum_{i\in R} n\cdot k_i\bigg) \,,
\ee
where the trace is over colors, $T$ denotes time ordering, $n^\mu = (1,0,0, 1)$ and $\bn^\mu=(1,0,0,-1)$ are light-cone vectors along the $\pm z$ directions, and $R,L$ specify the $\pm \hat z$ hemispheres which we refer to as right and left hemispheres. The Wilson lines $Y_{n,\bn}$ are exponentials of soft gluons,
\be
  Y_n^\dagger = P\exp\left[ig\int_0^\infty ds\, n\cdot  A_s(ns)\right]
   \,,\qquad
  Y_\bn = \overline P \exp\left[-ig\int_0^\infty ds\,
  \bn\cdot A_s(\bn s)\right]\,, 
\ee 
where $P$ denotes path ordering for color matrices and $\overline P$ denotes
anti-path ordering.  When deriving factorization theorems in SCET, soft Wilson
lines are generated for each jet direction from rotating the collinear fields in
the SCET Lagrangian to decouple soft gluon interactions with collinear quarks
and gluons~\cite{Bauer:2001yt}. Equation~(\ref{hemisoftdef}) is referred to as a
dijet soft function because $Y_n^\dagger$ and $Y_\bn$ Wilson lines appear from
the $n$-jet and $\bn$-jet, and is referred to as a hemisphere soft function
because 
it involves kinematic
variables $\la,\lb$ restricted to hemispheres.
The measurement can be represented by the action of an operator,
\be \label{measop}
  \widehat{\cM}(\la,\lb) = \delta(\la - \widehat P_L) 
  \delta(\lb - \widehat P_R) \quad , \quad \widehat{\cM}(\la,\lb)\ket{X} \equiv \cM(\la,\lb)\ket{X}\,,
\ee
where the projection operators acting on a final state give
\be
\widehat P_{L}\ket{X} = \bigg(\sum_{i\in L} \bar n \cdot k_i\bigg)\ket{X} \quad , \quad \widehat P_{R}\ket{X} = \bigg(\sum_{i\in R}  n \cdot k_i\bigg)\ket{X}\,,
\ee
and the measurement function $\cM(\la,\lb)$ is therefore given by
\be \label{measfn}
  \cM(\la,\lb) = \delta\bigg(\la - \sum_{i\in L} \bn \cdot k_i\bigg) 
  \delta\bigg(\lb - \sum_{i\in R} n\cdot k_i\bigg) \,.
\ee
It is possible to  construct the projection operators explicitly from the energy-momentum tensor, which then allows the sum over states $X_S$ in \eq{hemisoftdef} to be removed \cite{Bauer:2008dt}. 

A more symmetric expression for the dijet hemisphere soft function can be obtained by converting to the $\overline 3$ representation~\cite{Bauer:2003di}.
Using $T (Y_\bn)^T = \overline Y_\bn^\dagger = P \exp \big(i g \int_0^\infty ds
\, \bn\cdot \overline A_s(\bn s) \big)$ and $\overline T (Y_\bn^\dagger)^T =
\overline P \exp \big(-i g \int_0^\infty ds \, \bn\cdot \overline A_s(\bn s)
\big)$, where $\overline A_\mu = \overline T^A A_\mu^A = - (T^A)^T A_\mu^A$,
gives
\be 
\label{hemisoftdef2}
 S(\la,\lb) =  \frac{1}{N_c} \bbbra{0} \Tr 
\overline Y_\bn^T\, Y_n \, \widehat{\cM}(\la,\lb) \, 
  Y_n^\dag \, \overline Y_\bn^*
\bbket{0} \,.
\ee
Due to the simple product structure of the position-space factorization theorem
in Eq.~(\ref{positionfactorization}), it is often easier to work with the
Fourier transform of $S(\la,\lb)$,
\be
\label{positionsoft}
\widetilde S(x_1,x_2) = \int\! d\la d\lb\, e^{-i\la x_1} e^{-i\lb x_2} S(\la,\lb)\,.
\ee
The definitions in Eqs.~(\ref{hemisoftdef}) and (\ref{positionsoft}) imply that
the functions $S$ and $\widetilde S$ are symmetric under interchange of the two
hemispheres,
\be
 S(\la,\lb) = S(\lb,\la) \,, \qquad
 \widetilde S(x_1,x_2) = \widetilde S(x_2,x_1) \,.
\ee
Note that the function $S(\la,\lb)$ has naive mass dimension $-2$, whereas
$\widetilde S(x_1,x_2)$ is dimensionless.

\subsection{Renormalization Group and Exponentiation Constraints}
\label{ssec:RGconstraints}

Renormalization group invariance of the factorized hemisphere cross section
implies a factorized structure for the soft function renormalization and its
renormalization group evolution~\cite{Hoang:2007vb}. Defining counterterms $Z_S$
in the $\overline {\rm MS}$ scheme, the renormalized soft function which appears
in the factorization theorem in Eq.~(\ref{positionfactorization}) is
\be
  \widetilde S(x_1,x_2,\mu) 
   = \widetilde Z_S^{-1}(x_1,\mu) \widetilde Z_S^{-1}(x_2,\mu) \widetilde S(x_1,x_2)^{\rm bare} \,,
\ee
and its inverse Fourier transform to momentum space, $S(\la,\lb,\mu)$, appears
in the factorization theorem in Eq.~(\ref{dijetcsfactorized}). Since $\widetilde
S(x_1,x_2,\mu)$ is dimensionless it can only be a function of $\mu x_1$, $\mu
x_2$, $x_1/x_2$, and $\alpha_s(\mu)$, where the $\mu$ dependence is uniquely
determined by the renormalization group.  From now on we will drop the
supercript ``bare'' for bare objects, which are always denoted without
dependence on $\mu$, and we will include $\mu$ as an argument for renormalized
functions. The renormalization group equation (RGE) for $\widetilde S$ is
\be
\label{eq:softRGE}
\mu \frac{d}{d\mu}\ln \widetilde
 S(x_1,x_2,\mu) = \gamma_S(x_1,\mu) + \gamma_S(x_2,\mu) 
 \,,
\ee
where the anomalous dimension is given by
\be
\label{anomdimdef}
 \gamma_S(x,\mu) = - \widetilde Z_S^{-1}(x,\mu) \mu \frac{d}{d\mu} \widetilde Z_S(x,\mu) = -
 \Gamma_{\rm cusp}[\alpha_s] \ln(i e^{\gamma_E} x\,\mu) + \gamma_S[\alpha_s] \,.
\ee
The solution of this RGE allows $\widetilde S$ at $\mu$ to be expressed in
terms of $\widetilde S$ at $\mu_0$,
\be
\label{softevolution1}
\widetilde S(x_1,x_2,\mu) = U_S(x_1,\mu,\mu_0) U_S(x_2,\mu,\mu_0) \widetilde S(x_1,x_2,\mu_0)\,,
\ee
where 
\be
\label{evolutionkernel}
U_S(x,\mu,\mu_0) = e^{K(\Gamma_{\rm cusp},\gamma_S,\mu,\mu_0)}\: \big(i e^{\gamma_E} x
\mu_0\big)^{\omega(\Gamma_{\rm cusp},\gamma_S,\mu,\mu_0)} \,,
\ee
and the functions $K$ and $\omega$ are given in Appendix~\ref{appx:anomdims}.

Following Ref.~\cite{Hoang:2007vb} the $\mu$ dependence in \eq{softevolution1}
can be entirely organized into the evolution factors by writing
\be
\label{softevolution2}
\widetilde S(x_1,x_2,\mu) = U_S(x_1,\mu,\mu_{x_1}) U_S(x_2,\mu,\mu_{x_2})
e^{\widetilde T(x_1,x_2)}\,,
\ee
where $\mu_{x_i} = (i e^{\gamma_E} x_i)^{-1}$ and the exponent $\widetilde T(x_1,x_2)$ is independent of $\mu$. In Ref.~\cite{Hoang:2008fs} Hoang and Kluth used non-Abelian exponentiation~\cite{Gatheral:1983cz,Frenkel:1984pz} to show that the last factor exponentiates in this way.
 With $\mu_0$ set equal to $\mu_{x_1}$ or $\mu_{x_2}$ respectively in the two evolution factors, the second factor in \eq{evolutionkernel} dependent on $\omega$ becomes unity, and the soft function \eq{softevolution2} takes the simple form
\be
\label{softsimpleform}
\widetilde S(x_1,x_2,\mu) =  e^{K_1(x_1,\mu) + K_2(x_2,\mu)} e^{\widetilde T(x_1,x_2)}\,,
\ee
where $K_i(x_i,\mu) = K(\Gamma_{\text{cusp}},\gamma_S,\mu,\mu_{x_i})$. 

Hoang and Kluth \cite{Hoang:2008fs}  also pointed out other properties satisfied by the function $\widetilde T(x_1,x_2)$. In particular, it is symmetric, $\widetilde T(x_1,x_2)= \widetilde T(x_2,x_1)$, and there are constraints
on the color factors in $\widetilde T(x_1,x_2)$ at each order in $\alpha_s$. For
instance, in the Abelian limit with $n_f=0$ light quarks the function
$\widetilde T(x_1,x_2)$ is one-loop exact. At two loops the only color
structures in $\widetilde T(x_1,x_2)$ are $C_F C_A$ and $C_F T_R n_f$, where for
$\rm SU(3)_{color}$ the fundamental and adjoint Casimirs are $C_F=4/3$ and $C_A=3$, and
$T_R=1/2$.  To order $\as^2$ we can write
\be\label{Ttildex1x2}
\widetilde T(x_1,x_2) = \frac{\as(\mu_{x_1})}{4\pi} t_1 + \frac{\as(\mu_{x_2})}{4\pi} t_1 + 2\frac{\as^2}{(4\pi)^2}\, t_2(x_1/x_2)\,,
\ee
where the one-loop constant $t_1$ is 
\be
t_1 = -C_F\frac{\pi^2}{2}\,.
\ee
The arguments of the $\as^2$ can be chosen as $\as(\mu_{x_1})\as(\mu_{x_2})$ or
any other symmetric combination of $\mu_{x_{1,2}}$ (e.g.
$\sqrt{\mu_{x_1}\mu_{x_2}}$) to the order we are working. The dimensionless
two-loop function $t_2(x_1/x_2)$ is unknown, and its computation is one of the
main goals of this paper. From the symmetry $x_1\leftrightarrow x_2$ we have
$t_2(b)=t_2(1/b)$.

We will present our result for the position  space soft functions by writing \eq{softsimpleform} as
\be \label{Sxfinalform}
\widetilde S(x_1,x_2,\mu) = \widetilde R(x_1,x_2,\mu) + \widehat S (x_1,x_2)\,,
\ee
where all terms containing logs of $(\mu x_i)$ are grouped together into
$\widetilde R(x_1,x_2,\mu)$ and all other terms are separated into $\widehat
S(x_1,x_2)$.  All quantities appearing in $\widetilde R$ or $\widehat S$ will be
expanded in powers of $\as(\mu)$ [in contrast to $\as(\mu_{x_i})$ in
\eq{Ttildex1x2}].  So $\widetilde R(x_1,x_2,\mu)$ contains cross terms between
the one-loop anomalous dimensions in $K_{1,2}$ and the one-loop part of
$\widetilde T(x_1,x_2)$, as well as terms generated by the running of
$\as(\mu_{x_i})$ in \eq{Ttildex1x2}. Similarly we will present the result for
the double cumulant momentum space soft function $\Scum(\la^c,\lb^c,\mu)$ defined in
Eq.~(\ref{Scumdefn}), by writing
\be
 \Scum(\la^c,\lb^c,\mu) = {\cal R}_c(\la^c,\lb^c,\mu) + \widehat \Scum(\la^c,\lb^c)\,,
\ee
where ${\cal R}_c(\la^c,\lb^c,\mu)$ contains all the logarithmic plus function
distributions in $\mu/\la^c$ or $\mu/\lb^c$, and $\widehat \Scum(\la^c,\lb^c)$
everything else.\footnote{Due to some constant $\mu$-independent terms ${\cal
    R}$ and $\widehat\cS_c$ are not separately related to the inverse Fourier
  transform of $\widetilde R$ and $\widehat S$ in Eq.~(\ref{Sxfinalform}), but
  can be obtained from these transforms by inspecting the results.}

\subsection{Numerical Results for Thrust and Heavy Jet Mass Projections}
\label{ssec:numresults}

Results for single differential event shape cross sections will depend on
various projections of the function $t_2(b)$.  In the literature results for
these projections have been obtained with the EVENT2
generator~\cite{Catani:1996jh,Catani:1996vz}.  Thrust $T$ in $e^+e^-\to {\rm
  jets}$ is only sensitive to the symmetric combination of hemisphere
masses in the dijet limit, $\tau\equiv 1-T = (m_1^2+m_2^2)/Q^2 +\cdots$, where
the ellipses denote terms that are power corrections to the dijet factorization
theorem. The relevant projection from $t_2(b)$ for thrust is
\be \label{s2thrust}
  t_2(1) = C_F C_A \, s_2^{[C_F C_A]} + C_F T_R n_f\,  s_2^{[n_f]} \,,
\ee
and these $s_2^{[a]}$'s contributes to the $\alpha_s^2 \delta(\tau)$ term in
momentum space.  Numerical determinations using EVENT2 in the literature include
\begin{align}
 \text{Becher \& Schwartz~\cite{Becher:2008cf}} &:&  
  &s_2^{[C_F C_A]} = -30.0 \pm 0.5  \,, 
  & & s_2^{[n_f]} = 21.5 \pm 0.5 \,,
  \nonumber \\
\text{Hoang \& Kluth~\cite{Hoang:2008fs}} &:&  
  & s_2^{[C_F C_A]} = -29.4 \pm 1.1 \,, 
  & & s_2^{[n_f]} = 21.9 \pm 1.5 \,,
 \nonumber \\
\text{Chien \& Schwartz~\cite{Chien:2010kc}} &:&  
  & s_2^{[C_F C_A]} = -28.9 \,, 
  & & s_2^{[n_f]} = 21.7  \,,
 \nonumber \\
\text{AHMSS~\cite{AHMSS}} &:&  
  & s_2^{[C_F C_A]} = -28.33 \pm 0.11  \,, 
  & & s_2^{[n_f]} = 21.82 {}^{+0.02}_{-0.11} \,.
\end{align}
Recently the two thrust constants have been computed directly in
Ref.~\cite{Monni} [MGL], giving
\begin{align} \label{s2Monni}
 & s_2^{[C_F C_A]} = -28.242 \pm 0.003\,, 
 & & s_2^{[n_f]} = 21.692 \pm 0.003 \,.
\end{align}
For our analysis we will treat these two constants as known
quantities.\footnote{For further discussion on analytic results for these
  constants see the Note Added at the end of \sec{sec:conclusions}.}

Another event shape of interest is the heavy jet mass, $\rho_H\equiv {\rm
  max}(m_1^2,m_2^2)/Q^2$. In this case the only projection of $t_2(b)$ that
appears at ${\cal O}(\alpha_s^2)$ is~\cite{Chien:2010kc}
\begin{align} \label{s2rho}
  \int_0^\pi \frac{d\theta}{\pi}\:  t_2(e^{i\theta}) 
   = C_F C_A \, s_{2\rho}^{[C_F C_A]} + C_F T_R n_f\,  s_{2\rho}^{[n_f]} \,. 
\end{align}
We will provide an alternate derivation of the integral moment appearing in
Eq.~(\ref{s2rho}) in~\sec{sec:projections}. Numerical EVENT2 results include
\begin{align}
   \text{Chien \& Schwartz~\cite{Chien:2010kc}}&:&  
  &s_{2\rho}^{[C_F C_A]} = -16.6  \,, 
  & & s_{2\rho}^{[n_f]} = 25.1 \,,
  \nonumber \\
\text{AHMSS~\cite{AHMSS}} &:&  
  &s_{2\rho}^{[C_F C_A]} = -16.79 \pm 0.46  \,, 
  & & s_{2\rho}^{[n_f]} = 25.15  {}^{+0.08}_{-0.05} \,.
\end{align}
(No errors are quoted for the numbers in \cite{Chien:2010kc}.)
One of the benefits of our analytic results for $t_2(b)$ is that one can compute
results for the constants appearing in various event shapes directly. In this
vein in \sec{ssec:hjmandthrust} we use our results to obtain analytic
results for the combination
\begin{align} \label{s2diff}
   \int_0^\pi \frac{d\theta}{\pi}\:  \big[ t_2(e^{i\theta})
  -t_2(1) \big]
   = \big( s_{2\rho}^{[C_FC_A]} - s_2^{[C_FC_A]}\big) 
 + \big( s_{2\rho}^{[n_f]} - s_2^{[n_f]}\big)\,.
\end{align}
Comparison of this combination with EVENT2 will provide a nontrivial
cross-check on our calculations and we quote results here for easy reference
\begin{align} \label{s2diffs}
 \text{CS~\cite{Chien:2010kc}} - \text{MGL~\cite{Monni}} &:&  
 & (s_{2\rho} - s_{2})^{[C_F C_A]} = 11.64   \,, 
  & & (s_{2\rho} - s_{2})^{[n_f]} =  3.41  \,.
\nn\\
 \text{AHMSS~\cite{AHMSS}} - \text{MGL~\cite{Monni}} &:&  
 & (s_{2\rho} - s_{2})^{[C_F C_A]} =  11.55 \pm 0.46   \,, 
  & & (s_{2\rho} - s_{2})^{[n_f]} =   3.46 {}^{+0.08}_{-0.05}  \,.
\end{align}

\subsection{Non-Global Logs from $S(x_1,x_2,\mu)$} \label{NGLinS}

In this section we make the connection between calculations of non-global
logarithms and logarithmic terms in the function $t_2(x_1/x_2)$.  The possible
form of $t_2(x_1/x_2)$ becomes simpler if we make a power expansion about the
non-global limit, $x_1\gg x_2$, which can at most yield logarithmic
singularities. The leading singularity is related to the non-global double
logarithm of Ref.~\cite{Dasgupta:2001sh}. This expansion therefore yields
\be \label{t2power}
 \lim_{x_1\gg x_2}\ t_2(x_1/x_2) 
  = s_2^{[2]} \,\ln^2\left(\frac{x_1}{x_2}\right) 
   + s_2^{[1]} \,\ln\left(\frac{x_1}{x_2}\right) 
   + s_2^{[0]} 
   + \ldots
\ee
where the terms in ellipses are suppressed by at least one power of $x_1/x_2$.
Since the limit is not symmetric in $x_1$ and $x_2$, the odd power
$\ln(x_1/x_2)$ can appear. In fact the coefficient $s_2^{[2]}$ can be extracted
from the non-global double logarithm computed in Ref.~\cite{Dasgupta:2001sh}. To
derive this relation note that the double cumulant invariant mass distribution
is
\begin{align}
 \frac{1}{\sigma_0} \Sigma(m_1^{c2},m_2^{c2})
  = 1 + \cdots + \frac{\alpha_s^2 s_2^{[2]}}{8\pi^2} \Big[
  \ln^2\!\frac{m_1^{c2}}{Q^2}  
  -2\ln\frac{m_1^{c2}}{Q^2} \ln\frac{m_2^{c2}}{Q^2} 
  + \ln^2\!\frac{m_2^{c2}}{Q^2} \Big] + \cdots \,, 
\end{align}
where only the term involving $s_2^{[2]}$ is shown. In the limit $Q^2 \gg m_2^{c2}
\gg m_1^{c2}$ the $\ln^2(m_1^{c2}/Q^2)$ term is leading, and must correspond with the
double logarithm of Ref.~\cite{Dasgupta:2001sh} since it was derived in
precisely this limit. The Dasgupta and Salam calculation therefore gives
$s_2^{[2]}=-2 C_F C_A \pi^2/3$, or for the coefficients of the two possible
color structures
\begin{align} \label{DSs22}
 s_2^{[2][C_F C_A]} & = - \frac{2\pi^2}{3} = -6.580 \,,
 & s_2^{[2][n_f]} & = 0 \,.
\end{align}

\subsection{Hoang-Kluth Ansatz}

Hoang and Kluth~\cite{Hoang:2008fs} argued that $t_2(x_1/x_2)$ should only
contain powers of $\ln(x_1/x_2)$ based on the expectation that event shape
distributions should contain only delta functions and plus distributions in the
dijet limit. The log powers must then be even powers to satisfy the
$x_1\leftrightarrow x_2$ symmetry and known LL results constrain the largest
power to $\ln^2(x_1/x_2)$ at ${\cal O}(\alpha_s^2)$. This led
Ref.~\cite{Hoang:2008fs} to make the ansatz
 \be
\label{HKansatz}
t_2^{\rm ansatz}(x_1/x_2) =  s_2^{[2]}\,\ln^2\left(\frac{x_1}{x_2}\right) + s_2\,, 
\ee
with two constants, $s_2$ for thrust (given by $t_2(1)$ and hence by
$s_2^{[C_FC_A]}$ and $s_2^{[n_f]}$ from Eq.~(\ref{s2thrust})), and $s_2^{[2]}$
for the double log coefficient. With this form the constant $s_2^{[2]}$ can be
determined by knowing the ${\cal O}(\alpha_s^2)$ $\delta$-function constants for
thrust and the heavy jet mass, via $s_2^{[2]}= 3(s_2-s_{2\rho})/\pi^2$.  The
numerical results in Eq.~(\ref{s2diffs}) give $s_2^{[2][C_F C_A]}=-3.5\pm 0.1$
and $s_2^{[2][n_f]}=-1.02\pm 0.04$.  These results differ from those in
Eq.~(\ref{DSs22}), so either the ansatz is incomplete or the calculation of
Ref.~\cite{Dasgupta:2001sh} is missing a source of double logarithms. Using our
full computation of $t_2(x_1/x_2)$ given in Sec.~\ref{sec:calculation} we will show that it is the ansatz in
Eq.~(\ref{HKansatz}) that is incomplete. This implies that non-logarithmic terms
contribute to the integral in Eq.~(\ref{s2diff}).  Expanding our result for  $t_2(x_1/x_2)$
for $x_1\gg x_2$ we will exactly reproduce the result for $s_2^{[2]}$ in
Eq.~(\ref{DSs22}), and derive non-zero results for the coefficients $s_2^{[1]}$
and $s_2^{[0]}$.

%%%%%%%%%%%%%%%%%%%%%%%%%%%%%%%%%%%%%%%%%%%%%%%%%%%%%%%%%%%%%%%%%%%%%%%%%%%%%%%%%%%%%%%%%%%%%%%%%%%%%%%%%
\section{Calculation of the Dijet Hemisphere Soft Function}
\label{sec:calculation}
%%%%%%%%%%%%%%%%%%%%%%%%%%%%%%%%%%%%%%%%%%%%%%%%%%%%%%%%%%%%%%%%%%%%%%%%%%%%%%%%%%%%%%%%%%%%%%%%%%%%%%%%%

In this section we calculate the dijet hemisphere soft function in both position
space, $\widetilde S(x_1,x_2,\mu)$, and momentum space, $S(\la,\lb,\mu)$, to
$\cO(\as^2)$. To make the $\alpha_s$ expansion we write
\begin{align}
 \widetilde S(x_1,x_2,\mu) &= 1 + \widetilde S_1(x_1,x_2,\mu) 
  + \widetilde S_2(x_1,x_2,\mu) + {\cal  O}(\alpha_s^3)\,,
  \nn\\
  S(\la,\lb,\mu) &= \delta(\la)\delta(\lb) +  S_1(\la,\lb,\mu) 
  +  S_2(\la,\lb,\mu) + {\cal  O}(\alpha_s^3)\,,
\end{align}
where $\widetilde S_i$ and $S_i$ are the ${\cal O}(\alpha_s^i)$ terms. For
nonzero $\ell_i$ or $x_i$ the final results for each $\widetilde S_i$ and $S_i$
are IR finite (see~\cite{Sterman:1978bj}). We regulate the UV divergences in
$d=4-2\epsilon$ dimensions and renormalize with $\overline {\rm MS}$.  In
momentum space the renormalization occurs through a convolution, while it is a
product in position space:
\begin{align}
S(\la,\lb) &= \int d\la' d\lb' \, Z(\la - \la', \lb - \lb', \mu) S^{(1)}_{\rm ren} (\la',\lb',\mu) \,, \nn \\
\widetilde S(x_1, x_2) &= \widetilde Z(x_1,x_2,\mu) \widetilde S(x_1,x_2,\mu) \,.
\end{align}
The counterterms factor into single-variable pieces,
\begin{align} \label{Zprod}
 \widetilde Z(x_1,x_2,\mu) &= \widetilde Z(x_1,\mu) \widetilde Z(x_2,\mu)\,,
 & Z(\la,\lb,\mu) &=Z(\la,\mu) Z(\lb,\mu)\,,
\end{align}
and have similar $\alpha_s$ expansion formulae, 
\begin{align}
  \widetilde Z(x,\mu) &= 1 + \widetilde Z_1(x_1,\mu) + \widetilde Z_2(x_1,\mu)+ {\cal  O}(\alpha_s^3) \,,
  \nn\\
  Z(\ell,\mu) &= \delta(\ell) + Z_1(\ell,\mu) + Z_2(\ell,\mu)+ {\cal  O}(\alpha_s^3)
  \,.
\end{align}

The calculations can be organized by the number of cut propagators,
\begin{align}
\label{origeq:cutdelta}
  \cC(k) = 2\pi\, \delta(k^2)\, \theta(k^0) \,.
\end{align}
At ${\cal O}(\alpha_s)$ we have zero or one $\cC(k)$ for the virtual and real
emission graphs respectively. The ${\cal O}(\alpha_s)$ matrix elements are
straightforward to calculate and we give the results in~\sec{sec:oneloop}. At
${\cal O}(\alpha_s^2)$ we have 0, 1, or 2 factors of $\cC$. The general
structure of the calculation is explored in~\sec{sec:twoloopsetup},
where we show that it suffices to compute the double cut graphs since the
remaining terms are determined entirely by the renormalization group and the
constant $s_2$.  Final results for $S(\la,\lb,\mu)$ and $\widetilde S(x_1,x_2,\mu)$ at
${\cal O}(\alpha_s^2)$ are given in~\sec{sec:twoloop}, while the details
of the calculation are described extensively in Appendix~\ref{appx:twoloopcalc}.

For easy reference we also record here the form of the hemisphere measurement
functions in momentum and position space, ${\cal M}^{[j]}_{\{k\}}(\la,\lb)$ and ${\cal
  M}^{[j]}_{\{k\}}(x_1,x_2)$ for $j=0,1,2$ final state particles with momenta $k_i$.  It
is the measurement functions that introduce the $\la,\lb$ or $x_{1,2}$
dependence into the soft function. With no final-state particle we have a tree-level or purely virtual contribution and Eq.~(\ref{measfn}) gives
\begin{align}
   {\cal M}^{[0]}(\la,\lb) &= \delta(\la) \delta(\lb) \,, 
   & {\cal M}^{[0]}(x_1,x_2) & =1 \,.
\end{align}
For one final-state particle of momentum $k$ we have in momentum space
\begin{align} \label{M1lalb}
  {\cal M}^{[1]}_k(\la,\lb) &= \delta(\la - k^-)\delta(\lb) \theta(k^+ -
  k^-) + \delta(\lb - k^+)\delta(\la) \theta(k^- - k^+) \nn\\
  & \equiv {\cal M}^{[L]}_k(\la)\delta(\lb) +
   \delta(\la) {\cal M}^{[R]}_k(\lb) \,,
\end{align}
where $k^+ \equiv n\cdot k$ and $k^-\equiv \bn\cdot k$,
and in position space
\begin{align} \label{M1x1x2}
  {\cal M}^{[1]}_k(x_1,x_2) &= e^{-i x_1 k^-} \theta(k^+ -
  k^-) + e^{-i x_2 k^+} \theta(k^- - k^+) \nn\\
  & \equiv {\cal M}^{[L]}_k(x_1) + {\cal M}^{[R]}_k(x_2) \,.
\end{align}
Eqs.~(\ref{M1lalb}) and (\ref{M1x1x2}) show that the measurement divides into
two terms, the contribution when the parton is in the left hemisphere plus the
contribution when the parton is in the right hemisphere.

For two final-state particles of momentum $k_1$ and $k_2$ we have contributions
from both particles in the left hemisphere, both in the right hemisphere, or one
in each hemisphere
\begin{align} \label{M2}
  {\cal M}^{[2]}_{k_1,k_2}(\la,\lb) &\equiv {\cal M}^{[LL]}_{k_1,k_2}(\la)
  \delta(\lb) + \delta(\la)  {\cal M}^{[RR]}_{k_1,k_2}(\lb)
  + {\cal M}^{[LR]}_{k_1,k_2}(\la,\lb) \,, 
  \nn\\
  {\cal M}^{[2]}_{k_1,k_2}(x_1,x_2) &\equiv {\cal M}^{[LL]}_{k_1,k_2}(x_1)
  +  {\cal M}^{[RR]}_{k_1,k_2}(x_2)
  + {\cal M}^{[LR]}_{k_1,k_2}(x_1,x_2) \,.
\end{align}
For two particles in the left hemisphere the functions are
\begin{align}
 {\cal M}^{[LL]}_{k_1,k_2}(\la) &\equiv
    \delta(\la-k_1^--k_2^-)\theta(k_1^+-k_1^-)\theta(k_2^+-k_2^-) \,,
   \nn\\
  {\cal M}^{[LL]}_{k_1,k_2}(x_1) &\equiv 
    e^{-i x_1(k_1^-+k_2^-)} \theta(k_1^+-k_1^-)\theta(k_2^+-k_2^-) \,,
\end{align}
and there is a trivial dependence on $\lb$ or $x_2$.
Similarly for two particles in the right hemisphere we have
\begin{align}
 {\cal M}^{[RR]}_{k_1,k_2}(\lb) &\equiv
    \delta(\lb-k_1^+-k_2^+)\theta(k_1^--k_1^+)\theta(k_2^--k_2^+) \,,
    \nn\\
 {\cal M}^{[RR]}_{k_1,k_2}(x_2) &\equiv 
    e^{-i x_2(k_1^++k_2^+)} \theta(k_1^--k_1^+)\theta(k_2^--k_2^+)\,,
\end{align}
with a trivial dependence on $\la$ or $x_1$.
For one particle in each hemisphere we have
\begin{align} \label{eq:CLR}
 {\cal M}^{[LR]}_{k_1,k_2}(\la,\lb) \equiv& \:\delta(\la - \kam) \delta(\lb - \kbp) 
 \theta(\kap - \kam) \theta(\kbm - \kbp)  \nn\\
  & + \delta(\la - \kbm)\delta(\lb - \kap) \theta(\kam - \kap) \theta(\kbp - \kbm)
  \,,\nn\\
  {\cal M}^{[LR]}_{k_1,k_2}(x_1,x_2) \equiv& \:
    e^{-ix_1\kam - ix_2 \kbp} \theta(\kap - \kam)\theta(\kbm - \kbp)
  \nn\\
  & + e^{-ix_1\kbm - ix_2 \kap}\theta(\kam - \kap) \theta(\kbp - \kbm) \,.
\end{align}
Diagrams involving ${\cal M}^{[LR]}$ are the only ones that can simultaneously
depend on $\la$ and $\lb$ (or $x_1$ and $x_2$) and hence on $\la/\lb$ (or $x_1/x_2$).  We will exploit this further
in \sec{sec:twoloopsetup}.  We will label the different graphs by the locations
of the cut partons.  We call the diagrams where the final-state partons are all
in the left hemisphere or all in the right hemisphere the \textit{same
  hemisphere} terms, and we call diagrams with one parton in the left hemisphere
and one parton in the right hemisphere the \textit{opposite hemisphere} terms.

Finally we summarize some notation that will be used extensively below. In
momentum space the logarithms are distributions
\begin{align} 
 {\cal L}_k(t) = \Big[\frac{\ln^k t}{t} \Big]_+ \,,
\end{align}
while for 
 cumulant momentum space the logarithms are 
\begin{align} \label{L1L2c} 
 \Lac \equiv \ln(\la^c/\mu )\,, \qquad
 \Lbc \equiv \ln(\lb^c/\mu ) \,.
\end{align}
In position space we denote the $\mu$ dependent logarithms as
\be \label{L1L2x}
 \Lax \equiv \ln(ie^{\gamma_E}x_{1} \mu )\,, \qquad
 \Lbx \equiv \ln(ie^{\gamma_E}x_{2} \mu ) \,.
\ee
The
non-global structure of the soft function will appear through the variables
\begin{align}
 r \equiv \frac{\ell_2}{\ell_1} \,,
  \qquad \text{or} \qquad   b \equiv \frac{x_1}{x_2} \,,
   \qquad \text{or} \qquad   a \equiv \frac{\lb^c}{\la^c}\,.
\end{align}

%%%%%%%%%%%%%%%%%%%%%%%%%%%%%%%%%%%%%%%%%%%%%%%%%%%%%%%%%%%%%%%%%%%%%%%%%%%%%%%%%%%%%%%%%%%%%%%%%%%%%%%%%
\subsection{$\cO(\as)$ Results}
\label{sec:oneloop}
%%%%%%%%%%%%%%%%%%%%%%%%%%%%%%%%%%%%%%%%%%%%%%%%%%%%%%%%%%%%%%%%%%%%%%%%%%%%%%%%%%%%%%%%%%%%%%%%%%%%%%%%%

The $\cO(\as)$ diagrams contribute to the $\cO(\as^2)$ terms, so we give them
explicitly.  Here the soft gluon can either be real or virtual.  Purely virtual
diagrams (no partons cut) in the soft function are scaleless in pure dimensional
regularization, so we do not need to compute them.  These
graphs convert all $1/\epsilon_{\rm IR}$'s in the real emission result into
$1/\epsilon_{\rm UV}$'s. A demonstration of this which
applies to our calculation here was given in Refs.~\cite{Hornig:2009vb,Hornig:2009kv}.  The measurement function $\cM_k^{(1)}(\ell_1,\ell_2)$
in~Eq.~(\ref{M1lalb}) divides the calculation into two terms, one coming from
the gluon in the left hemisphere and one coming from the gluon in the right
hemisphere.

The bare one-loop soft function in momentum space and position space in
dimensional regularization are
\begin{align}
S_1(\la,\lb) &= 4g^2 C_F \left(\frac{e^{\gamma_E}\mu^2}{4\pi}\right)^{\e} 
\int \frac{d^D k}{(2\pi)^D} \frac{1}{k^+ k^-} \, \cC(k) \cM^{(1)}_k (\la,\lb)
  \nn\\
 &= \frac{\as(\mu)C_F}{\pi} \frac{(e^{\gamma_E} \mu^2)^{\e}}{\Gamma(1-\e)} \frac{1}{\e} \left[ \la^{-1-2\e} \delta(\lb) + \lb^{-1-2\e} \delta(\la) \right] \,,
\nn\\
\widetilde S_1(x_1,x_2) &= \frac{\as(\mu)C_F}{\pi} \frac{(e^{-\gamma_E})^{\e}}{\Gamma(1-\e)} \frac{\Gamma(-2\e)}{\e} \big[ (ie^{\gamma_E} \mu x_1)^{2\e} + (ie^{\gamma_E} \mu x_2)^{2\e} \big] \,.
\end{align}
At one-loop $Z_1(\la,\lb,\mu)=Z_1(\la,\mu)\delta(\lb) +\delta(\la) Z_1(\lb,\mu)$
and $\widetilde Z_1(x_1,x_2,\mu) = \widetilde Z_1(x_1,\mu) + \widetilde Z_1(x_2,\mu)$ and the momentum and
position space counterterms are
\begin{align} \label{Z1}
Z_1(\la,\mu) &= \frac{\as(\mu) C_F}{\pi} \left\{ -\frac{1}{2\e^2}\delta(\la)
 + \frac{1}{\e}  \frac{1}{\mu}\cL_0 \Big(\frac{\la}{\mu} \Big) \right\} 
  \,, \nn \\
\widetilde Z_1(x_1,\mu) &= \frac{\as(\mu)C_F}{\pi}\left\{-\frac{1}{2\e^2} -
    \frac{1}{\e} \Lax  \right\} \,,
\end{align}
which will feed into the $\mu$-dependent two-loop calculation. The remainder is
the one-loop renormalized soft function in momentum and position space
\begin{align}
S_1(\la,\lb,\mu) &= \frac{\as(\mu) C_F}{\pi} \left\{ -\frac{2}{\mu}
  \cL_1\left(\frac{\la}{\mu}\right)\delta(\lb) - \frac{2}{\mu}
  \cL_1\left(\frac{\lb}{\mu}\right)\delta(\la) + \frac{\pi^2}{12}
  \delta(\la)\delta(\lb) \right\} \,, 
 \nn\\
\widetilde S_1 (x_1,x_2,\mu) &= \frac{\as(\mu)C_F}{\pi}\left\{ -(\Lax^2 + \Lbx^2) - \frac{\pi^2}{4}\right\} \,.
\end{align}

%%%%%%%%%%%%%%%%%%%%%%%%%%%%%%%%%%%%%%%%%%%%%%%%%%%%%%%%%%%%%%%%%%%%%%%%%%%%%%%%%%%%%%%%%%%%%%%%%%%%%%%%%
\subsection{Structure of the $\cO(\as^2)$ Real and Virtual Terms}
\label{sec:twoloopsetup}
%%%%%%%%%%%%%%%%%%%%%%%%%%%%%%%%%%%%%%%%%%%%%%%%%%%%%%%%%%%%%%%%%%%%%%%%%%%%%%%%%%%%%%%%%%%%%%%%%%%%%%%%%

\begin{table}[t]
  \centering
\begin{tabular}{l c c c c c}
  \hline \hline
  \multicolumn{6}{l}{\hspace{1.cm}\includegraphics[width=0.93\textwidth]{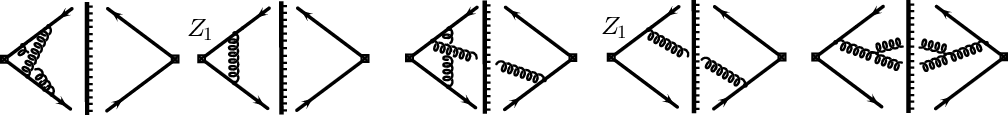}}
\\ 
Type:  &\hspace{0.3cm} $V_2$ 
       &\hspace{1.2cm} $V_1 Z_1$ 
       &\hspace{0.8cm} $V_1 R_1$ 
       &\hspace{0.8cm} $Z_1 R_1$ 
       &\hspace{0.5cm} $R_2$ \\ 
Color: &\hspace{0.3cm} all 
       &\hspace{1.2cm} $C_F^2$   
       &\hspace{0.8cm} $\{C_F^2,C_FC_A\}$ 
       &\hspace{0.8cm} $C_F^2$ 
       &\hspace{0.5cm} all \\
$x_1,x_2$: &\hspace{0.3cm} none
      &\hspace{1.2cm} single 
      &\hspace{0.8cm} single 
      &\hspace{0.8cm} both 
      &\hspace{0.5cm} both \\  
  \hline\hline
  \end{tabular}
  \caption{Examples of various types of loop graphs which enter at ${\cal
      O}(\alpha_s^2)$ and their general properties. 
    $V_2$ and $R_2$ are the ${\cal O}(\alpha_s^2)$ purely virtual and purely real graphs,
    $V_1R_1$ are the mixed virtual-real graphs, and $Z_1$ is the one-loop
    counterterm. For each class we indicate which of the three possible color structures $C_F^2,C_F C_A, C_F T_R n_F$ appear, and whether
    the graphs depend  simultaneously on $x_{1,2}$ (in position space), on only a single variable at a
    time, or neither variable. The two-loop counterterm $Z_2$ is discussed in
    the text.}
\label{tab:graphs}
\end{table}

The structure of the various types of $\cO(\as^2)$ diagrams for the calculation of
$\widetilde S(x_1,x_2)$ are shown in Table~\ref{tab:graphs}. In dimensional
regularization the bare two-loop virtual graphs and one-loop virtual graphs with
a counterterm are scaleless and need not be considered. The remaining terms can
all be divided into single hemisphere contributions that depend on a function of
$x_1$ plus the same function of $x_2$, and opposite hemisphere contributions
that depend simultaneously on $x_{1,2}$ and contribute non-global terms. The
bare and counterterm graphs can all be split into the form
\begin{align} \label{Ssplit} 
 \widetilde S(x_1,x_2) = 
   \big[ \widetilde S^{\rm same}(x_1)
    + \widetilde S^{\rm same}(x_2) \big]
    + \widetilde S^\opp(x_1,x_2) 
\end{align}
as follows. Consider the two-loop
counterterm $\widetilde Z_2(x_1,x_2,\mu)$, from  Eq.~(\ref{Zprod})
\begin{align} \label{Z2decom}
\widetilde  Z_2(x_1,x_2,\mu) =  \widetilde Z_2(x_1,\mu) + \widetilde Z_2(x_2,\mu) + \widetilde Z_1(x_1,\mu)  \widetilde Z_1(x_2,\mu) 
 \,,
\end{align}
which is the appropriate form for Eq.~(\ref{Ssplit}), and shows that only $\widetilde Z_1$
contributes to $\widetilde S^\opp$. Graphs that have only a single parton
crossing the cut involve $\cM_k^{[1]}(x_1,x_2)$ from Eq.~(\ref{M1x1x2}) and are
part of the single hemisphere terms. This includes the mixed virtual-real graphs
denoted by $V_1 R_1$ in Table~\ref{tab:graphs}. For the double cut real emission
graphs the measurement function $\cM^{[2]}_{k_1,k_2}(x_1,x_2)$ in Eq.~(\ref{M2})
splits the result into a sum of single hemisphere and opposite hemisphere terms,
\begin{align} \label{R2decom}
  R_2(x_1,x_2) = \big[ R_2^{\rm same}(x_1) +  R_2^{\rm same}(x_2) \big] + 
   R_2^{\opp}(x_1,x_2) \,,
\end{align}
where $ R_2^{\rm same}(x_1)$ involves $\cM^{[LL]}_{k_1,k_2}(x_1,x_2)$, $ R_2^{\rm same}(x_2)$ involves $\cM^{[RR]}_{k_1,k_2}(x_1,x_2)$, and
$ R_2^{\opp}(x_1,x_2)$ involves $\cM^{[LR]}_{k_1,k_2}(x_1,x_2)$. Using results
from~\sec{sec:oneloop} the $R_1\widetilde Z_1$ results can also be directly
manipulated into the required form in Eq.~(\ref{Ssplit}).

Next consider the color structures of various opposite hemisphere contributions.
From Eq.~(\ref{Z2decom}) and Table~\ref{tab:graphs} we see that for $\widetilde
S^\opp(x_1,x_2)$ both $\widetilde Z_2$ and $\widetilde Z_1 R_1$ only contribute to $C_F^2$. Hence
these terms are just part of the exponentiation of the one-loop result, and are
not needed for our computation of the unknown terms with color structures $C_F
C_A$ and $C_F T_R n_f$ which appear in $t_2(x_1/x_2)$ (see~\sec{ssec:RGconstraints}).  Thus the only term we need to consider is
the double cut real emission graphs $R_2^\opp$, and there are no UV divergences
or counterterms for the color structures of interest in our calculation.

Before proceeding to analyze results of our computation of $R_2^\opp(x_1,x_2)$,
we must address one important issue, namely that only the sum of $\widetilde S^\opp$
and $\widetilde S^{\rm same}$ terms in Eq.~(\ref{Ssplit}) is infrared finite. From
our computation discussed in~\sec{sec:twoloop} and
Appendix~\ref{appx:twoloopcalc} we find the infrared divergent terms in
dimensional regularization are
\begin{align} \label{R2oppIR}
 R_2^\opp(x_1,x_2) &= \frac{\alpha_s^2(\mu)}{(2\pi)^2} \bigg[  C_F
   C_A \Big(\frac{\pi^2}{6\epsilon^2}+ \frac{11\pi^2-3}{18\epsilon}\Big)
   + C_F T_R n_f \, \frac{3-2\pi^2}{9\epsilon} \bigg]
  \big( i x_1 i x_2 \mu^2 e^{2 \gamma_E} \big)^{2\epsilon} 
  \nn\\
  & \quad +\cdots \,.
\end{align}
Here the ellipses denote IR finite terms and terms with the $C_F^2$ color
structure. The divergences in Eq.~(\ref{R2oppIR}) cancel against same hemisphere
contributions, and hence determine the IR structure in $\widetilde S^{\rm
  same}$. In dimensional regularization the graphs contributing to $\widetilde
S^{\rm same}(x_1,\mu)$ give powers of $(i x_1\mu)^\epsilon$ determined by
dimensional analysis and whether the graph involves counterterms, times
functions of $\epsilon$. Since the same hemisphere graphs involving $Z_1$ do not
involve the color structures shown in Eq.~(\ref{R2oppIR}), the IR divergent
contributions which cancel those in Eq.~(\ref{R2oppIR}) can only come from
$R_2^{\rm same}$ and $V_1R_1$. They are therefore uniquely determined to be
\begin{align} \label{R2sameIR}
\widetilde S_2^{\rm same}(x_1,x_2) &= -\frac{\alpha_s^2(\mu)}{(2\pi)^2} \bigg[  C_F
   C_A \Big(\frac{\pi^2}{6\epsilon^2}+ \frac{11\pi^2-3}{18\epsilon}\Big)
   + C_F T_R n_f \,\frac{3-2\pi^2}{9\epsilon} \bigg] \nn\\
&\quad \times 
\frac{(ix_1\mu e^{\gamma_E})^{4\epsilon}+(ix_2\mu e^{\gamma_E})^{4\epsilon}}{2}
   +\cdots \,.
\end{align}
Adding and expanding Eqs.~(\ref{R2oppIR}) and (\ref{R2sameIR}) the
terms involving these IR divergences cancel, but leave a finite remainder
\begin{align} \label{R2fullIR}
R_2^\opp(x_1,x_2) + \widetilde S_2^{\rm same}(x_1,x_2) 
 &= -\frac{\alpha_s^2(\mu)}{4\pi^2} \, C_F C_A \frac{\pi^2}{3} 
 \ln^2\Big(\frac{x_1}{x_2}\Big) + \cdots \,.
\end{align}
Thus the $1/\epsilon$ terms in $R_2^\opp(x_1,x_2)$ with the form shown in Eq.~(\ref{R2oppIR})
can be dropped if we replace them by the double logarithm in
Eq.~(\ref{R2fullIR}). The calculation of the opposite hemisphere infrared finite
terms in the ellipses then proceeds without further complications.  

Our result in Eq.~(\ref{R2fullIR}) yields precisely the coefficient
$s_2^{[2]}=-2\pi^2/3$ of the non-global logarithm appearing in $t_2(x_1/x_2)$  as discussed in~\sec{NGLinS} and first computed in Ref.~\cite{Dasgupta:2001sh}. The
procedure discussed here to determine the non-global logarithm was first
presented at the SCET 2011 workshop~\cite{scetchris}, including the fact that it
can be computed by an effective theory that refactorizes scales in the soft
function, and has ultraviolet divergences with the same coefficient as the above
infrared divergences.

Accounting for the cancellation of IR divergences, the full renormalized soft
function can be arranged into what we will refer to as renormalized same
hemisphere contributions that depends on $\mu$ and one of the $x$ variables, and
a renormalized opposite hemisphere contribution that depends on $x_1/x_2$,
\begin{align} \label{Ssplitren} 
 \widetilde S_2(x_1,x_2,\mu) = 
   \big[ \widetilde S_2^{\rm same}(x_1,\mu)
    + \widetilde S_2^{\rm same}(x_2,\mu) \big]
    + \widetilde S_2^\opp(x_1/x_2,\alpha_s(\mu)) \,.
\end{align}
Here the double log in Eq.~(\ref{R2fullIR}) is included in $\widetilde
S_2^\opp(x_1/x_2,\alpha_s(\mu))$.  We present our full result for $\widetilde
S_2^{\rm opp}(x_1/x_2,\alpha_s(\mu))$ in the next section.  In dimensional
regularization the graphs contributing to $\widetilde S_2^{\rm same}(x_1,\mu)$
give powers of $(i x_1\mu)^\epsilon$ times functions of $\epsilon$, and hence
only depend on the $\mu$ dependent logs $\widetilde L_1$ in Eq.~(\ref{L1L2x}). At ${\cal
  O}(\alpha_s^2)$ the general result is
\begin{align} \label{Ssameform}
  \widetilde S_2^{\rm same}(x_1,\mu) = a_4 \Lax^4 + a_3 \Lax^3 
    + a_2 \Lax^2 + a_1 \Lax + a_0 \,. 
\end{align}
Since the logarithms from infrared divergences appear in $S^{\rm opp}$ all the
$\Lax$ terms in Eq.~(\ref{Ssameform}) are determined by the known anomalous
dimensions for the soft function, through the RGE solution discussed in~\sec{ssec:RGconstraints}. Hence the only thing not determined by the
above general considerations and our opposite hemisphere calculation of $R_2$
is the $C_F C_A$ and $C_FT_Rn_f$ color factor terms in the constant $a_0$. These
two terms contribute directly to the two thrust constants $s_2^{[a]}$ in
Eq.~(\ref{s2thrust}).  We will present our final result as an analytic
function of $\Lax$, $\Lbx$, and $b=x_1/x_2$, plus $s_2^{[C_FC_A]}$ and
$s_2^{[C_Fn_f]}$ whose numerical values are given in~\sec{ssec:numresults}.

%%%%%%%%%%%%%%%%%%%%%%%%%%%%%%%%%%%%%%%%%%%%%%%%%%%%%%%%%%%%%%%%%%%%%%%%%%%%%%%%%%%%%%%%%%%%%%%%%%%%%%%%%
\subsection{Final Dijet Hemisphere Soft Functions}
\label{sec:twoloop}
%%%%%%%%%%%%%%%%%%%%%%%%%%%%%%%%%%%%%%%%%%%%%%%%%%%%%%%%%%%%%%%%%%%%%%%%%%%%%%%%%%%%%%%%%%%%%%%%%%%%%%%%%

In~\sec{sec:twoloopsetup} we showed that up to a constant the entire $\cO(\as^2)$
dijet hemisphere soft function can be computed from the opposite hemisphere
double real emission graphs and the known $\cO(\as^2)$ anomalous dimensions. 
\begin{figure}[t]{
    \includegraphics[width=0.99\textwidth]{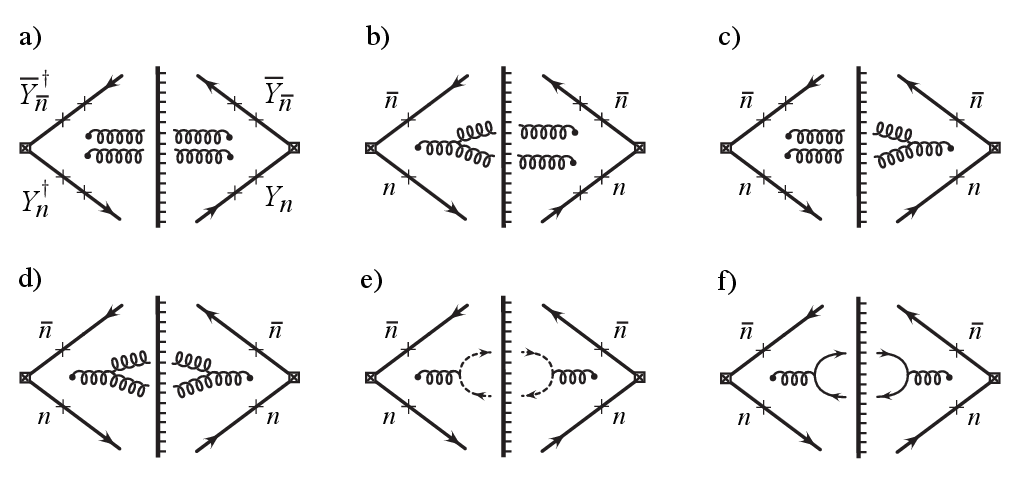} { \caption[1]{$\cO(\as^2)$
        opposite hemisphere diagrams.  The endpoints of the gluons can be
        attached to the points on the Wilson lines labeled by a `x' in any
        order.  Figure (a) gives the $\cI$ diagrams, (b) and (c) give the $\cT$
        diagrams, (d) gives the $\cG$ diagrams, (e) the $\cH$ diagrams with
        ghosts, and (f) the $\cQ$ diagrams with massless quarks.}
  \label{fig:diagrams}} }
\end{figure}
For this calculation there are five classes of diagrams shown
in~\fig{fig:diagrams}: independent emission ($\cI$) diagrams, diagrams with a
single three-gluon vertex ($\cT$), and vacuum polarization diagrams with a gluon
loop ($\cG$), quark loop ($\cQ$), or ghost loop ($\cH$).  Note that the $C_F^2$
color structure only appears in $\cI$, and $C_F T_R n_f$ only appears in $\cQ$,
whereas the $C_FC_A$ color structure appears in all classes except $\cQ$. For
any graph whose squared matrix element is $\cA_i$, the opposite hemisphere contribution is
\be 
 \int \meas{k_1} \meas{k_2} \cA_i(k_1,k_2) {\cal
  M}^{[LR]}_{k_1,k_2} \mathcal{C}(k_1)\mathcal{C}(k_2) \,,
\ee 
where $\cC(k)$ is the cut propagator, Eq.~(\ref{origeq:cutdelta}), $\cA_i=\{\cI,\cT,\cG,\cH,\cQ\}$ and the
measurement function ${\cal M}_{k_1,k_2}^{[LR]}$ is given in Eq.~(\ref{eq:CLR})
for both position space and momentum space.  We present details of their
calculation using Feynman gauge in both position and momentum space in
Appendix~\ref{appx:twoloopcalc}. The results appear in the following
equations:
\begin{align}
\cI &: \;\; \textrm{Eqs.([\ref{eq:Iposition}, \ref{eq:T+Iposition}], [\ref{eq:Imomentum} \& \ref{eq:T+Imomentum}]) }
\,, 
& \cT &: \;\;\textrm{Eqs.(\ref{eq:T+Iposition},  \ref{eq:T+Imomentum})}\,, 
& \cG &: \;\; \textrm{Eqs.(\ref{eq:G+Hposition},\ref{eq:G+Hmomentum})}\,,  
\nn \\
 \cH &: \;\; \textrm{Eqs.(\ref{eq:G+Hposition},\ref{eq:G+Hmomentum})}\,,  
& \cQ &: \;\; \textrm{Eqs.(\ref{eq:Qposition},\ref{eq:Qmomentum})}  \,.
\end{align}
We have explicitly checked that our final result is unchanged if the gluon
propagators in~\fig{fig:diagrams} are taken in a general covariant gauge.
The gauge parameter cancellation occurs individually for the $\cT$, $\cG+\cH$,
and $\cQ$ terms (and provides a non-trivial cross check on the relative overall
signs of $\cG$ and $\cH$).

Next we present final results for the renormalized soft function that includes
contributions from both the same hemisphere and opposite hemisphere terms, using
the approach described in~\sec{sec:twoloopsetup}. We first discuss
position space and then the double cumulant distribution in momentum space.
\eqs{finalposition}{finalcumulant} are the main results of this paper.

%%%%%%%%%%%%%%%%%%%%%%%%%%%%%%%%%%%%%
\subsubsection{Result in Position Space}

\begin{figure}[t!]{    
\begin{center}
    \includegraphics[width=.7\textwidth]{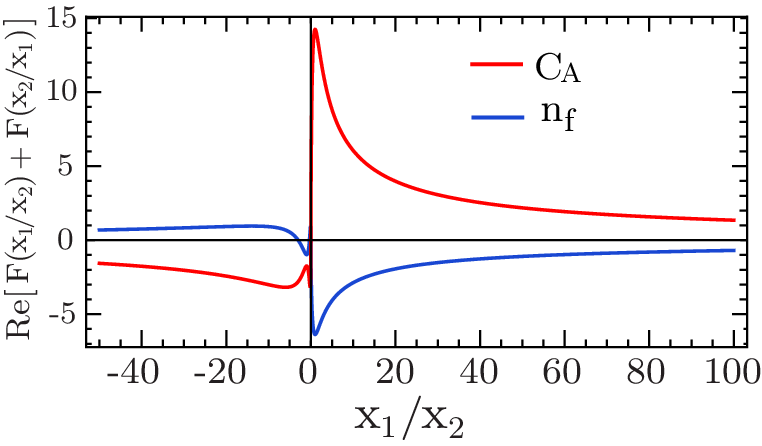} 
  \end{center} 
  \vspace{-2em}
  {\caption[1]{The non-logarithmic non-global function in position space. We
      show the functions ${\rm Re}[F(b)+F(1/b)]$, where $b=x_1/x_2$, that appear
      in the position-space soft function \eq{finalposition} for the color
      structures $C_F C_A$ (solid red) and $C_F T_R n_f$ (solid blue).}
  \label{fig:FQN}} }
\end{figure}

In position space we find
\begin{align} 
\label{finalposition}
 & \widetilde S(x_1,x_2,\mu) = 1 - \frac{\as(\mu)C_F}{4\pi} \pi^2 + \widetilde
 R(x_1,x_2,\mu) 
 + \frac{\as^2(\mu)}{4\pi^2}\bigg[ C_F^2 \frac{\pi^4}{8} + \frac{1}{2} t_2\Big(\frac{x_1}{x_2}\Bigr)
\bigg]\,, 
\end{align}
where
\begin{align}
\label{finalt2pos}
t_2\Big(\frac{x_1}{x_2}\Big) &=  
  -  C_F C_A \frac{2\pi^2}{3} \: \ln^2\Big(\frac{x_1}{x_2}\Big)  \\ 
 &\quad + 2 \ln\Big(\frac{x_1/x_2+x_2/x_1}{2}\Big)
   \Big( C_F C_A\frac{11\pi^2\minus 3\minus 18\zeta_3}{9}
   + C_F T_R n_f \frac{6-4\pi^2}{9} \Big)
 \nn \\
&\quad + 2C_F T_R n_f \Big[ F_Q\Big(\frac{x_1}{x_2}\Big) 
 + F_Q\Big(\frac{x_2}{x_1}\Big)
 -2F_Q(1) \Big]  \nn \\
 &\quad + 2C_F C_A \Big[
  F_N\Big(\frac{x_1}{x_2}\Big)  + F_N\Big(\frac{x_2}{x_1}\Big)-2F_N(1)\Big] 
+  C_F C_A s_2^{[C_F C_A]} 
  + C_F T_R n_f s_2^{[n_f]} \, , \nn
  \end{align}
  determining the non-global function appearing in \eq{Ttildex1x2}.
Here and throughout this paper $x_1$ and $x_2$ have a small imaginary
components, and should be regarded as $x_1 - i0^+$ and $x_2 - i0^+$.  Values for
the constants $s_2^{[a]}$ are given in \sec{ssec:numresults}. The functions $F_{Q.N}$
are
\begin{align} \label{FQFN}
  F_Q(b) &= \frac{2\ln b  }{3 (b\!-\!1)} \!-\! \frac{b\,\ln^2 b}{3(b\!-\!1)^2}
 \!-\!  \frac{3\minus 2\pi^2}{9}  \ln\!\Big(b\!+\!\frac{1}{b}\Bigr) 
  \plus \frac23 \ln^2b\:  \ln(1\!-\!b) \plus \frac83\ln b\, \Li_2 (b)  \!-\! 4 \Li_3(b) 
 \nn  , \\
  F_N(b) &=  -\frac{\pi^4}{36} \!-\! \frac{\ln b  }{3 (b\!-\!1)}
  + \frac{b\,\ln^2 b}{6(b\!-\!1)^2} 
  + \frac{3\minus 11\pi^2\!+\! 18\zeta_3}{18} \ln\Big(b\plus \frac{1}{b}\Bigr)
  \!-\! \frac{11}{6} \ln^2\! b\, \ln(1\minus b)
  \!+\! \frac{\ln^4 b}{24} \nn\\
 &\quad - \frac{\pi^2}{3}  \Li_2(1\minus b)+ \big[\!\Li_2(1\minus b)\!\big]^2 
  \!-\!\frac{22}{3} \ln b \: \Li_2(b) + 2\ln b \Li_3(1\minus b) +11\Li_3(b)  
,
\end{align}
and the formula for the $\mu$-dependent terms, $\widetilde R(x_1,x_2,\mu)$, will
be given below. The result in Eq.~(\ref{finalt2pos}) is written so that the
double and single logarithmic singularities for $b=x_1/x_2\to 0$ and $b\to\infty$
are separated out in the first two lines. The more complicated structures in
$F_Q(b)+F_Q(1/b)$ and $F_N(b)+F_N(1/b)$ are bounded on the real $b$ axis and have real parts going to zero for $b\to 0$ and $b\to\infty$. For negative $\Real b$, the functions also have an imaginary part.  The real parts  are illustrated in Fig.~\ref{fig:FQN}.

Taking the $x_1\gg x_2$ limit discussed in Eq.~(\ref{t2power}) yields terms
$s_2^{[i]} \ln^i(x_1/x_2)$ at ${\cal O}(\alpha_s^2)$ and our result in
Eq.~(\ref{finalt2pos}) determines the coefficients 
\begin{align} \label{s2s}
 s_2^{[2]} &= -\frac{2\pi^2}{3} C_F C_A \,,
 \quad
 s_2^{[1]}  = 2 \Big[ C_F C_A\frac{(11\pi^2\minus 3\minus 18\zeta_3)}{9}
   + C_F T_R n_f \Big(\frac{6-4\pi^2}{9}\Big) \Big] \,,
 \\
 s_2^{[0]} &= - s_2^{[1]}\, \ln 2 \minus  4 C_F C_A F_N(1)
 \minus  4 C_F T_R n_f  F_Q(1)
 + C_F C_A s_2^{[C_F C_A]} + C_F T_R n_f
 s_2^{[n_f]} , \nn
\end{align}
where 
\begin{subequations}
\begin{align}
2F_Q(1) &= \frac{2}{3} + \left(\frac{4\pi^2 }{9} - \frac{2}{3}\right)\ln 2 - 8\zeta_3 \,, \\
2F_N(1) &= -\frac{1}{3}-\frac{\pi^4}{18} + \left(\frac{1}{3} - \frac{11\pi^2}{9}  + 2\zeta_3\right)\ln 2 + 22\zeta_3 \,.
\end{align}
\end{subequations}
In the left panel of Fig.~\ref{fig:fFQNLog} we plot the double log (dotted),
single log (dashed), and non-log (solid) ${\cal O}(\alpha_s^2)$ non-global
contributions to $\widetilde{S}(x_1,x_2,\mu)$ as a function of $x_1/x_2$.
Results are shown separately for the $C_F C_A$ and $C_F T_R n_f$ color
structures with overall signs for the plot chosen so that the former are
negative and latter positive, which involves plotting the single NGL terms
$\propto\ln(b+1/b)$ with the opposite sign.  The solid lines are the
non-logarithmic functions $\bigl[F_{N,Q}(b)+F_{N,Q}(1/b)-2F_{N,Q}(1) +
\frac{1}{2}s_2^{[C_F C_A,n_f]}\bigr]$, the dashed lines are the single NGL
terms, and the dotted black line is the double NGL $\propto\ln^2 b$ (nonzero
only for $C_F C_A$), where $b=x_1/x_2$. For values in the vicinity of $x_1=x_2$
the contributions from all sources are comparable in size and the non-analytic
non-log dependences are important. As we already remarked, the $F$-dependent
terms exhibit a smooth bounded behavior.

Expanding Eq.~(\ref{softsimpleform}) and inserting the known anomalous
dimensions the $\widetilde{L}_{1,2}=\ln(ix_{1,2}e^{\gamma_E}\mu)$ dependent 
part of the soft function, $\widetilde R(x_1,x_2,\mu)$, is
\begin{align} \label{Rposition}
 \widetilde R(x_1,x_2,\mu) =  
&  - \frac{\as(\mu)C_F}{\pi} \left( \Lax^2 + \Lbx^2\right) 
  + \frac{\as^2(\mu)}{16\pi^2} 
   \Biggl\{C_F^2\left(8\Lax^4 + 8\Lbx^4 + 16 \Lax^2 \Lbx^2\right)  \\
 &+\left(- \frac{88}{9} C_F C_A + \frac{32}{9} C_F T_R n_f\right)(\Lax^3 + \Lbx^3)
 \nn \\
 &
 + \left[4\pi^2 C_F^2 -4 C_F C_A \left(\frac{67}{9}-\frac{\pi^2}{3}\right) +
   \frac{80}{9}C_F T_R n_f \right](\Lax^2 + \Lbx^2) \nn \\
 &+ \biggl[C_F C_A \left(-\frac{808}{27} - \frac{22\pi^2}{9} +
   28\zeta_3\right) 
+ C_F T_R n_f\left(\frac{224}{27} + \frac{8}{9}\pi^2\right)\biggr] (\Lax+\Lbx) \Biggr\}\,.\nn
\end{align}

\subsubsection{Result in Momentum Space}
\label{sssec:MomSpaceCumulant}

\begin{figure}[t!]{ 
 \includegraphics[width=0.48\textwidth]{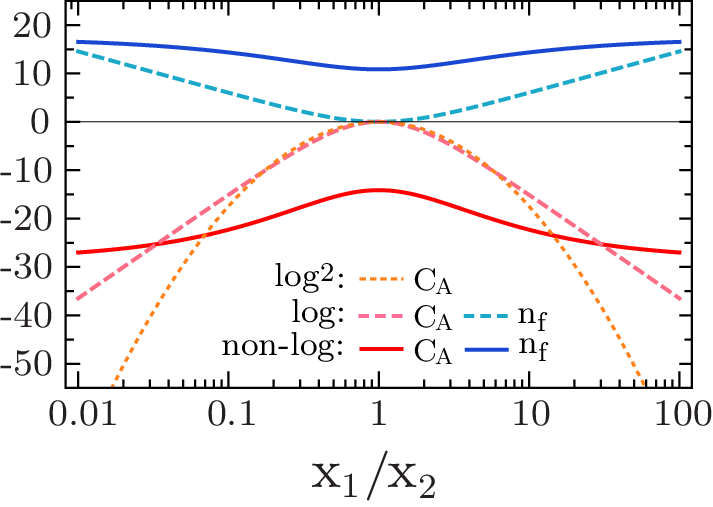}
 \includegraphics[width=0.48\textwidth]{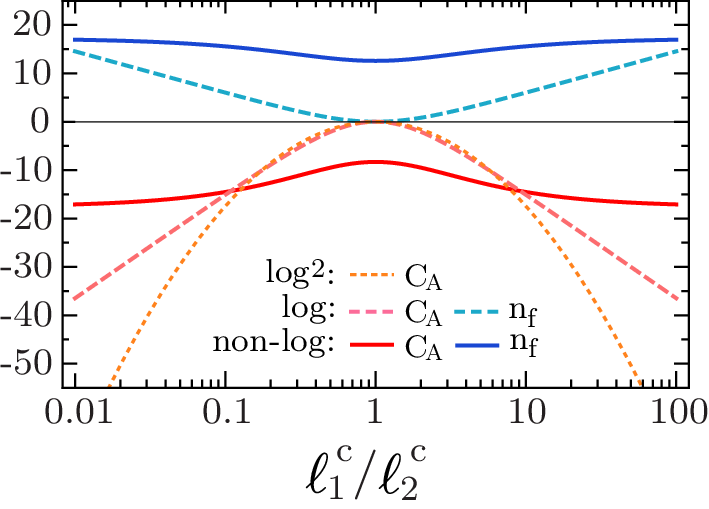}
      \vspace{-1em}
 {\caption[1]{Non-global
     logarithms and non-global non-logarithms in position space as a function of
     $x_1/x_2$ (left panel) and in cumulant momentum space as a function of
     $\la^c/\lb^c$ (right panel). We plot separately the $\log^2$ (dotted),
     $\log$ (dashed), and non-log (solid) parts of the various functions that
     appear in the soft functions $\widetilde S(x_1,x_2,\mu)$ and
     $\Scum(\la^c,\lb^c,\mu)$ (ignoring the $\mu$ dependent terms in $\widetilde
     R$ and ${\cal R}_c$). The single log curves have been plotted with the opposite sign.}
      \label{fig:fFQNLog}} }
\end{figure}

Our complete result for the bare two-loop opposite hemisphere piece of the
momentum space soft function $S(\la,\lb)$ is given in \eq{Soppfinalmom}.  In
\sec{ssec:doublecumulant} we describe how this result can be used to obtain the
double cumulant $\Scum(\la^c,\lb^c,\mu)$.  The final result for the renormalized
momentum space double cumulant distribution is
\begin{align}
\label{finalcumulant}
 & \Scum(\la^{\,c},\lb^{\,c},\mu) = \theta(\la^c) \theta(\lb^c) \Bigg[ 1 
 - \frac{\as(\mu) C_F}{4\pi} \pi^2
 + {\cal R}_c(\la^{\,c},\lb^{\,c},\mu) 
 + \frac{\as(\mu)^2}{4\pi^2} \biggl\{
 - \frac{\pi^2}{3}C_F C_A \ln^2\Big(\frac{\la^{\,c}}{\lb^{\,c}}\Big) 
  \nn \\
&\quad + \left[C_F C_A\frac{11\pi^2-3-18\zeta_3}{9}+ C_F T_R n_f
  \frac{6-4\pi^2}{9} \right] \ln \left(\frac{\lac/\lbc + \lbc/\lac}{2}\right)
 \nn \\
&\quad + C_F C_A \bigg[
f_N\Big(\frac{\lac}{\lbc}\Big) + f_N\Big(\frac{\lbc}{\lac}\Big) -
2f_N(1)\biggr] \!+ C_F T_R n_f\bigg[ f_Q\Big(\frac{\lac}{\lbc}\Big) +
f_Q\Big(\frac{\lbc}{\lac}\Big) \!-\! 2f_Q(1)\biggr]  
\!\nn\\
&\quad + C_F^2 \frac{\pi^4}{8} + \frac12 C_F C_A s_{2\rho}^{[C_F C_A]} 
  + \frac12 C_F T_R n_f s_{2\rho}^{[n_f]} \biggr\} \Bigg]\,,
\end{align}
where the logarithmic dependence is isolated on the first two lines and the
remainder depends on the constants $s_{2\rho}^{[C_FC_A,n_f]}$,
$\mu$-dependent terms ${\cal R}_c(\la^c,\lb^c,\mu)$ to be discussed below, and
the functions
\begin{align} \label{fQfN}
f_Q(a) &\equiv \left( \frac{2 \pi^2}{9}-\frac{2}{3(a+1)}\right) \ln(a)-\frac{4}{3} \ln(a) \Li_2(-a)+4\Li_3(-a) -  \frac{1}{9} ( 3-2\pi^2)\ln\Big(a+\frac{1}{a}\Big),
\nn\\
f_N(a) &\equiv -4\text{Li}_4\Big(\frac{1}{a+1}\Big)
  -11\, \text{Li}_3(-a)
  +2\text{Li}_3\Big(\frac{1}{a+1}\Big) \ln\Big[\frac{a}{(a+1)^2}\Big]
 \nn \\ & 
\quad +\text{Li}_2\Big(\frac{1}{a+1}\Big) \bigg\{\pi^2-\ln ^2(a+1)
 -\frac{1}{2} \ln (a)
  \ln\Big[\frac{a}{(a+1)^2}\Big] +\frac{11}{3} \ln(a) \bigg\}
\nn\\ &
 \quad +\frac{1}{24} \bigg\{22 \ln\Big[\frac{a}{(a+1)^2}\Big]
  -6 \ln \Big( 1+ \frac{1}{a}\Big) \ln (1+a)+
  \pi^2\bigg\} \ln ^2(a) -\frac{(a-1) \ln (a)}{6 (a+1)}
\nn\\ &
\quad  +\frac{5\pi^2}{12}  \ln\Big(1+\frac{1}{a}\Big) \ln (1+a)
  -\frac{11 \pi ^4}{180}
 - \frac{(11\pi^2\minus 3\minus 18\zeta_3)}{18}\ln\Bigl(a+\frac{1}{a}\Bigr)\,. 
\end{align}
The combinations $f_Q(a)+f_Q(1/a)$ and $f_N(a)+f_N(1/a)$ appearing in
Eq.~(\ref{finalcumulant}) are bounded functions of $a$, vanishing as $a\to 0$ or
$a\to\infty$.

\begin{figure}[t!]{    
\begin{center}
    \includegraphics[width=0.75\textwidth]{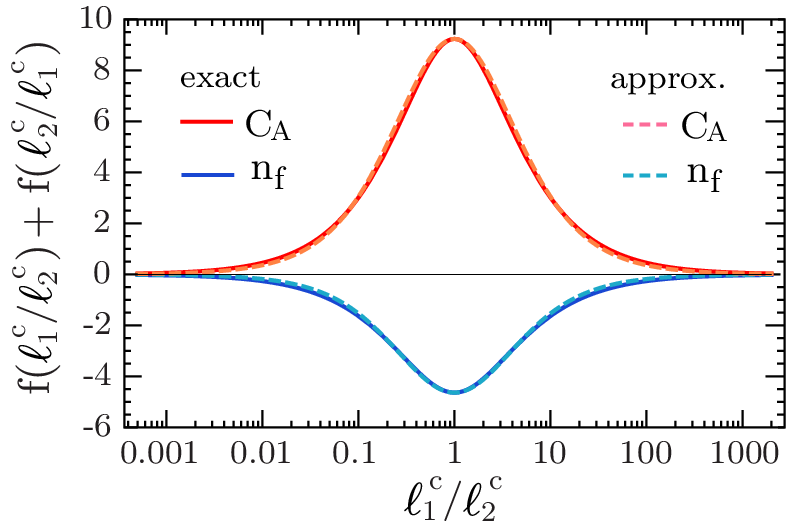}
\end{center}
\vspace{-2em}
        { \caption[1]{The non-logarithmic non-global function in momentum space. We show the functions $f(a)+f(1/a)$, where $a=\lbc/\lac$,  that appear in the momentum-space double cumulant \eq{finalcumulant} for the color structures $C_F C_A$ (solid red) and $C_F T_R n_f$ (solid blue). The simple approximate forms \eq{fQNsimple} of these functions are the dotted lines.}
  \label{fig:fQN}} }
\end{figure}

Taking the $\la^c\gg \lb^c$ limit of Eq.~(\ref{finalcumulant}) we see that the
non-global logs in momentum space have the same coefficients
as those in position space. We find this limit of 
%these terms to give
the non-Abelian terms in Eq.~(\ref{finalcumulant}), analogous to the $t_2$ function in Eqs.~(\ref{finalposition}--\ref{finalt2pos}), gives
\begin{align}
  \frac{\alpha_s^2}{8\pi^2} \bigg[
  s_{2c}^{[2]} \,\ln^2\left(\frac{\la^c}{\lb^c}\right) 
   + s_{2c}^{[1]} \,\ln\left(\frac{\la^c}{\lb^c}\right) 
%   - C_F^2 \frac{11\pi^4}{180}
   + s_{2c}^{[0]} + \ldots \bigg],
\end{align}
where $s_{2c}^{[2]} = s_2^{[2]}$ and $s_{2c}^{[1]} = s_2^{[1]}$ from
Eq.~(\ref{s2s}). The constant $s_{2c}^{[0]}$, meanwhile, is given by
\be
s_{2c}^{[0]} = - s_{2c}^{[1]}\, \ln 2 -  4C_F C_A  f_N(1)
 -  4 C_F T_R n_f   f_Q(1)
 + C_F C_A s_{2\rho}^{[C_F C_A]} + C_F T_R n_f
 s_{2\rho}^{[n_f]} 
\ee
where
\begin{align}
  2f_{Q}(1) &= -6\zeta_3 - \frac{2}{9}(3-2\pi^2)\ln 2\,,
\\
 2f_N(1)  &=  -8 \text{Li}_4\Big(\frac{1}{2}\Big)
  + \zeta_3\Big(\frac{33}{2} \minus 5\ln 2\Big)
   + \frac{\ln 2 \minus \ln^4 2}{3}
   +\frac{2
   \pi ^4}{45}+\frac{\pi^2}{3}  \Big(\ln^2 2-\frac{11}{3} \ln 2\Big)\,.
 \nn
\end{align}
In the right panel of Fig.~\ref{fig:fFQNLog} we plot the double log (dotted),
single log (dashed), and non-log (solid) non-global contributions to
$\Scum(\la^c,\lb^c,\mu)$.  The overall signs for the plot are chosen so that the
$C_F C_A$ color structures are negative and $C_F T_R n_f$ are positive, which
involves plotting the single NGL $\propto \ln(a+1/a)$, where $a=\lbc/\lac$, with
the opposite sign.  The solid lines are the non-logarithmic functions
$\bigl[f_{N,Q}(a)+f_{N,Q}(1/a)-2f_{N,Q}(1) + \frac{1}{2}s_{2\rho}^{[C_F
  C_A,n_f]}\bigr]$. Comparing the left and right panels
of Fig.~\ref{fig:fFQNLog} the log terms are identical, but the non-log terms
differ.

The symmetric sum of complicated functions in Eq.~(\ref{fQfN}) that appears in \eq{finalcumulant} is well
approximated for all $a$ by a simple function, taking the form
\be 
\label{fQNsimple}
f_{Q,N}(a)+f_{Q,N}(1/a) \simeq
2f_{Q,N}(1) \frac{4a}{(1+a)^2} \, .
\ee
The comparison between the exact forms in Eq.~(\ref{fQfN}) and the approximate
forms in \eq{fQNsimple} is shown in \fig{fig:fQN}, displaying remarkable agreement.

The constant terms $s_{2\rho}^{[C_FC_A]}$ and $s_{2\rho}^{[n_f]}$ in
\eq{finalcumulant} are given by the thrust constants $s_2^{[C_FC_A]}$ and
$s_{2}^{[n_f]}$, plus analytic constants.  In \sec{ssec:HJMthrustP} we
calculate the differences $s_{2\rho}^{[a]} - s_2^{[a]}$ analytically, and find
\begin{align} \label{s2ps2diff}
  s_{2\rho}^{[n_f]} &= s_2^{[n_f]} +   4 \zeta_3 -\frac{4}{3} \,,   \\
 s_{2\rho}^{[C_FC_A]} &= s_2^{[C_FC_A]} + 
  \frac{2}{3} +\frac{19 \pi^4}{45}+\frac{2 \pi^2}{3} \ln^2 2-\frac{2}{3}\ln^4 2
  -16 \Li_4\left( \frac{1}{2}\right) -11 \zeta_3 -14\zeta_3  \ln 2 
  \,.\nn
\end{align}
Thus only the constants $s_2^{[C_FC_A,n_f]}$ remain as unknowns in both of our final results
\eqs{finalposition}{finalcumulant}, and for them we use the numerical values
quoted in Eq.~(\ref{s2Monni}) from Ref.~\cite{Monni}.

The $\ln(\lac/\mu)$ and $\ln(\lbc/\mu)$ dependent part ${\cal
  R}_c(\lac,\lbc,\mu)$, obtainable by taking the inverse Fourier
transform of \eq{Rposition}, is given by
\begin{align}
 \label{Rcumulant}
 & {\cal R}_c(\la^c,\lb^c,\mu) = 
 -\frac{\as(\mu)C_F}{\pi} \Bigl(\Lac^2+ \Lbc^2 - \frac{\pi^2}{3}\Bigr) 
+ \frac{\as^2(\mu)}{(4\pi)^2}
 \biggl\{8C_F^2\!\left(\Lac^2 +  \Lbc^2\right)^2 + \Big(\frac{88}{9}C_F C_A 
  \nn \\
  & - \frac{32}{9}C_F T_R n_f\Big)
 \left(\Lac^3 +  \Lbc^3\right)   
+ \Bigl[-\frac{20\pi^2}{3} C_F^2 + C_F C_A\Big(\frac{4\pi^2}{3} -
   \frac{268}{9}\Big) + \frac{80}{9}C_F T_R n_f\Bigr]  \left(\Lac^2 +    \Lbc^2\right)  \nn \\
   &
   + \Bigl[64\zeta_3 C_F^2 + C_F C_A\Bigl(\frac{808}{27} -
   \frac{22\pi^2}{9}- 28 \zeta_3\Bigr) 
- C_F T_R n_f\Bigl(\frac{224}{27} - \frac{8\pi^2}{9} \Bigr)\Bigr]  \left(\Lac+\Lbc\right) \nn \\
& - C_F^2 \frac{28\pi^4}{45} + C_F C_A\Bigl(\frac{352\zeta_3}{9} + \frac{268\pi^2}{27} - \frac{4\pi^4}{9}\Bigr) - C_FT_R n_f \Bigl(\frac{128\zeta_3}{9} + \frac{80\pi^2}{27}\Bigr)  \biggr\},
\end{align}
where $\Lac^n = \ln^n(\la^c/\mu)$, and $\Lbc^n = \ln^n(\lb^c/\mu)$.
Eq.~(\ref{finalcumulant}) will be used in the double cumulant comparison to
EVENT2 data done in~\sec{sec:event2}.

%%%%%%%%%%%%%%%%%%%%%%%%%%%%%%%%%%%%%%%%%%%%%%%%%%%%%%%%%%%%%%%%%%%%%%%%%%%%%%%%%%%%%%%%%%%%%%%%%%%%%%%%%
\section{Projection onto Other Observables}
\label{sec:projections}
%%%%%%%%%%%%%%%%%%%%%%%%%%%%%%%%%%%%%%%%%%%%%%%%%%%%%%%%%%%%%%%%%%%%%%%%%%%%%%%%%%%%%%%%%%%%%%%%%%%%%%%%%

Having obtained the explicit analytic result for the dijet soft function, we can
make use of it in several ways.  In~\sec{ssec:doublecumulant} we give further
details about the projection of the momentum space result used to obtain the
renormalized double cumulant, $\Scum(\la^c,\lb^c,\mu)$ in
Eq.~(\ref{finalcumulant}), since the procedure is useful for other projections.
Additionally, in~\sec{ssec:hjmandthrust} we project the distribution onto two
classes of $e^+e^-$ event shapes $\tau_\alpha$ and $\rho_\alpha$ originally
defined in \cite{Hoang:2008fs}, which we call asymmetric thrust\footnote{This is
  to be distinguished from the hadronic event shape $\tau_A$ for $pp$ collisions
  called asymmetric thrust in  \cite{Kelley:2010qs}.} and asymmetric heavy jet
mass, respectively.  Asymmetric thrust can be
defined from the hemisphere masses $m_1^2$ and $m_2^2$ as

\be
\tau_{\alpha} = \frac{2}{1+\alpha}\frac{\alpha m_1^2 + m_2^2}{Q^2} \,,
\ee
where $\alpha > 0$ is a dimensionless parameter. In the dijet limit
$\tau_\alpha$ corresponds to the original thrust for $\alpha=1$.  The asymmetric
heavy jet mass is
\be
\rho_{\alpha} = \frac{2}{1+\alpha}\max\left(\alpha\frac{m_1^2}{Q^2}, \frac{m_2^2}{Q^2}\right) \,,
\ee
and corresponds to the heavy jet mass for $\alpha=1$.  The
$d\sigma/d\tau_\alpha$ and $d\sigma/d\rho_\alpha$ distributions contain
non-global logarithms in $\alpha$, and we will show that the full non-global
structure of their cumulants bears close relation to the non-global structure in
the position-space dijet cross section and the double cumulant momentum-space
cross section.

As discussed in~\ssec{numresults}, the projection of the non-global structure
onto the two-loop constants needed for the heavy jet mass and thrust is of interest.
These observables have recently been used in fits of $\alpha_s$ from $e^+e^-$
collider data, placing importance on the knowledge of these
constants~\cite{Becher:2008cf,Abbate:2010xh,Chien:2010kc,Abbate:2010vw}.
In~\sec{ssec:HJMthrust} we project our non-global results onto heavy jet mass
and thrust and determine the difference between the cumulants of the heavy jet
mass and thrust distributions, \eq{s2diff}.
We compare our analytic result,~\eq{eq:ourProj}, to a numerical
projection from our position-space result and to a numerical extraction from
EVENT2, \eq{s2diffs}, both of which provide a nontrivial cross-check of our
calculations.

\subsection{The Double Cumulant of the Non-global Terms}
\label{ssec:doublecumulant}

To determine the double cumulant of the non-global terms, we compute the
opposite hemisphere cumulant
\be
{\cal S}_c^{\opp}(\la^{\,c},\lb^{\,c}) = \int_{-\infty}^{\la^{\,c}} d\la \int_{-\infty}^{\lb^{\,c}} d\lb \, \Sopp(\la,\lb) \,,
\ee
The function $\Sopp(\la,\lb)$ only has support for $\la,\lb\ge 0$. The relation
of this double cumulant soft function to the position-space soft function is
\begin{align}
  {\cal S}_c^{\opp}(\la^{\,c},\lb^{\,c}) &= 
 -  \int_{-\infty}^\infty \frac{dx_1}{2\pi}\frac{dx_2}{2\pi} 
  \exp(i \la^c x_1 + i \lb^c x_2)
 \frac{\widetilde S^{\opp}(x_1,x_2) }{(x_1\,x_2)} 
 \,  \,,
\end{align}
where $x_i \equiv x_i-i0^+$ in the above equation. Comparing this result to Eq.~(\ref{eq:asymHJMproj}) below with $\alpha=1$ we see that
the diagonal projection $\Sopp(\la^c,\la^c)$ is equal to the heavy jet mass
cumulant with $\la^c = Q \rho_H$, which explains the appearance of the heavy jet
mass constants $s_{2\rho}^{[C_FC_A,n_f]}$ in our final result in
Eq.~(\ref{finalcumulant}).

For the analytic computation of the dependence on $\la^c/\lb^c$ of the
opposite hemisphere terms  $\cS_c^{\opp}(\lac,\lbc)$ it is more convenient to start with the bare momentum-space result.  We use the general form of the opposite hemisphere results
from~\eqs{eq:generalmomspace}{eq:Fdef}:
\be
\label{eq:softform}
 S^{\opp \, [i]}(\la,\lb) 
  = A\,C_i \, \mu^{4\e} s^{-1-2\e} \big[ F_0 + F_1(r) \big] \,,
\ee
where the variables
\begin{align} \label{sr}
s=\la\lb \,,\qquad r=\frac{\lb}{\la} \,,
\end{align}
the color factor $C_i = C_F C_A$ or $C_F T_R n_f$, $F_0$ and $F_1(r)$ depend on the class of diagram (see \fig{fig:diagrams}) contributing to $S^{[i]}_\opp$ and with $\mu$
in $\overline{\textrm{MS}}$ the prefactor
\be
A = \left(\frac{\alpha_s}{2\pi}\right)^2 \frac{(e^{\gamma_E})^{2\e}}{\Gamma(1-\e)^2} = \left(\frac{\alpha_s}{2\pi}\right)^2 \Big( 1 - \frac{\pi^2}{6} \e^2 + \cO(\e^3) \Big) \,.
\ee
To evaluate the double cumulant we change variables to $s$ and $r$:
\be
\label{eq:dicumulantsrintegral}
{\cal S}_c^{\opp }(\la^{\,c},\lb^{\,c}) = \int_0^{\lb^{\,c}/\la^{\,c}} \frac{dr}{2r} \int_0^{(\la^{\,c})^2 r} ds \, \Sopp(\la,\lb) + \int_{\lb^{\,c}/\la^{\,c}}^{\infty} \frac{dr}{2r} \int_0^{(\lb^{\,c})^2 /r} ds \, \Sopp(\la,\lb) \,.
\ee
Plugging in the form in~\eq{eq:softform}, we can perform the $s$ integral.  Then
expanding to collect terms, we find
\begin{align}
{\cal S}_c^{\opp\, [i]}(\la^{\,c},\lb^{\,c}) &= C_i \left\{ \frac{1}{4\e^2} A F_0 \left(\frac{\mu^2}{\la^{\,c}\lb^{\,c}}\right)^{2\e} - \left(\frac{\as}{2\pi}\right)^2\frac{1}{4\e} \left(\frac{\mu^2}{\la^{\,c}\lb^{\,c}}\right)^{2\e} \int_0^{\infty} \frac{dr}{r} F_1(r)  \right. \\
& \left. + \frac12 \left(\frac{\as}{2\pi}\right)^2\int_0^{\la^{\,c}/\lb^{\,c}} \! \frac{dr}{r} \ln\left(r \frac{\lb^{\,c}}{\la^{\,c}} \right) F_1(r) + \frac12 \left(\frac{\as}{2\pi}\right)^2\int_0^{\lb^{\,c}/\la^{\,c}} \! \frac{dr}{r} \ln\left(r \frac{\la^{\,c}}{\lb^{\,c}} \right) F_1(r)  \right\} . \nn
\end{align}
The final two terms contain only finite $\mu$-independent terms which contribute to $\cS$.  The first two terms contain $\mu$-dependent terms, some of which cancel with the same hemisphere terms in the same way as explained in \sec{sec:twoloopsetup}.  The only term that does not cancel, but combines in a non-trivial way, is the double logarithmic term.  Putting these terms together, the cumulant of the non-global terms $\cS$ is 
\begin{align}
\label{eq:dicumulant}
{\cal S}_c^{{\rm NG} \, [i]}(\la^{\,c},\lb^{\,c}) 
 &= \left(\frac{\as}{2\pi}\right)^2 C_i \, \bigg\{ -\frac12 F_0^{(0)} \ln^2 \left(\frac{\la^c}{\lb^c}\right) \\
&\qquad + \frac12 \int_0^{\la^{\,c}/\lb^{\,c}} \frac{dr}{r} \ln\left(r \frac{\lb^{\,c}}{\la^{\,c}} \right) F_1^{(0)}(r) + \frac12 \int_0^{\lb^{\,c}/\la^{\,c}} \frac{dr}{r} \ln\left(r \frac{\la^{\,c}}{\lb^{\,c}} \right) F_1^{(0)}(r) + \cS_0 \bigg\} \,, \nn
\end{align}
where $\cS_0$ is the constant term in the non-global terms and is given by
\be
\label{eq:S0dicumulant}
\cS_0 = -\frac{\pi^2}{24}F_0^{(0)} + \frac{F_0^{(2)}}{4} - \frac{1}{4}\int_0^\infty\frac{dr}{r}F_1^{(1)}(r)\,,
\ee  
and $F_{0,1}^{(n)}$ is the coefficient of $\epsilon^n$ in the expansion of
$F_{0,1}$.  We can perform these integrals analytically for the total momentum
space result from all graphs, given by \eq{Soppfinalmom}. Note that our
calculation of ${\cal S}_c^{\opp \, [C_F C_A]}$ gives
\be
F_0^{(0)} = \frac{2\pi^2}{3}  \,.
\ee
The final result for the double cumulant with both opposite and same hemisphere
terms is given above in \eq{finalcumulant}.

\subsection{Asymmetric Heavy Jet Mass and Asymmetric Thrust}
\label{ssec:hjmandthrust}

In this section we give the projection from both the position and momentum space opposite hemisphere terms onto asymmetric heavy jet mass and asymmetric thrust. We will use the $\alpha = 1$ results of these projections to extract the difference in the constant terms of heavy jet mass and thrust.  Projecting both forms of the hemisphere soft function result, position and momentum space, onto the asymmetric heavy jet mass and asymmetric thrust cumulants provides a nontrivial check of the consistency of our results and uncovers interesting relations for the hemisphere soft function.  We present the relevant projections starting from the hemisphere soft function in both spaces. 

The asymmetric thrust cumulant is defined from the momentum space and
position space hemisphere mass soft function by the projections~\cite{Hoang:2008fs}
\begin{align}
\label{eq:asymthrustproj}
\Sigma_{\tau_{\alpha}} (\tau_{\alpha}^c) &= \int_0^{\tau_{\alpha}^c}
d\tau_{\alpha} \int d m_1^2 d m_2^2\, \delta\left( \tau_{\alpha} -
  \frac{2}{1+\alpha}\frac{\alpha m_1^2 + m_2^2}{Q^2} \right) \frac{d^2
  \sigma}{dm_1^2 dm_2^2} \,, 
 \nn\\
\Sigma_{\tau_{\alpha}} (\tau_{\alpha}^c) &= -i \int \frac{dy}{2\pi} 
  \, \exp\Big[i Q^2 \tau_\alpha^c\, y \frac{(1\plus\alpha)}{2}\Big]\: \, 
 \frac{\widetilde \sigma(\alpha y,y)}{y-i0^+} \,.
\end{align}
Similarly, the asymmetric heavy jet mass distribution is defined from the
momentum space and position space hemisphere mass distributions
by~\cite{Hoang:2008fs}
\begin{align}
\label{eq:asymHJMproj}
\Sigma_{\rho_{\alpha}} (\rho_{\alpha}^c) &= \int_0^{\rho_{\alpha}^c}
d\rho_{\alpha} \int d m_1^2 d m_2^2 \, \delta\left( \rho_{\alpha} -
  \frac{2}{1+\alpha}\frac{\max(\alpha m_1^2,m_2^2)}{Q^2}\right) \frac{d^2
  \sigma}{dm_1^2 dm_2^2} \,, 
\nn\\
\Sigma_{\rho_\alpha}(\rho_{\alpha}^c) &= 
 -  \int \frac{dy_1}{2\pi}\frac{dy_2}{2\pi} 
  \exp\Big[ iQ^2\rho_{\alpha}^c(y_1 + y_2)\frac{(1\plus\alpha)}{2} \Big] 
 \frac{\widetilde \sigma(\alpha y_1,y_2) }{(y_1 - i0^+)(y_2 - i0^+)} 
 \,  \,.
\end{align}

\subsubsection{Projection of Non-global Terms in Position Space}

Let's consider the contribution to the two asymmetric event shapes from the
non-global term. In position space we can determine this contribution by
inserting $\widetilde \sigma(y_1,y_2) = \alpha_s^2/(8\pi^2)
t_2(y_1/y_2)$ into Eqs.~(\ref{eq:asymthrustproj}) and (\ref{eq:asymHJMproj}).

For asymmetric thrust this gives 
\begin{align}
\label{eq:asymthrustpos}
\Sigma_{\tau_\alpha}^{[t_2]}(\tau_{\alpha}^c) &= \Big(\frac{\as^2}{8\pi^2}\Big)
 (-i) \int \frac{dy}{2\pi} 
  \, \exp(i Q^2 \tau_\alpha^c\, y)\: \, 
 \frac{t_2(\alpha)}{y-i0^+} \nn\\
 &= \Big[\frac{\as^2}{8\pi^2} \: t_2(\alpha)\Big]\: \theta(\tau_\alpha^c)  \,.
\end{align}
This result tells us that the non-global dependence on $\alpha$ of asymmetric
thrust, given by $t_2(\alpha)$, is equivalent to the non-global dependence on
$x_1/x_2$ in the position space hemisphere soft function, given by
$t_2(x_1/x_2)$.  We will see a similar correspondence for asymmetric heavy jet
mass in the momentum space projection.

\begin{figure}[t!]{    
 \hspace{0.1cm}
 \includegraphics[width=.45\textwidth]{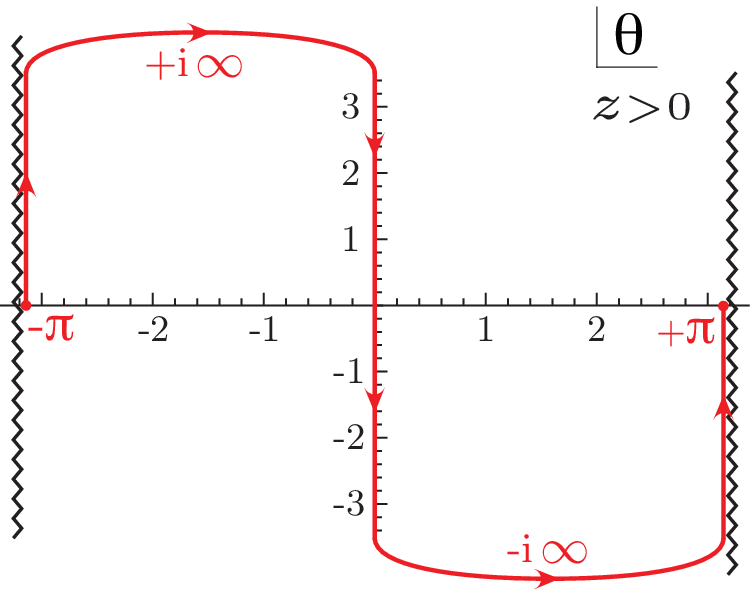} 
 \hfill
 \includegraphics[width=.45\textwidth]{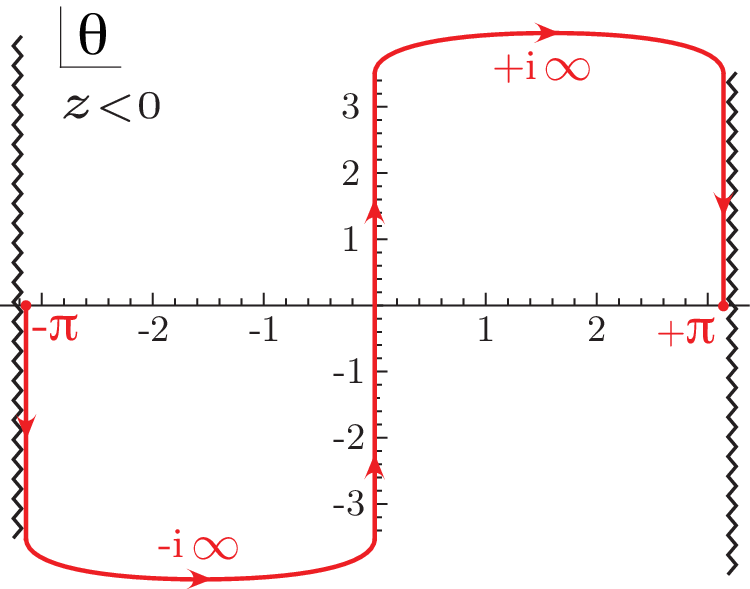} 
 \hspace{0.1cm}
 {\caption[1]{Original integration contours for $\theta$ in the complex plane
     after the change of variable in Eq.~(\ref{eq:changevar}).}
  \label{fig:contour}} }
\end{figure}

For asymmetric heavy jet mass the fact that $t_2$ only depends on the ratio of
positions motivates making the following change of variables in
Eq.~(\ref{eq:asymHJMproj}):
\begin{align} \label{eq:changevar}
i \theta = \ln \frac{y_1-i0^+}{y_2-i0^+} , \quad z = y_1+y_2-i0^+  \quad
\Longrightarrow    \quad 
\frac{d y_1 d y_2}{(y_1-i0^+)(y_2-i0^+)} = \frac{i d\theta\, dz}{(z-i0^+)}    \,.
\end{align}
With the new variables $z$ is integrated along the real axis,
$-\infty<z<\infty$, and $\theta$ is integrated along a contour in the complex
plane shown in Fig.~\ref{fig:contour}, which depends on the sign of $z$.
For either sign of $z$ this contour can be deformed to the real
axis without encountering cuts or poles, so that the limits become $-\pi<
\theta< \pi$.  This gives
\begin{align}
\label{eq:asymHJMpos}
\Sigma_{\rho_\alpha}^{[t_2]}(\rho_{\alpha}^c) &= 
 \frac{\as^2}{8\pi^2} \int_{-\pi}^{+\pi} \frac{d\theta}{2\pi}
 \int_{-\infty}^{+\infty} \frac{dz}{2\pi}\:
 \frac{(-i)}{z-i0^+} \exp\left(i Q^2\rho_{\alpha}^c z\right) \, t_2\big(\alpha e^{i\theta}\big) \nn \\
& = \Big[ \frac{\as^2}{8\pi^2}  \int_0^{\pi} \frac{d\theta}{\pi} t_2\big(\alpha
e^{i\theta}\big) \Big]\: \theta(\rho_{\alpha}^c)\,.
\end{align}
The projection again just gives a $\theta$-function of $\rho_\alpha^c$, but this
time a more complicated projection from $t_2$ is needed to get the coefficient.

Note that the calculation used to obtain Eqs.~(\ref{eq:asymthrustpos}) and
(\ref{eq:asymHJMpos}) is also valid for terms in the cross section  at higher
orders in $\alpha_s$ as long as they are only functions of $x_1/x_2$.

\subsubsection{Projection of Non-global Terms in Momentum Space}
\label{sssec:momentumspace}

For both asymmetric heavy jet mass and asymmetric thrust, we consider the
projection for the bare opposite hemisphere terms and remove the $\mu$-dependent
pieces to obtain the projection for the non-global terms.  To perform the
projection, we change variables from $\la,\lb$ to $s$ and $r$ in Eq.~(\ref{sr}).
For asymmetric heavy jet mass, the cumulant for the opposite hemisphere terms
is
\be
\label{eq:rhocumulantsrintegral}
{\Scum}_{\rho_\alpha}^{\opp} (\rho_{\alpha}^c) = \int_0^{\rho_{\alpha}^c} d\rho_{\alpha} \int d\la d\lb \, \delta\left(\rho_{\alpha} - \frac{2}{1+\alpha} \max \Big(\alpha \frac{\la}{Q}, \frac{\lb}{Q}\Big) \right) \Sopp(\la,\lb,\mu) \,.
\ee
Using the general form in Eq.~(\ref{eq:softform}), the cumulant in terms of
integrals over $s$ and $r$ is
\begin{align}
\Scum_{\rho_\alpha}^{\opp \, [i]} (\rho_{\alpha}^c) &=  \theta(\rho_{\alpha}^c)\biggl\{\int_0^{\alpha} \frac{dr}{2r} \int_0^{Q^2\rho_{\alpha}^{c2} ((1+\alpha)/2\alpha)^2 r} ds \,A\,C_i \,\mu^{4\e} s^{-1-2\e} \left( F_0 + F_1(r) \right) \nn \\
& \qquad\quad  + \int_{\alpha}^{\infty} \frac{dr}{2r} \int_0^{Q^2\rho_{\alpha}^{c2} ((1+\alpha)/2)^2 / r} ds \, A \,C_i \,\mu^{4\e} s^{-1-2\e} \left( F_0 + F_1(r) \right) \biggr\} \,.
\end{align}
From this formula we see the $\rho_\alpha$ cumulant is just the momentum-space double cumulant \eq{eq:dicumulantsrintegral} with $\lac = Q\rho_\alpha^c(1+\alpha)/(2\alpha)$ and $\lbc = Q\rho_\alpha^c(1+\alpha)/2$. 
Performing the $s$ integral in \eq{eq:rhocumulantsrintegral}, expanding up to ${\cal O}(\epsilon^0)$, and canceling $1/\epsilon$ and $\mu$-dependent terms against same hemisphere pieces as in \sec{sec:twoloopsetup} and \eq{eq:dicumulant}, we find
\begin{align}
\label{eq:asymHJMmom}
\Scum_{\rho_\alpha}^{ {\rm  NG} \, [i]} (\rho_{\alpha}^c) =  \theta(\rho_{\alpha}^c) \left(\frac{\as}{2\pi}\right)^2C_i \biggl\{\!&-\frac12 F_0^{(0)} \ln^2\alpha  + \mathcal{S}_{0}  \\
& + \frac12  \int_0^{1/\alpha} \frac{dr}{r} \ln(r\alpha) F_1^{(0)}(r) + \frac12 \int_0^{\alpha} \frac{dr}{r} \ln(r/\alpha) F_1^{(0)}(r) \biggr\} \,. \nn
\end{align}
From this result we find the non-global structure in the last two terms is
equivalent to the corresponding non-global structure in the double
cumulant,~\eq{eq:dicumulant}, with the replacement $\lb^{\,c}/\la^{\,c} \to
\alpha$.  The constant $\cS_{0}$ is the same as in \eq{eq:S0dicumulant}.

We can go through the same procedure for asymmetric thrust, whose cumulant is
\be
\Scum_{\tau_\alpha}^{\opp} (\tau_{\alpha}^c) = \int_0^{\tau_{\alpha}^c} d\tau_{\alpha} \int d\la d\lb \, \delta\left(\tau_{\alpha} - \frac{2}{1+\alpha} \frac{\alpha \la + \lb}{Q} \right) \Sopp(\la,\lb,\mu) \,.
\ee
In terms of integrals over $s$ and $r$,
\be
\Scum_{\tau_\alpha}^{\opp} (\tau_{\alpha}^c) = \int_0^{\infty} \frac{dr}{2r} \int_0^{s^c} ds \, \Sopp(\la,\lb,\mu) \,,
\ee
where
\be
s^c = \frac{(Q\tau_{\alpha}^c)^2 (1+\alpha)^2}{4\alpha [\alpha r + 1/(\alpha r) + 2]} \,.
\ee
Evaluating the $s$ integral, expanding to finite terms, and canceling $\mu$- and $\epsilon$-dependent terms against same-hemisphere contributions as in \sec{sec:twoloopsetup} and \eqs{eq:dicumulant}{eq:asymHJMmom},
\begin{align}
\label{eq:asymthrustmom}
\Scum_{\tau_\alpha}^{ {\rm  NG} \, [i]} (\tau_{\alpha}^c) &= \theta(\tau_{\alpha}^c) \left(\frac{\as}{2\pi}\right)^2 C_i \biggl\{-\frac12 F_0^{(0)} \ln^2\alpha  - \frac12  \int_0^{\infty}\! \frac{dr}{r} \ln\left(2+\alpha r + \frac{1}{\alpha r}\right) F_1^{(0)}(r)  + \mathcal{S}_{0\tau} \biggr\} , 
\end{align}
where $\cS_{0\tau} = \cS_0 - (\pi^2/6)F_0^{(0)}$.
Although the non-global structure in this result is less obvious, we know from
the position space result that the non-global structure in $\alpha$ for
$\tau_{\alpha}$ is functionally equivalent to the non-global structure in
$t_2(x_1/x_2)$, the position space form of the non-global terms in the soft
function.

To conclude, we have shown that the non-global structure in the hemisphere soft function in position space and momentum space is reproduced by the asymmetric thrust and asymmetric heavy jet mass observables respectively.

\subsection{Heavy Jet Mass and Thrust}
\label{ssec:HJMthrust}

For heavy jet mass $\rho_H$ and thrust $T=1-\tau$ we will compute the difference
of the cumulants of the non-global terms at $\cO(\alpha_s^2)$.  This corresponds
to the projection of $t_2(x_1,x_2)$ in position space, \eq{Ttildex1x2}. From
Eqs.~(\ref{eq:asymthrustpos}) and (\ref{eq:asymHJMpos}) with $\alpha=1$ this
simply gives constant terms in the cumulants, which are $\delta(\rho_H)$ and
$\delta(\tau)$ terms in the corresponding distributions. Since we will take the
difference of heavy jet mass and thrust we only need opposite hemisphere terms.
We examine the projections in both momentum space and position space.  The
results are the same, providing an internal consistency check on our position
and momentum space results, and more importantly the analytic results for the
projection match the numerical extraction of these constants from EVENT2, which
were discussed in~\sec{ssec:numresults}.  We call $\Sigma_H$ the heavy jet mass
cumulant and $\Sigma_\tau$ the thrust cumulant, and add a superscript NG to
refer to keeping only the terms in the cumulants induced by the non-global
structure at ${\cal O}(\alpha_s^2)$.

\subsubsection{Projection From Position Space}
\label{ssec:HJMthrustX}

Taking the $\alpha = 1$ result for the asymmetric heavy jet mass projection
from~\eq{eq:asymHJMpos} to find the projection for the non-global
terms and plugging in the result \eq{finalt2pos} for $t_2(x_1/x_2)$, we find
\begin{equation}
\begin{split}
\SigmaNG_H (\rho) & = \frac{\alpha_s^2}{8\pi^2} \theta(\rho) \int_0^\pi \frac{d
  \theta}{\pi} t_2(e^{i \theta}) \\
  &=  \frac{\alpha_s^2}{8\pi^2} \, \theta(\rho) \,
  \Big[ C_F C_A s_{2\rho}^{[C_FC_A]} + C_F T_R n_f s_{2\rho}^{[n_f]}\Big]
  \,, 
\end{split}
\end{equation}
where we used $ t_2(e^{i \theta}) = t_2(e^{-i\theta})$.  Similarly, we can take the $\alpha=1$ result for the asymmetric thrust projection from~\eq{eq:asymthrustpos}:
\be
\SigmaNG_{\tau} (\tau) = \frac{\as^2}{8\pi^2} \theta(\tau) t_2(1) =  \frac{\alpha_s^2}{8\pi^2} \, \theta(\tau) \,
  \Big[ C_F C_A s_{2}^{[C_FC_A]} + C_F T_R n_f s_{2}^{[n_f]}\Big] \,.
\ee
Therefore the difference in cumulants is
\begin{align} \label{cumdiffposn}
\SigmaNG_H (\rho) -\SigmaNG_{\tau} (\rho) 
 &= \frac{\alpha_s^2}{8\pi^2} \theta(\rho)
\int_0^\pi \frac{d \theta}{\pi} \left ( t_2(e^{i\theta})-t_2(1) \right)  
 \\
 &=  \frac{\alpha_s^2}{8\pi^2} \theta(\rho) \Big[
  C_F C_A \left(s_{2_\rho} - s_2 \right)^{[C_FC_A]} 
  + C_F T_R n_f \left(s_{2_\rho} - s_2 \right)^{[n_f]}\Big] \nn \,,
\end{align}
which provides an alternate derivation of~\eq{s2diff}.  We will use the
momentum space projection to determine the analytic value for the difference in
cumulants, as the integrals are more straightforward.

\subsubsection{Projection from Momentum Space}
\label{ssec:HJMthrustP}

In momentum space, we can take the $\alpha = 1$ results for asymmetric heavy jet mass and asymmetric thrust, given in~\eqs{eq:asymHJMmom}{eq:asymthrustmom} respectively.  When we take the difference, 
what remains is the difference in constants for the two distributions:
\begin{align}
\SigmaNG_{H} (\rho) - \SigmaNG_{\tau} (\rho) &=  \left(\frac{\alpha_s}{2\pi}\right)^2 \theta(\rho) \, C_i \left( \frac{\pi^2}{6} F_0 + 2 \int_0^1 \frac{dr}{r} \ln(1+r) F_1(r) \right) \\
& = \frac{\alpha_s^2}{8\pi^2} \theta(\rho) 
   \Big[ C_F C_A (s_{2\rho}-s_2)^{[C_FC_A]} 
 + C_F T_R n_f (s_{2\rho}-s_2)^{[n_f]}\Big]
  \,, \nn
\end{align}
where the $s_{2_\rho}^{[a]}$ and $s_2^{[a]}$ constants are defined in
\eqs{s2thrust}{s2rho}.  Note that $-\frac12 F_0$ gives the coefficient of the
non-global double log term in the hemisphere soft function. Using the results in
Appendix~\ref{appx:twoloopcalc} we find
\begin{align}
\label{eq:ourProj}
\SigmaNG_{H} (\rho) - \SigmaNG_{\tau} (\rho) &=\frac{\alpha_s^2}{8\pi^2}
\theta(\rho) \left[   C_F T_R n_f \left( 4 \zeta_3 -\frac{4}{3} \right)
\right. +C_F C_A \biggl( \frac{2}{3} +\frac{19 \pi^4}{45}
+\frac{2 \pi^2}{3} \ln^2 2
 \nn\\
 &\qquad\qquad\quad  -\frac{2}{3}\ln^4 2 
  -16 \Li_4\left( \frac{1}{2}\right) -11 \zeta_3
  -14 \zeta_3 \ln 2 \biggr)  \biggr] \,.
\end{align}
This yields the analytic result quoted above in Eq.~(\ref{s2ps2diff}), and
agrees with a numerical evaluation of the projection in Eq.~(\ref{cumdiffposn})
using our analytic position space result.  Numerically Eq.~(\ref{eq:ourProj})
gives
\be
(s_{2\rho} -s_{2})^{[C_F C_A]}=11.6352, \quad \quad (s_{2\rho} - s_{2})^{[n_f]}=3.4749 \,
\ee
which can be compared to the EVENT2 extractions in \eq{s2diffs}.  There is
excellent agreement between our analytic results and the EVENT2 extractions,
providing a strong consistency check on our calculation.

The constant $s_{2\rho} - s_2$ has contributions from both the non-global logarithms and the remaining non-global terms.  For each color structure, the contributions from each set of terms is
\be
\begin{array}{c c c}
& \;\;\,C_F C_A & \;\;\;\;\;C_F T_R n_f \\
\textrm{double log: } & \quad\;\;\,  43.293 \quad & \quad - \\
\textrm{single log: }  & \quad\;\;\, 39.586 \quad & \quad -15.791 \\
\textrm{non-log: } & \quad -59.609 \quad & \quad \;\;\;\;\; 22.741 \,.
\end{array}
\ee
We see that the non-global logs do not dominate the contribution, and the full set of non-global terms is needed to accurately determine the constant.

%%%%%%%%%%%%%%%%%%%%%%%%%%%%%%%%%%%%%%%%%%%%%%%%%%%%%%%%%%%%%%%%%%%%%%%%%%%%%%%%%%%%%%%%%%%%%%%%%%%%%%%%%
\section{Two Dimensional Comparison with EVENT2}
\label{sec:event2}
%%%%%%%%%%%%%%%%%%%%%%%%%%%%%%%%%%%%%%%%%%%%%%%%%%%%%%%%%%%%%%%%%%%%%%%%%%%%%%%%%%%%%%%%%%%%%%%%%%%%%%%%%

The Monte Carlo EVENT2~\cite{Catani:1996jh,Catani:1996vz} allows us to study the
structure of the hemisphere jet mass distribution.  EVENT2 contains the matrix
elements necessary to compute dijet event shapes at $\cO(\alpha_s^2)$, and it
allows us to numerically compute the non-global contributions in the hemisphere
mass distribution. We use the program to numerically compute the distribution of
$m_2^2$ making the cut $m_1^2 < m_1^{c2}$:
\be
\label{eq:EVENT2dist}
\frac{d\sigma}{dm_2^2} (m_1^{c2}) = \int_0^{m_1^{c2}} dm_1^2 \frac{d^2\sigma}{dm_1^2 dm_2^2} \,.
\ee
In the region $m_2^2 \ll m_1^{c2} \ll Q^2$, the dijet factorization theorem we
use here applies, and we can compare the results of EVENT2 in this regime to our
calculation.  In~\sec{ssec:SoftNGL} we compare the soft non-global logarithms
from EVENT2 with our calculation and show there is excellent agreement.

When $m_2^2 \ll m_1^{c2} \sim Q^2$, our factorization theorem does not apply and
new contributions arise.  In~\sec{ssec:hardglobal} we consider the logarithmic
structure in this hard region and compare the leading double log to EVENT2.

\subsection{Soft Non-Global Logarithms}
\label{ssec:SoftNGL}

In this section we perform a direct comparison of our hemisphere soft function with the
Monte Carlo program EVENT2.  The distribution in~\eq{eq:EVENT2dist} is related to the double cumulant in $m_1^2$ and $m_2^2$,
$\Sigma(m_1^{c2},m_2^{c2})$.  For a bin in $m_2^2$ between $m_{2,\min}^2$ and
$m_{2,\max}^2$, the value of the distribution in the bin will be
\begin{align} \label{eq:dcumdiff}
&\Sigma\big(m_1^{c2}, m_{2,\max}^2\big)
 - \Sigma\big(m_1^{c2}, m_{2,\min}^2\big) \nn \\
&\quad = \int_0^{m_1^{c2}} dm_1^2 \int_0^{m_{2,\max}^2} dm_2^2 \, \frac{d^2\sigma}{dm_1^2\,dm_2^2}  - \int_0^{m_1^{c2}} dm_1^2 \int_0^{m_{2,\min}^2} dm_2^2 \, \frac{d^2\sigma}{dm_1^2\,dm_2^2} \,.
\end{align}
This double cumulant is for the full distribution, meaning it is a sum of the
global terms in the distribution, the non-global terms that we have determined,
and terms that are power suppressed by $m_i^2/Q^2\ll 1$.  To test our analytic
results we will make use of the double cumulant for the non-global terms given
in~\eq{finalcumulant}.  The region we are interested in to avoid power
corrections and obtain sensitivity to non-global logs is 
\be
\label{eq:event2PS}
m_2^{2} \ll m_1^{c2} \ll Q^2\,.
\ee
As $m_1^c\to Q$, we become inclusive over a region of phase space,  $m_2^2 \ll m_1^{c2} \sim Q$, where additional logs of $m_2^2/Q$ that were power suppressed for $m_1^c\ll Q$ begin to contribute.  These additional logs are non-singular in $m_1$, e.g., $m_1 \ln(m_2/Q)$.  As we vary
$m_1^{c2}$ in the program, we observe the effects of such terms, and we can turn
them off by choosing $m_1^{c2} \ll Q^2$.

In our comparison to EVENT2 we focus on the $C_F C_A$ and $C_F T_R n_f$ color
structures, as the form of the $C_F^2$ color structure is constrained by
exponentiation.  To compare our calculation with the soft non-global logarithms
in the double hemisphere mass distribution, we subtract the full dependence in
these color structures for the double hemisphere mass distribution from EVENT2.

The level of numeric accuracy in the doubly differential distribution required
to resolve the contribution of the non-global terms is significant.  To achieve
this accuracy for the $C_F C_A$ terms we ran EVENT2 with $1.82\times 10^{12}$
events.  Additionally for the $C_F C_A$ color structure, we chose the parameters
${\rm CUTOFF} = 10^{-15}$ and ${\rm NPOW1}={\rm NPOW2}=3$.  The choice of CUTOFF
allows us to reliably sample $m_2^2$ down to $\log_{10} m_2^2/Q^2 = 10^{-7}$ and
the choices of NPOW1 and NPOW2 significantly reduce the statistical
uncertainties compared to the default choices ${\rm NPOW1}={\rm NPOW2}=2$.  For
the $C_F T_R n_f$ color structure, we found that taking ${\rm NPOW1}={\rm
  NPOW2}=4$ was necessary to obtain reliable results, and with this parameter
choice we ran $0.253\times 10^{12}$ events.  This choice of parameters, though,
introduces large systematic and statistical errors into the large $m_2^2$ region
of the distribution.  It appears to be more delicate to compare the results of
EVENT2 for this color structure with our calculation\footnote{The four-particle phase space generator in EVENT2 is optimized to sample the singular regions of phase space for a $q\bar{q}gg$ final state \cite{Seymour}. This may be a reason why the $C_F T_R n_f$ results, which have contributions from matrix elements with a $q\bar{q} q\bar{q}$ final state, are less consistent.}.

The full set of terms contributing to the distribution can be divided up into
groups.  In the limit we are working,  the cumulant terms
can be divided up as
\be
\Sigma = \Sigma^{\rm global} + \Sigma^{\ln^2} + \Sigma^{\ln} 
 + \Sigma^{\rm NG} + \Sigma^{\rm p.c.} \,.
\ee
The global singular terms determined by RG evolution in the dijet hemisphere
factorization theorem, are in $\Sigma^{\rm global}$.  The non-global double and
single log terms are in $\Sigma^{\ln^2}$ and $\Sigma^{\ln}$, and the remaining
non-global non-log terms are contained in $\Sigma^{\rm NG}$. Here $\Sigma^{\ln}$
and $\Sigma^{\rm NG}$ are determined by our calculation.  Additionally, there
are power corrections to the dijet limit that sit in $\Sigma^{\rm p.c.}$ and
contribute in EVENT2, but which are not known analytically.
\begin{figure}[t!]{
\includegraphics[width=0.99\textwidth]{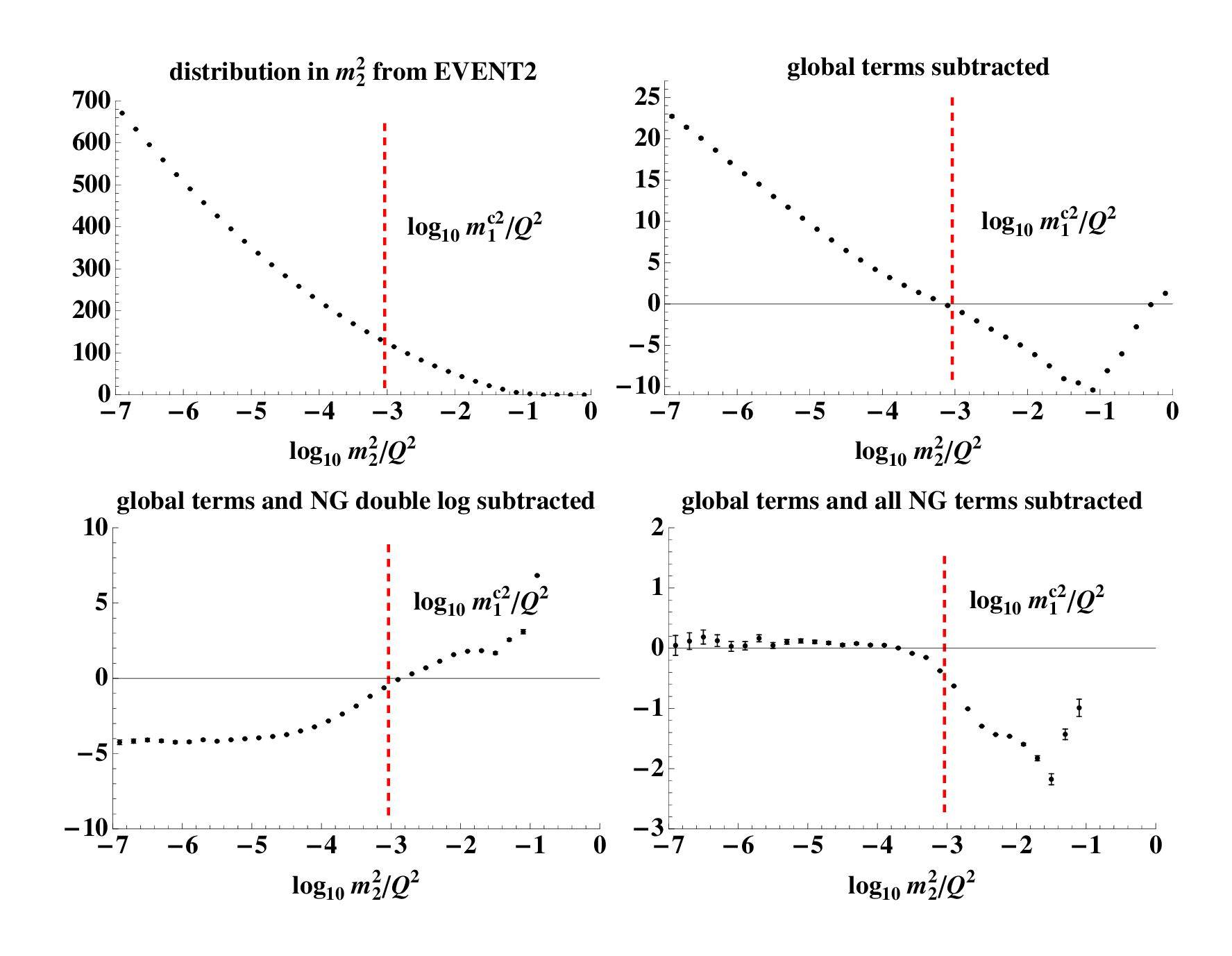}
{ \caption[1]{The $\cO(\alpha_s^2)$, $C_F C_A$ distribution from EVENT2 with
    various sets of terms removed.  The upper left plot shows the distribution
    directly from EVENT2.  The upper right plot has the known global terms
    subtracted, and the lower left plot has both the global terms and the known
    non-global double log in the cumulant subtracted.  The lower right plot has the global terms and the non-global terms from our calculation subtracted.  The red line marks the value of $m_1^c$ used.  Note the changing vertical scale between the plots.}
\label{fig:event2}}
}
\end{figure}

For both of the color structures $C_F C_A$ and $C_F T_R n_f$ we will show the
distribution in $m_2^2$ from EVENT2 with a cut on $m_1^2$, ${d\sigma_{\rm
    EV2}}/{dm_2^2} (m_1^{c2})$ \,, with various terms subtracted.  We choose the
cut in $m_1$ to be
\be
m_1^{c2}/Q^2 = 9.2\times 10^{-4} \,, \qquad \log_{10}(m_1^{c2}/Q^2) = -3.0 \,.
\ee
\begin{figure}[t!]{
    \includegraphics[width=0.99\textwidth]{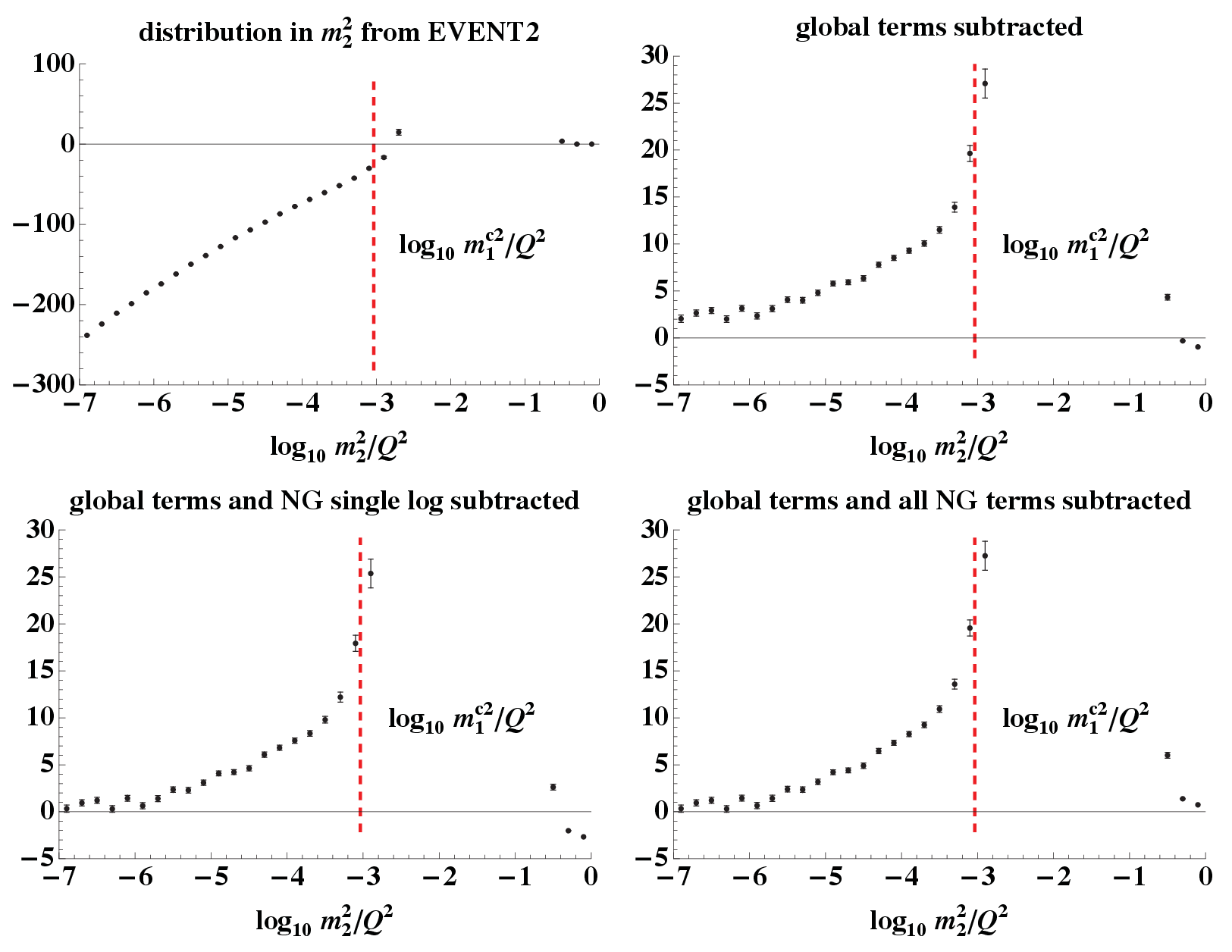} { \caption[1]{The
        $\cO(\alpha_s^2)$, $C_F T_R n_f$ distribution from EVENT2 with various
        sets of terms removed, as in~\fig{fig:event2}.  The upper left plot
        shows the distribution directly from EVENT2.  The upper right plot has
        the known global terms subtracted.  The lower left plot has the single
        log in the cumulant subtracted.
        The lower right plot has the global terms and all the non-global terms from
        our calculation subtracted.  The red line marks the value of $m_1^c$
        used.  Note the changing vertical scale between the plots.  The high
        region of the distribution has large numerical errors; see the
        discussion in the text for more details.}
\label{fig:event2NF}}
}
\end{figure}
Specifically, in~\figs{fig:event2}{fig:event2NF} we plot the
distributions
\begin{align}
\label{eq:event2dists}
\text{(top-left)}\qquad &\frac{d\sigma_{\rm EV2}}{dm_2^2} (m_1^{c2}) \,, 
 \\
\text{(top-right)}\qquad &\frac{d\sigma_{\rm EV2}}{dm_2^2} (m_1^{c2})
  - \frac{d\sigma_{\rm global}}{dm_2^2} (m_1^{c2}) \,, 
 \nn \\
\text{(bottom-left)}\qquad &\frac{d\sigma_{\rm EV2}}{dm_2^2} (m_1^{c2})
  - \frac{d\sigma_{\rm global}}{dm_2^2} (m_1^{c2}) -
 \,\,\,\,\frac{d\sigma_{\ln^2}}{dm_2^2} (m_1^{c2}) 
 \qquad \text{for }C_FC_A\,, \nn \\
&\frac{d\sigma_{\rm EV2}}{dm_2^2} (m_1^{c2})
  - \frac{d\sigma_{\rm global}}{dm_2^2} (m_1^{c2}) -
 \,\,\,\,\frac{d\sigma_{\ln}}{dm_2^2} (m_1^{c2}) 
 \qquad\ \text{for }C_FT_Rn_f\,, \nn \\
\text{(bottom-right)}\qquad &\frac{d\sigma_{\rm EV2}}{dm_2^2} (m_1^{c2}) -
 \,\,\frac{d\sigma_{\rm global}}{dm_2^2} (m_1^{c2}) - \frac{d\sigma_{\ln^2}}{dm_2^2} (m_1^{c2}) -
\frac{d\sigma_{\ln}}{dm_2^2} (m_1^{c2}) - \! \frac{d\sigma_{\rm NG}}{dm_2^2} (m_1^{c2}) \,. \nn 
\end{align}
In~\fig{fig:event2} we show the distributions in~\eq{eq:event2dists} for the
$C_F C_A$ color structure.  Note that a non-zero flat region in the distribution
indicates the presence of a single log of $m_2^{c2}$, while a line with non-zero
slope indicates the presence of a double log of $m_2^{c2}$.  We can clearly see
that after we subtract the global terms a double log remains.  Once we subtract
this double log coming from $\Sigma^{\ln^2}$, the remaining distribution still has
a single log of $m_2^{c2}$ in $\Sigma^{\ln}$.  Once we remove it and the remaining non-logarithmic non-global terms that are in $\Sigma^{\text{NG}}$, the residual distribution for $m_2^2 \ll m_1^{c2}$ for the
$C_F C_A$ terms is very small compared to the size of the non-global terms.  We
expect that power corrections in $m_1^{c2}/Q^2$ account for the remaining tiny
discrepancy from zero.

The nice agreement with EVENT2 for the $C_F C_A$ color structure is a very
strong check on our calculation.  The $m_2^2$ dependence of the $C_F C_A$
distribution has been fully accounted for by our calculation of the non-global
terms, up to a very small correction likely arising from power corrections in
$m_1^{c2}/Q^2$.

In~\fig{fig:event2NF} we show the distributions in~\eq{eq:event2dists} for the
$C_F T_R n_f$ color structure.  Note that there is no double log term to
subtract for this color structure, so the lower left plot subtracts the single
log.  Recall that the parameter choice ${\rm NPOW1}={\rm NPOW2}=4$, needed for
reliability in the small $m_2^2$ region, has introduced large statistical and systematic errors
in the large $m_2^2$ region, to the point that there is a clear deviation of the
EVENT2 results away from the true distribution for $m_2^2/Q^2 > 10^{-3}$.  However, we still see consistency 
between our calculation and the EVENT2 results in the small $m_2^2$ region.  
These uncertainties at large $m_2^2$ make the result less definitive as the $C_F C_A$ 
color structure, but it nonetheless gives us confidence in our results.

\subsection{Logarithms in the Hard Regime of One Hemisphere}
\label{ssec:hardglobal}

We have seen that our calculation of the soft non-global logarithms matches well
to the results of EVENT2.  These soft non-global logs appear in the region of
phase space where both jet masses are small, $m_{1,2}^2\ll Q^2$, and hence
showed up as elements of the leading-order dijet factorization theorem.  When
one hemisphere jet mass is large, of order the hard scale,
\begin{align} \label{eq:hardregion}
m_2^2 \ll m_1^{c2} \sim Q^2 \,,
\end{align} 
the structure of the logarithmic terms in $m_2$ is different.  In this regime,
the left hemisphere has two or more energetic jets that give a large value of
$m_1$ and the right hemisphere has one collimated jet that gives a small value
of $m_2$.  From the point of view of the dijet factorization theorem in
\eq{dijetcsfactorized} the new logarithmic terms that contribute for the region
in \eq{eq:hardregion} are power-suppressed terms, of the form $(m_1^2 / Q^2)^j
\ln^k (m_2^2 / Q^2)$ with $j\ge 1$.\footnote{Using SCET in the dijet limit these
  terms are described by factorization theorems involving power-suppressed hard,
  jet, and soft functions, order-by-order in $j$ (see  e.g.~\cite{Lee:2004ja}).
  When $N$ jets in the left hemisphere contribute to $m_1$, these terms can also
  be described by an $(N+1)$-jet factorization theorem in SCET (see 
  e.g.~\cite{Ellis:2010rw,Bauer:2008jx,Ellis:2009wj}).} We will refer to these
terms as coming from the hard regime of $m_1$.

The logarithms of $m_2$ in this hard regime have both global and non-global
sources.  Global logarithms come from collinear emissions from the jet in the
right hemisphere and soft emissions from all energetic jets.  Non-global
logarithms come from secondary soft emissions from soft gluons, as in our
calculation of non-global logarithms in the dijet regime.  These non-global
logarithms will be different than in the dijet regime, and require additional
calculations.  Since non-global logarithms in the hard regime start at
$\cO(\alpha_s^3)$ we will only consider global logarithms from the hard regime here.

To validate the above picture we will consider $\cO(\alpha_s^2)$ logarithms in
the hard regime and compare them to the results from EVENT2.  Specifically, we
will consider the double cumulant distribution with $m_1^c \sim Q$ and $m_2^c
\ll Q$.  The soft non-global logarithms that we have calculated dominate in the
small $m_1$ region, and are only part of the contributions to the double
cumulant when $m_1^c\sim Q$.  For this hard regime in $m_1^c$ we will compute
the double logarithmic dependence on $m_2^c$ for the $C_F C_A$ color structure.

At $\cO(\alpha_s^2)$, the hard regime logarithms come from collinear and soft
emissions from a configuration of three hard partons (a quark, anti-quark, and
gluon).  To have a large value of $m_1$, two of these partons must be in the
left hemisphere.  The tree-level cross section for this 3-parton configuration
is the well-known
\be \label{eq:sigma3tree}
\frac{d\sigma_3}{d x_q d x_{\bar q}} = \sigma_0 \frac{\alpha_s C_F}{2\pi} \frac{x_q^2 + x_{\bar q}^2}{(1-x_q)(1-x_{\bar q})} \,,
\ee
where $x_{q, \bar{q}} = 2E_{q,\bar{q}} / Q$ are the energy fractions of the
quark and anti-quark, and $x_g=2E_g/Q=2-x_q-x_{\bar{q}}$ is the energy fraction of
the gluon.  To determine the double logarithms in $m_2$, we consider
an emission of a soft gluon from any of the 3 jets or a collinear gluon from the
jet in the right hemisphere. For the double $m_2$ logarithm the sum of soft
emissions depends only on the color of the parton in the right hemisphere. Thus
the sum of all soft and collinear contributions yields a multiplicative factor
to~\eq{eq:sigma3tree},
\be
- \frac{\alpha_s}{2\pi} \, C_i\, 
 \ln^2 \Big( \frac{m_2^{c2}}{Q^2} \Big) \,,
\ee
\begin{figure}[t!]{ 
\begin{center}
 \includegraphics[width=0.7\textwidth]{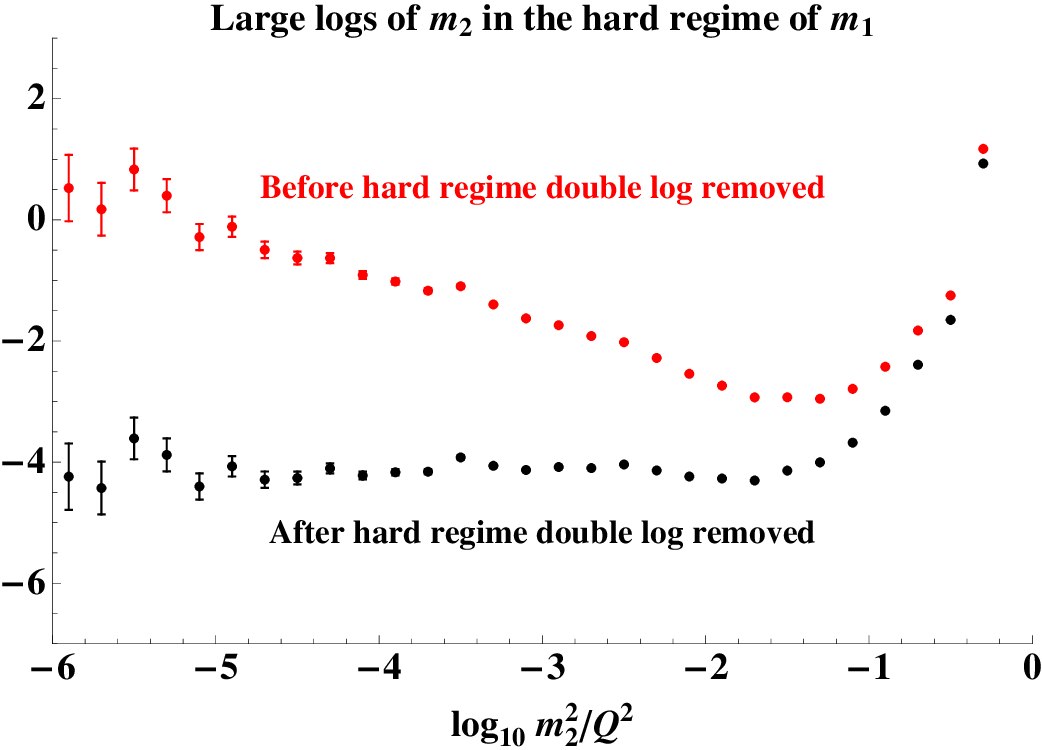}
\end{center} 
\vspace{-1em}
{\caption[1]{The EVENT2 distribution for the $C_F C_A$ color
    structure in $m_2^2 / Q^2$ before and after subtracting the double logarithm
    of $m_2^{c2}/Q^2$ from the regime of hard $m_1^{c2}$ given in~\eq{eq:harddoublelog}, for $m_1^{c2} / Q^2 =
    0.2$.  In both distributions we have subtracted the global dijet terms and the
    soft non-global terms.  After we subtract the additional double logarithm coming from the regime of hard $m_1^{c2}$, the
    remaining distribution at small $m_2$ is flat, and therefore  contains
    only a single logarithm of $m_2^{c2}/Q^2$.}
 \label{fig:harddoublelog}}
 }
\end{figure}
where $C_i$ is the Casimir invariant of the jet in the right hemisphere ($C_F$
for quarks or anti-quarks, $C_A$ for gluons).  Therefore for the $C_F C_A$
double logarithm the gluon jet is in the right hemisphere, and quark and
antiquark jets are in the left hemisphere.  All that remains is to convert the
distribution differential in $x_q$ and $x_{\bar q}$ in \eq{eq:sigma3tree} to a
cumulant in $m_1^2$.  Up to power corrections in $m_2^2 / Q^2$ we can use $m_1^2
= Q^2 (x_q + x_{\bar q} - 1)$ to determine $x_{\bar q}$.  The gluon being alone
in a hemisphere requires $2-x_q-x_{\bar{q}}=x_g>x_{q,\bar q}$ which yields the
limits on $x_q$. We also multiply by $1/2$ since we are requiring that the left
hemisphere mass be larger than the right, which restricts the gluon's angular
phase space. Thus in the hard regime for $m_1^{c2}$, the $C_F C_A$ double
logarithm in $m_2^{c2}$ is
\begin{align} \label{eq:harddoublelog}
 \Sigma^{\rm hard}(m_1^{c2},m_2^{c2})
 &= \frac{\sigma_0}{Q^4} \frac{\alpha_s^2 C_F C_A}{(2\pi)^2} \frac{1}{2}
 \!\int_0^{m_1^{c2}} \!\!\!\! d m_1^2
 \int_{2 \frac{m_1^{2}}{Q^2}}^{1 - \frac{m_1^{2}}{Q^2}} dx_q \:\,
  \frac{x_q^2 \plus (1 \minus x_q \plus m_1^2 / Q^2)^2}
 {(1\minus x_q)(x_q \minus m_1^2 / Q^2)} 
\left( - \ln^2\frac{m_2^{c2}}{Q^2} \right) \nn \\
%
% older form
%
%&= -\frac{\sigma_0}{Q^2} \left(\frac{\alpha_s}{2\pi}\right)^2 \frac{C_F C_A}{2}
% \Bigg\{ - \frac32 \frac{m_1^{c2}}{Q^2}
% + 3 \Big(\frac{m_1^{c2}}{Q^2} \Big)^2 
%   \nn \\
%& \quad 
%+ \Big( 2+\frac{m_1^{c2}}{Q^2} \Big) \frac{m_1^{c2}}{Q^2} 
%  \ln\frac{m_1^{c2}}{Q^2\minus 2m_1^{c2}}
%  + \bigg[ \frac54  - 4\ln \Big(2 - 2 \frac{m_1^{c2}}{Q^2} \Big) \bigg] 
%  \ln \Big(1 - 2 \frac{m_1^{c2}}{Q^2} \Big) \nn \\
%& \quad
%  +  \frac{\pi^2}{3} - 4 \Li_2 \Big( 1 - \frac{m_1^{c2}}{Q^2}\Big) 
%  - 4\Li_2 \Big( 2 \frac{m_1^{c2}}{Q^2} - 1\Big) \Bigg\} 
% \ln^2\Big( \frac{m_2^{c2}}{Q^2} \Big) 
%\nn\\
%
% new form
%
&= -\frac{\sigma_0}{Q^2} \left(\frac{\alpha_s}{2\pi}\right)^2 \frac{C_F C_A}{2}
 \frac{m_1^{c2}}{Q^2}  \Bigg\{ - \frac32 
 +  \frac{3m_1^{c2}}{Q^2} 
  + \Big( 2+\frac{m_1^{c2}}{Q^2}\Big)\ln\Big(\frac{m_1^{c2}}{Q^2\minus 2m_1^{c2}}\Big)
    \nn \\
& \quad 
+  \frac{Q^2}{m_1^{c2}} \bigg[ \frac{5}{4}  
  \ln \Big(1 - 2 \frac{m_1^{c2}}{Q^2} \Big) 
 + 4 \ln \Big(1 - \frac{m_1^{c2}}{Q^2} \Big) 
  \ln \Big(\frac{m_1^{c2}}{Q^2} \Big)
 + 4 \Li_2 \Big( \frac{m_1^{c2}}{Q^2}\Big) 
 \nn \\
& \qquad 
 - 4\Li_2 \Big(  \frac{2m_1^{c2}}{Q^2} \Big) 
  + 2\Li_2 \Big( \frac{4m_1^{c2}(Q^2-m_1^{c2})}{Q^4} \Big)\bigg] \Bigg\} 
 \ln^2\Big( \frac{m_2^{c2}}{Q^2} \Big) \,.
\end{align}
The overall $m_1^{c2}/Q^2$ shows that this result is power suppressed for
$m_1^{c2}\ll Q^2$, and in this limit the terms in $\{\cdots\}$ reduce to $-2
\ln(m_1^{c2}/Q^2)$.  If we set $m_1^{c2}/Q^2$ to the kinematic upper limit of
$1/ 3$, then \eq{eq:harddoublelog} reduces to the result in~\eq {eq:hardlogs}
with $m_2^{c2} / Q^2 = \rho_R$.

We can compare this additional double logarithm of $m_2^{c2}/Q^2$ from the regime of hard $m_1^{c2}$ to EVENT2 in the same way that
we compared the soft non-global logarithms.  In~\fig{fig:harddoublelog} we show
the effect of subtracting the $C_F C_A$ double log from the EVENT2 distribution
for the mixed distribution-cumulant cross section, $d\sigma/dm_2^2(m_1^{c2})$,
computed as in \eq{eq:dcumdiff}.  Choosing a value of $m_1^{c2} = 0.2\, Q^2$, we
first subtract the dijet global logarithms and soft non-global terms in $m_2^{c2}$
given by \eq{finalcumulant}, and plot the result as the red (upper) points
in~\fig{fig:harddoublelog}. We then subtract the double logarithm in $m_2^{c2}$
from the hard $m_1^{c2}$ regime given in \eq{eq:harddoublelog} and plot the result as the
black (lower) points in the figure.  It is clear that after subtracting the
double logarithm the remaining distribution contains just single logarithms of
$m_2^{c2}$, and therefore \eq{eq:harddoublelog} correctly accounts for 
double logarithms from the hard $m_1^{c2}$ regime.

\section{Conclusions, Simple Generalizations, and Outlook}
\label{sec:conclusions}

We have calculated the dijet hemisphere soft function to order $\as^2$ in both
position and cumulant momentum space, $\widetilde S(x_1,x_2,\mu)$ and
$\Scum(\la^c,\lb^c,\mu)$. In doing so, we uncovered the full non-global
structure of these functions, that is, the dependence on the ratios $x_1/x_2$ or
$\lac/\lbc$ which is not fixed by renormalization group invariance of the
factorized dijet cross section \eq{dijetcsfactorized}. We re-derived the
non-global double logarithm whose presence in related observables was pointed
out by \cite{Dasgupta:2001sh,Dasgupta:2002dc}, obtained new results for the single logarithmic
terms, as well as determining the full non-logarithmic non-global structure in
the dijet soft function and related observables. 
\eqs{finalposition}{finalcumulant} are the main results of this paper.  These
results go significantly beyond previous discussion of non-global structures in
jet cross sections at ${\cal O}(\alpha_s^2)$. 

The discovery of these non-global structures resolves the question of the
structure of the dijet soft function. The ansatz of Ref.~\cite{Hoang:2007vb}
that only double logarithms of $x_1/x_2$ and a constant could exist in
$\widetilde S(x_1,x_2,\mu)$ due in part to $x_1\leftrightarrow x_2$ symmetry, is
incomplete. Many other functions of $x_1/x_2$ appear, including a single
logarithm $\ln(x_1/x_2+x_2/x_1)$ that is symmetric under $x_1\leftrightarrow
x_2$. 

A key piece of our strategy to simplify calculation of the dijet soft function
was to take advantage of renormalization group properties and infrared
finiteness of the full soft function $\widetilde S(x_1,x_2,\mu)$ or
$S(\la,\lb,\mu)$ so that all the non-global dependence at $\cO(\as^2)$ could be
extracted only from those diagrams in which two particles in the final state
enter opposite hemispheres of the dijet event.  This procedure determines all
the terms in the soft function except for the constant term, which we take from
a recent calculation of this constant in the $\cO(\as^2)$ thrust
distribution by \cite{Monni}. Cancellation of IR divergences between same and
opposite hemisphere final-state configurations leaves over logs of $x_1/x_2$ or
$\la/\lb$, illuminating how NGLs find their origin in the structure of a soft
function \cite{scetchris}.

In this paper we explored non-global logarithms that arise due to the
sensitivity of an observable to soft radiation at two different scales in
sharply divided regions, i.e. $m_1$ and $m_2$ in opposite hemispheres.  These
observables can naturally be taken to be similar, $m_1\sim m_2$, or
parametrically disparate, $m_1\gg m_2$. The NGLs of $\ln(m_1/m_2)$ become large
in the latter case, and are distinguished from global logs $\ln(m_i/Q)$ which
are the only large logs in the former case.  Ref.~\cite{Dasgupta:2002dc}
distinguished observables with soft radiation measured everywhere, from those in
which soft radiation is probed with a single scale in only a portion of phase
space, e.g.  $m_1$ in one hemisphere (the NGLs originally studied in
\cite{Dasgupta:2001sh}).  We found in the case of hemisphere masses that NGLs of
the latter type can be derived from those where soft radiation is measured
everywhere, by integrating the variables for the unmeasured regions up to the
hard regime.  Thus from an effective field theory point of view which focuses
on hierarchies of scales, we find it more economical to think of NGLs as being
of a single type. We consider the measurement of independent observables like
$m_1$ and $m_2$ that cover all of phase space, but separately covering
sharply divided phase space regions. These kinematic variables can naturally be
taken similar ($m_1\sim m_2$), or can be taken as disparate ($m_1\gg m_2$). In
the former case it suffices to treat NGLs in fixed order perturbation theory. In
the latter case we have large NGLs that appear from soft radiation that probes
the sharply divided regions, and NGL resummation~\cite{Dasgupta:2001sh} must be
carried out.  Effectively we consider NGLs with all variables unintegrated and
derive  NGLs for other cases by integration.  Keeping the variables
unintegrated facilitates the task of distinguishing logs of global and
non-global origin.

We performed several cross-checks of our results, by considering general
covariant gauge, by projecting the dijet soft function onto various observables
and comparing to EVENT2 predictions for those observables. These comparisons
revealed existing EVENT2 extractions for the difference between heavy jet mass
and thrust soft function constants to be in excellent agreement with our
analytic result. We also compared our prediction for the double hemisphere mass
distribution to our own EVENT2 runs, and again found good agreement. Finally we
considered the regime of one hemisphere mass $m_1$ becoming hard and generating
additional global logs of $m_2/Q$. Our calculation of the additional double log
in this regime also matched our EVENT2 runs very well. We also verified the
above-mentioned relation between observables containing NGLs with soft radition
measured everywhere, versus measured only in a region of phase space, 
by integrating $m_1$ all the way up to its kinematic limit and reproducing the
global and non-global double logs of $\rho_R$ in \eqs{DSNGL}{eq:hardlogs} at
$\mathcal{O}(\as^2)$.

State-of-the-art fits for $\alpha_s$ from $e^+e^-\to {\rm jet}$ data currently
rely on N$^3$LL resummations of event shape
distributions~\cite{Becher:2008cf,Abbate:2010xh,Chien:2010kc} and thus on the
fixed-order $\cO(\as^2)$ dijet soft function. Our new results make possible
improved extractions of the strong coupling by removing a source of uncertainty
in the perturbative calculation. 

The $\cO(\as^2)$ soft function in position space, in particular, makes possible
a fully analytic N$^3$LL resummation of the doubly differential cross section
\eq{positionfactorization} in position space when $x_1/x_2\sim 1$, and is also a
crucial ingredient in the N$^3$LL resummation of the doubly differential
momentum space distribution \eq{dijetcsfactorized} or cumulant
\eq{dijetcumfactorized}.

One simple generalization of our results is to the case of hadron colliders. For
$pp$ and $p\bar p$ collisions an incoming dijet hemisphere soft function appears
for the event shape called beam thrust $\tau_B$ (or $0$-jettiness), which in the
dijet limit $\tau_B\ll 1$ provides a veto on central jets.  The only difference
between our soft function in Eq.~(\ref{hemisoftdef2}) and the incoming one are
that the Wilson lines are incoming from $-\infty$ rather than outgoing to
$+\infty$.  This reflects itself in the signs of $i0^+$ factors in the eikonal
denominators in the momentum space amplitudes.  Since for our opposite
hemisphere calculation the signs of those $i0^+$ factors drop out, our analytic
results for the $x_1/x_2$ and $\lac/\lbc$ dependence immediately carries over
to the incoming case. In Ref.~\cite{Stewart:2010qs} it was proven that the $\mu$-dependent logarithms are also the same for the outgoing and incoming dijet
hemisphere soft functions.  Hence at ${\cal O}(\alpha_s^2)$ the only thing that
can differ between the two cases is the thrust constant $s_2^{[C_F C_A,n_f]}$.
(It would be interesting to compute these constants directly for the incoming
case.) Thus our results already make possible an improvement of predictions for
jet observables in hadron collisions.

Another simple generalization of our results are to Wilson lines in different
color representations, such as octets for incoming or outgoing gluon dijets. It
is straightforward to see that our computations obey Casimir scaling, so we can
simply modify the $C_F$'s to get results for other representations. For the
$\cG$, $\cH$, and $\cQ$ graphs only traces that obey Casimir scaling appear. 
For the $\cI$ and $\cT$ graphs non-abelian exponentiation and the
generator commutation relations guarantee that Casimir scaling also applies.

Looking ahead, relating non-global logarithms to the soft function in a
factorization theorem in the context of effective field theory as we have done
also opens the door to the development of renormalization group techniques to
resum non-global logarithms. This would make it possible to go beyond the
resummation of NGLs in the conjectured form given in \cite{Dasgupta:2001sh}
which is parameterized in terms of coefficients determined numerically in the
large-$N_C$ limit. It is important to keep in mind that such NGLs (and other
non-global structures as we found here) will appear not only in the single or
double-hemisphere mass distributions, but also in most jet cross sections in
which jets are defined with one of the well known jet algorithms, which leave
the regions inside and outside these jets sensitive to different soft scales
\cite{Banfi:2010pa,scetchris}.

\paragraph{Note added:} While this paper was being finalized,
Ref.~\cite{Kelley:2011ng} appeared calculating the dijet hemisphere soft
function in momentum space, $\Scum(\la^c,\lb^c,\mu)$. Our result in
Eq.~(\ref{finalcumulant}) exactly agrees numerically with theirs, even though
the two have different algebraic forms.  In our algebraic result the logarithmic
singularities are separated out and the remaining polylog dependent
contributions [our $f(\lac/\lbc)+f(\lbc/\lac)$ given by \eq{fQfN}] are bounded.
An analog of our two dimensional position space result $S(x_1,x_2,\mu)$
in~\eq{finalposition} was not obtained in Ref.~\cite{Kelley:2011ng}. On the other hand, unlike our paper Ref.~\cite{Kelley:2011ng} does in addition compute
analytically the thrust constants, finding $s_2^{[C_F
  C_A]}=-\frac{1070}{81}-\frac{871}{108}\pi^2 + \frac{7}{15}\pi^4 +
\frac{143}{9}\zeta_3$ and $s_2^{[n_f]}=\frac{40}{81}+\frac{77}{27}\pi^2 -
\frac{52}{9}\zeta_3$. These agree numerically with the first four digits of the
result from Ref.~\cite{Monni} quoted in~\eq{s2Monni}. In the publication
corresponding to~\cite{Monni}, which is Ref.~\cite{Monni:2011gb}, analytic
results were also obtained for the thrust constants, and these agree with those of
Ref.~\cite{Kelley:2011ng}.

%%%%%%%%%%%%%%%%%%%%%%%%%%%%%%%%%%%%%%%%%%%%%%%%%%%%%%%%%%%%%%%%%%%%%%%%%%%%%%%%%%%%%%%%%%%%%%%%%%%%%%%%%

\acknowledgments The authors would like to thank the University of California at
Berkeley, Lawrence Berkeley National Laboratory, Massachusetts Institute of
Technology, University of Washington and Harvard University for hospitality
during portions of this work, and CL would like to thank the Aspen Center for
Physics for the same. We thank M. Dasgupta, G. Salam, and P.~Pietrulewicz for insightful comments and M. Seymour for advice on EVENT2. IS thanks M.~Schwartz for useful discussion. This work is supported in part by the Offices of Nuclear and High Energy Physics of the U.S. Department of Energy under Contracts
DE-FG02-96ER40956, DE-FG02-94ER40818, DE-SC003916, and DE-AC02-05CH11231.  The
work of JW was supported in part by a LHC Theory Initiative Postdoctoral
Fellowship, under the National Science Foundation grant PHY-0705682, and that of
IS by the Alexander von Humboldt foundation.
\appendix

%%%%%%%%%%%%%%%%%%%%%%%%%%%%%%%%%%%%%%%%%%%%%%%%%%%%%%%%%%%%%%%%%%%%%%%%%%%%%%%%%%%%%%%%%%%%%%%%%%%%%%%%%
\section{Anomalous Dimensions}
\label{appx:anomdims}
%%%%%%%%%%%%%%%%%%%%%%%%%%%%%%%%%%%%%%%%%%%%%%%%%%%%%%%%%%%%%%%%%%%%%%%%%%%%%%%%%%%%%%%%%%%%%%%%%%%%%%%%%

In~\sec{ssec:RGconstraints} we reviewed how the $\mu$-dependent terms in the soft function can be constrained using renormalization group invariance of the cross section \eq{dijetcsfactorized}. Here we record the soft anomalous dimensions that are needed to carry this out explicitly to $\mathcal{O}(\as^2)$. 

The exponents $K,\omega$ appearing in the evolution factors $U_S(x,\mu,\mu_0)$
in \eq{evolutionkernel} are defined by
\begin{subequations}
\label{Komegadefs}
\begin{align}
K(\Gamma_{\text{cusp}},\gamma_S,\mu,\mu_0)  &=  \int_{\mu_0}^\mu \frac{d\mu'}{\mu'}\left(-2\Gamma_{\text{cusp}}[\alpha_s(\mu')]\ln\frac{\mu'}{\mu_0} + \gamma_S[\alpha_s(\mu')]\right), \\
\omega(\Gamma_{\text{cusp}},\mu,\mu_0)  &= -2\int_{\mu_0}^\mu\frac{d\mu'}{\mu'}\Gamma_{\text{cusp}}[\alpha_s(\mu')]\,,
\end{align}
\end{subequations}
where the cusp and non-cusp parts of the anomalous dimension
$\Gamma_{\text{cusp}}$ and $\gamma_S$ are defined by \eq{anomdimdef}. We define
the expansions of the anomalous dimensions in $\as$ as
\be
\label{Gammagamma}
\Gamma_{\text{cusp}}[\alpha_s] = \sum_{k=0}^\infty \left(\frac{\as}{4\pi}\right)^{k+1} \Gamma_{\text{cusp}}^k \quad , \quad \gamma_S[\alpha_s] = \sum_{k=0}^\infty \left(\frac{\as}{4\pi}\right)^{k+1} \gamma_S^k\,.
\ee
At $\cO(\as)$ and $\cO(\as^2)$ \cite{Korchemsky:1987wg,Korchemskaya:1992je}, the cusp anomalous dimension is given by 
\begin{subequations}
\label{cusp01}
\begin{align}
\Gamma_{\text{cusp}}^0 &= 4 C_F \\
\Gamma_{\text{cusp}}^1 &= 4 C_F C_A \left(\frac{67}{9}-\frac{\pi^2}{3}\right) - C_F T_R n_f \frac{80}{9} \, ,
\end{align}
\end{subequations}
and the non-cusp anomalous dimension is given by
\begin{subequations}
\label{noncusp01}
\begin{align}
\gamma_S^0 &= 0 \\
\gamma_S^1 &= C_F C_A \left(-\frac{808}{27} + \frac{11\pi^2}{9} + 28\zeta_3\right) + C_F T_R n_f\left(\frac{224}{27} - \frac{4}{9}\pi^2\right)\,.
\end{align}
\end{subequations}
We also need the coefficients of the beta function,
\be
\label{betafunc}
\beta[\as] = -2\as\sum_{k=0}^\infty \left(\frac{\as}{4\pi}\right)^{k+1} \beta_k\,,
\ee
where
\begin{subequations}
\label{beta01}
\begin{align}
\beta_0 &= \frac{11}{3}C_A - \frac{4}{3}T_R n_f \,, \\
\beta_1 &=  \frac{34}{3}C_A^2 - \frac{20}{3}C_A T_R n_f - 4 C_F T_R n_F\,.
\end{align}
\end{subequations}
In terms of these beta function coefficients, the two-loop running coupling is given by
\be
\label{2loopalpha}
\frac{1}{\as(\mu')} = \frac{1}{\as(\mu)} + \frac{\beta_0}{2\pi}\ln\frac{\mu'}{\mu} + \frac{\beta_1}{4\pi\beta_0}\ln\left[1 + \frac{\beta_0}{2\pi}\as(\mu)\ln\frac{\mu'}{\mu}\right]
\ee
These are all the quantities needed to determine the $\mu$-dependent pieces of the soft function to $\cO(\as^2)$.\footnote{To achieve NNLL accuracy in the resummed soft function \eq{softevolution1} it is actually necessary to know the cusp anomalous dimension and beta function at $\cO(\as^3)$ as well, but in this paper we are only concerned with the fixed-order result.}

The quantities $K,\omega$ to $\cO(\as^2)$ written out explicitly are
\begin{subequations}
\label{Komegaexplicit}
\begin{align}
K(\Gamma_{\text{cusp}},\gamma_S,\mu,\mu_0) &= \frac{\as(\mu)}{4\pi} \left(-\Gamma_{\text{cusp}}^0\ln^2\frac{\mu}{\mu_0} + \gamma_S^0 \ln\frac{\mu}{\mu_0}\right) \\
&\quad + \left(\frac{\as(\mu)}{4\pi}\right)^2 \left[-\frac{2}{3}\Gamma_{\text{cusp}}^0\beta_0\ln^3\frac{\mu}{\mu_0} + (\gamma_S^0\beta_0 - \Gamma_{\text{cusp}}^1)\ln^2\frac{\mu}{\mu_0} + \gamma_S^1 \ln\frac{\mu}{\mu_0}\right] \nn \\
\omega(\Gamma_{\text{cusp}},\mu,\mu_0) &= \left(\frac{\as(\mu)}{4\pi}\right)\left(-2 \Gamma_{\text{cusp}}^0 \ln\frac{\mu}{\mu_0}\right) \\
&\quad - \left(\frac{\as(\mu)}{4\pi}\right)^2 \left(2\Gamma_{\text{cusp}}^0\beta_0\ln^2\frac{\mu}{\mu_0} + 2\Gamma_{\text{cusp}}^1\ln\frac{\mu}{\mu_0}\right)\nn \,.
\end{align}
\end{subequations}
These expressions can be used to determine the parts of the soft function dependent on logs of $\mu$, \eq{Rposition} in position space and \eq{Rcumulant} for the double cumulant in momentum space.

%%%%%%%%%%%%%%%%%%%%%%%%%%%%%%%%%%%%%%%%%%%%%%%%%%%%%%%%%%%%%%%%%%%%%%%%%%%%%%%%%%%%%%%%%%%%%%%%%%%%%%%%%
\section{ $\cO(\as^2)$ Diagram Results for the Opposite Hemisphere Soft Function}
\label{appx:twoloopcalc}
%%%%%%%%%%%%%%%%%%%%%%%%%%%%%%%%%%%%%%%%%%%%%%%%%%%%%%%%%%%%%%%%%%%%%%%%%%%%%%%%%%%%%%%%%%%%%%%%%%%%%%%%%

Here we present the details of our calculation of the $\cO(\as^2)$ opposite
hemisphere soft function, in both position and momentum space.  The opposite
hemisphere soft function in momentum space is given by
\be
\label{eq:Sopp}
\Sopp(\la,\lb) = \sum_j \int \meas{k_1} \meas{k_2} \cA_j(k_1,k_2) {\cal
  M}^{[LR]}_{k_1,k_2}(\la,\lb) \mathcal{C}(k_1)\mathcal{C}(k_2) \,,
\ee
where $\cC(k)$ is the cut propagator, Eq.~(\ref{origeq:cutdelta}), we sum over
the squared matrix elements from different classes of diagrams $\cA_j$ where $j=\{
\cI,\cT,\cG,\cH,\cQ \}$ (see~\fig{fig:diagrams}), and the momentum space
measurement function is
\begin{align}
 {\cal M}^{[LR]}_{k_1,k_2}(\la,\lb) =& \:\delta(\la - \kam) \delta(\lb - \kbp)  \theta(\kap - \kam) \theta(\kbm - \kbp)  \nn\\
  & + \delta(\la - \kbm)\delta(\lb - \kap) \theta(\kam - \kap) \theta(\kbp - \kbm) \,. \nn
\end{align}
Below we will denote the integrated contribution of each diagram $j$ to  $\Sopp$ in \eq{eq:Sopp} by the symbols $\cI,\cT,\cG,\cH,\cQ$ themselves.
The position space opposite hemisphere soft function is given by replacing the
measurement function above by its position space analog, ${\cal
  M}_{k_1,k_2}^{[LR]}(x_1,x_2)$ in Eq.~(\ref{eq:CLR}). The integrated result analogous to \eq{eq:Sopp} for each diagram's contribution to $\widetilde S^{\text{opp}}(x_1,x_2)$ will be denoted  $\widetilde \cI, \widetilde\cT,\widetilde\cG,\widetilde\cH,\widetilde\cQ$. More generally, the
measurement function for other observables, ${\cal M}_{k_1,k_2}$, can be used
with our results for the double cut squared matrix elements, $\cA_j$, to obtain
the contribution to the soft function for other observables. 

The observable and hemisphere geometry determine some natural variables for the two-loop calculation.  The measurement function and the diagrams are symmetric in
$k_1 \leftrightarrow k_2$, and therefore we focus only on the case where $k_1$
contributes to $\la$ and $k_2$ contributes to $\lb$.  The initial matrix
elements are straightforward to write in light-cone coordinates, and the
measurement function $\cM^{[LR]}_{k_1,k_2}$ in~\eq{eq:CLR} implements the
conversion
\be
\label{eq:transformstart}
\kam \to \la \,, \qquad \kbp \to \lb \,.
\ee
It is natural to then scale out by these variables.  We find that a pair of
variables particularly useful for the calculations are
\be
z \equiv \sqrt{\frac{\kap}{\kam}\frac{\kbm}{\kbp}} 
 = \sqrt{\frac{\kap}{\la}\frac{\kbm}{\lb}} \,, \qquad u
  \equiv \sqrt{\frac{\kap/\kam}{\kbm/\kbp}} 
 = \sqrt{\frac{\kap\lb}{\kbm\la}} \,.
\ee
The phase space cuts constrain $z > 1$, $u > 0$.  Physically, the ratios
$\kap/\kam$ and $\kbm/\kbp$ are related to the polar angles of the partons with
respect to their jet directions.  For example, $\kap/\kam = \tan^2 (\theta_1/2)$,
where $\theta_1$ is the polar angle of the parton with respect to the jet
(thrust) axis.

The $z$ variable measures how close to the boundary the combination of the two
partons are: as $z\to1$, both partons move towards the hemisphere boundary, and
as $z\to\infty$, the angular separation between the partons grows.  $u$ measures
the relative angles of the two partons. It is worth noting that divergences from
$z\to 1$ play a role in the appearance of non-global logarithms. 

With the exception of the $\cI$ matrix elements that are straightforward to
evaluate, we find a common set of transformations are useful to simplify the
integral form of the matrix elements.  The matrix elements depend on the set of
variables 
\be
 \{ \kap, \kam, \kbp, \kbm, \phi\} \,, 
\ee 
where $\phi$ is the angle between $k_1$ and $k_2$ in the plane transverse to the
thrust axis. Our matrix elements depend only on $\cos\phi$, which arises from
$k_1^\perp\cdot k_2^\perp$.  Starting with the two-loop measure, we can perform
some of the integrals to obtain
\begin{align}
\int \meas{k_1} \meas{k_2} \cC(k_1)\, \cC(k_2) &= \frac{1}{(16\pi^2)^2} \frac{(e^{\gamma_E})^{2\e}}{\Gamma(1-\e)^2}  \int_0^{\infty} d\kap d\kam d\kbp d\kbm (\kap\kam\kbp\kbm)^{-\e} \nn \\
&\quad \times  \left(\frac{\pi^{1/2}\Gamma(\frac12-\e)}{\Gamma(1-\e)}\right)^{-1}\int_0^{\pi} d\phi \sin^{-2\e} \phi \,.
\end{align}
The measurement function then allows us to perform the $\kam$ and $\kbp$
integrals.  For the opposite hemisphere contributions shifting to $z$ and $u$
requires the change of variables
\be
\int_0^{\infty} \frac{d\kap}{\kam} \frac{d\kbm}{\kbp} \theta(\kap > \kam) \theta(\kbm > \kbp) = \int_1^{\infty} dz \int_{1/z}^{z} du \frac{2z}{u} \,.
\ee
This change of variables is convenient because in all the diagrams the $u$
integration is independent of $\e$ and can be easily performed.  Additionally,
the $\phi$ dependent terms in the matrix elements, which give hypergeometric
functions when integrated, are independent of $u$.  This means that we can
simultaneously perform the $u$ and $\phi$ integrals, leaving only a
single integral over $z$ to be performed in each diagram.  Standard techniques can then be used to evaluate
these $z$ integrals.

In addition to a common strategy for evaluating the integrals, several common
functions arise in the calculation. Using the $z, u$ variables, a function of
$r$ that appears in nearly all the diagrams is 
\be
\label{eq:gfunc}
g(z,r) \equiv \ln\left(\frac{(1+r)^2 z}{(z+r)(1+z r)}\right) \,.
\ee
This function is symmetric in $r\to1/r$ and scales as $r$ for small $r$.  In
addition, there are two common hypergeometric functions that arise from
integrations over $\phi$:
\begin{align}
\left(\frac{\pi^{1/2}\Gamma(\frac12-\e)}{\Gamma(1-\e)}\right)^{-1}\int_0^{\pi} d\phi \, \sin^{-2\e}\phi \,\frac{1}{1 + z^2 - 2z\cos\phi} &= \frac{1}{z^2 + 1} f_1 \big(z^2 \big) \,,  \\
\left(\frac{\pi^{1/2}\Gamma(\frac12-\e)}{\Gamma(1-\e)}\right)^{-1}\int_0^{\pi} d\phi \, \sin^{-2\e}\phi \,\frac{1}{(1 + z^2 - 2z\cos\phi)^2}  &= -\frac{1}{(z^2+1)^2}\left[f_1\big(z^2 \big) -2 f_2 \big(z^2 \big) \right] \,, \nn
\end{align}
where
\begin{align}
\label{eq:hypgeo}
f_1(z) &\equiv  \F\left(\frac12,1,1-\e,\frac{4z}{(1+z)^2}\right) \,, \nn \\
f_2(z) &\equiv  \F\left(\frac12,2,1-\e,\frac{4z}{(1+z)^2}\right) \,.
\end{align}
It is necessary to expand these functions to $\cO(\e^2)$:
\begin{align}
\label{eq:transformend}
f_1(z) &= (1+z)\,z^{2\e}\,(z-1)^{-1-2\e}\left[1 + 2\e^2 \Li_2(1/z)\right] + \cO(\e^3) \,, \nn \\
f_2(z) &= (1+z)\,z^{2\e}\,(z-1)^{-3-2\e} \bigg[1+z^2 + 2\e(1+z) \nn \\ 
&\qquad\qquad\qquad - 2\e^2 (z^2-1) \ln\left(\frac{z-1}{z}\right) + 2\e^2 (1+z^2)\Li_2\left(\frac{1}{z}\right) \bigg] + \cO(\e^3)
\end{align}
Note that we have summed up certain terms at higher orders in $\e$ in these
expansions.  Throughout the calculation we made use of the \texttt{HypExp}
package \cite{Huber:2005yg,Huber:2007dx}.  The divergent structure in $\e$ of
the vacuum polarization diagrams is quite complex, and careful treatment is
necessary to ensure the divergences are regulated and terms are not forgotten.
The difficulty in evaluating these diagrams comes from the hypergeometrics;
using certain hypergeometric forms we can see more clearly the regulation of the
divergences in the integrand about $z=1$.

\subsection{Form of the Matrix Elements in Momentum and Position Space}

Before giving results for the matrix elements, we discuss the general form of
the soft function.  We determine the matrix elements in both momentum and
position space.  In position space the soft function is a simple function of the
variables $x_{1}$ and $x_2$, which are conjugate to $\la,\lb$.  In momentum
space the $\la,\lb$ dependence gives distributions (rather than functions), but
the result does not take its simplest form in these variables.  Instead, we use
variables
\be
s = \la\lb \,, \qquad r = \frac{\lb}{\la} \,, \qquad \int_0^{\infty} d\la\, d\lb = \int_0^{\infty} ds\, dr \, \frac{1}{2r} \,.
\ee
The distributions in terms of these variables are simple: all matrix elements
contain a distribution of $s$ and non-singular functions of $r$.  The general
form of the opposite hemisphere terms in momentum space is
\be
\label{eq:Soppsum}
\Sopp(\la,\lb) = \sum_i S^{\opp \, [i]}(\la,\lb) \,,
\ee
where
\be
\label{eq:generalmomspace}
S^{\opp \, [i]} (\la,\lb) = A\, C_i \,s^{-1-2\e} \cF(r)
\ee
and $C_i$ is a color structure, ($C_F^2$, $C_F C_A$, or $C_F T_R n_f$), $\cF(r)$ depends on the class of diagram, $(
\cI,\cT,\cG,\cH,\cQ )$, and in $\overline{\textrm{MS}}$
\be
A \equiv \left(\frac{\alpha_s}{2\pi}\right)^2 \frac{(e^{\gamma_E}\mu^2)^{2\e}}{\Gamma(1-\e)^2} \,.
\ee
The transformation from $s,r$ space to the position-space matrix element is
\be
\label{eq:S2Fourierform}
\widetilde S^{\opp \, [i]}(x_1,x_2) = A C_i
 \int_0^\infty ds\, dr\, \frac{1}{2r} \, 
  e^{-i \sqrt{s}( x_1/\sqrt{r} + x_2\sqrt{r})} s^{-1-2\e} \cF(r)\,.
\ee
Convergence of the Fourier transform requires $x_{1,2}$ to have a small negative imaginary component:
\be
\label{eq:i0prescription}
x_1 - i0^+ \,, \qquad x_2 - i0^+ \,.
\ee
Because both $s$ and $r$ are positive, the sign of $i0^+$ relative to $x_{1,2}$
is unchanged for the entire Fourier transform.  We will often write just
$x_{1,2}$ with the prescription in~\eq{eq:i0prescription} implied.

In all matrix elements the function $\cF(r)$ is non-singular, meaning it can
be split into a pure constant plus a nontrivial function of $r$,
\be
\label{eq:Fdef}
\cF(r) = F_0 +  F_1(r)\,,
\ee
where $F_1(r)$ is symmetric in $r\to 1/r$, and vanishes as $r\to 0,\infty$.  The $s$ integral can be done immediately, giving
\be
\widetilde S^{\opp \, [i]}(x_1,x_2)  =  A C_i\, \Gamma(-4\e)\int_0^\infty \frac{dr}{r}\left[F_0 + F_1(r)\right]  \left[i\left(\frac{x_1}{\sqrt{r}} + x_2\sqrt{r}\right)\right]^{4\e}\,.
\ee
The $r$ integral in the $F_0$ piece can also be done immediately, using
\be
\int_0^\infty \frac{dr}{r}\left[i\left(\frac{x_1}{\sqrt{r}} + x_2\sqrt{r}\right)\right]^{4\e} = (ix_1 i x_2)^{2\e}\frac{\Gamma(-2\e)^2}{\Gamma({-4\e})}\,.
\ee
Then each contribution to the opposite hemisphere soft function can be written
\be
\label{eq:S2Fourierformb}
\widetilde S^{\opp \, [i]}(x_1,x_2) = A C_i (ix_1 ix_2)^{2\e} \left\{\Gamma(-2\e)^2 F_0 + \Gamma(-4\e)\int_0^\infty \frac{dr}{r} F_1(r) \left(\sqrt{\frac{b}{r}} + \sqrt{\frac{r}{b}}\right)^{4\e} \right\}\,,
\ee
where
\be
b \equiv \frac{x_1 - i0^+}{x_2 - i0^+} \,.
\ee
The function $F_1(r)$ always vanishes sufficiently quickly as $r\to 0$ and $r\to\infty$ so that the $r$ integral introduces no additional powers of $1/\e$, and thus the integrand can be expanded and truncated at $\cO(\e)$. Splitting the integral into the regions $0<r<1$ and $1<r<\infty$, and using the symmetry of $F_1(r)$ in $r\to 1/r$, we can rewrite the last integral as
\be
\int_0^1 \frac{dr}{r} F_1(r) \left[2 + 4\e\ln\left(r+\frac{1}{r} + b + \frac{1}{b}\right) + \cO(\e^2)\right]\,.
\ee
The last integral displays logarithmic dependence on $b$ and $1/b$, which can be extracted by
\be
\ln\left(r+\frac{1}{r} + b + \frac{1}{b}\right) =  \ln\left(2 + b + \frac{1}{b}\right) + \ln\left(\frac{r+\frac{1}{r} + b + \frac{1}{b}}{2+b+\frac{1}{b}}\right).
\ee
Then \eq{eq:S2Fourierformb} can be written
\be
\label{eq:S2FourierFinal}
\begin{split}
\widetilde S^{\opp \, [i]}(x_1,x_2) = AC_i (ix_1 ix_2)^{2\e} &\Biggl\{\Gamma(-2\e)^2 F_0 + \Gamma(-4\e) \left[2 + 4\e\ln\left(2 + b + \frac{1}{b}\right)\right] \int_0^1 \frac{dr}{r} F_1(r) \\
& +4\e\, \Gamma(-4\e) \int_0^1\frac{dr}{r} F_1(r) \ln\left(\frac{r+\frac{1}{r} + b + \frac{1}{b}}{2+b+\frac{1}{b}}\right) \Biggr\}\,.
\end{split}
\ee 
In each term, the $F_{0,1}$ also have expansions in $\e$, and should
be kept/truncated to the appropriate order.  In particular, we must keep terms
of up to order $\e^2$ in $F_0$, but only to order $\e$ in $F_1(r)$
(and only $\e^0$ in the last integral). Note that in the last integral the
integrand now vanishes at both endpoints, $r\to 0$ (due to $F_1(r)$) and
$r\to 1$ (due to the log).

We now give the result for each group of diagrams in both position and momentum
space, as well as the intermediate step where only the $z$ integral remains.

\subsection{$\cI$ and $\cT$ diagrams}

The matrix element for the independent emission diagrams in \fig{fig:diagrams}(a) is a sum of contributions with color factors $C_F^2$ and $C_F C_A$, $\cA_\cI = \cA_{\cI,C_F^2} + \cA_{\cI,C_F C_A}$. The $C_F^2$ piece is
\be
\label{eq:CFsqdiagram}
\cA_{\cI,C_F^2} = 8 g^4 \mu^{4\e}C_F^2 \frac{1}{k_1^+ k_1^-}\frac{1}{k_2^+ k_2^-}\,,
\ee
This diagram is the product of one-loop diagrams with one gluon in each hemisphere (neglecting the delta function in the one-loop diagrams for the hemisphere with no parton).  In position space, the result of the diagram's contribution to $\widetilde S^{\text{opp}}$ is
\be
\label{eq:Iposition}
\widetilde{\cI}_{C_F^2} = A\, {C_F^2} (ix_1 ix_2)^{2\e} \,\Gamma(-2\e)^2 \frac{4}{\e^2} \,.
\ee
The result in momentum space is
\be
\label{eq:Imomentum}
\cI_{C_F^2} = A\,{C_F^2} (\la\lb)^{-1-2\e} \frac{4}{\e^2} \,.
\ee
The $C_F C_A$ contribution to the independent emission matrix element is 
\be
\label{eq:Idiagram}
\cA_{\cI, C_F C_A} = -2g^4 C_F C_A \mu^{4\e} \frac{(\kap + \kbp)(\kam + \kbm) + \kap\kam + \kbp\kbm}{(\kap + \kbp)(\kam + \kbm)\kap\kam\kbp\kbm} \,.
\ee
To evaluate this diagram we use the variables $u = \kap/\kbp$ and $v = \kbm/\kam$.  An intermediate step in the calculation is
\be
\cI_{C_F C_A} = - A\,{C_F C_A} s^{-1-2\e} \left[ \frac{1}{\e^2} + \int_{1/r}^{\infty} du \int_r^{\infty} dv\, (uv)^{-1-\e} \frac{u+v}{(1+u)(1+v)} \right] \,.
\ee
These integrals can be evaluated exactly in $\e$.  The leading divergences in $\cI_{C_F C_A}$ cancel with the $\cT$ diagrams, so we present the results for the sum of the diagrams.

The matrix element for the single 3-gluon vertex $\cT$ diagrams in \fig{fig:diagrams}(b) and (c) is
\begin{align}
\label{eq:Tdiagram}
\cA_{\cT} &= 2g^4 C_F C_A \,\mu^{4\e}  \frac{\kap\kbm + \kam\kbp}{2k_1\cdot k_2}  \left[1 + \frac{\kap\kam + \kbp\kbm}{(\kap+\kbp)(\kam+\kbm)}\right]\frac{1}{\kap\kam\kbp\kbm} \,.
\end{align}
Using the variables and transformations described in
Eqs.~(\ref{eq:transformstart}) -- (\ref{eq:transformend}), the intermediate step
with the remaining $z$ integral  in the calculation for the diagrams' contribution to $\Sopp$ is
\begin{align}
\cT &= A\,{C_F C_A} s^{-1-2\e} \int_1^{\infty} dz\, z^{-1-\e} f_1(z) \left[2\ln z + \frac{z+1}{z-1} g(z,r) \right] \,.
\end{align}

Since the leading divergences in the $\cI_{C_F C_A}$ and $\cT$ diagrams cancel, we sum them.  In position space, the result for their contribution to $\widetilde S^{\rm opp}$ is
\begin{align}
\label{eq:T+Iposition}
\widetilde{\cI}_{C_F C_A} + \widetilde{\cT} &= A\,{C_F C_A}(ix_1 ix_2)^{2\e}  \biggl\{ \Gamma(-2\e)^2 \left(\frac{2\pi^2}{3} + 4\zeta_3 \e + \frac{14\pi^4}{45}\e^2\right) \nn\\
&+ 4 \Gamma(-4\e)\left[\zeta_3 - \frac{\pi^2}{3} - 2 \e\left(\zeta_3 + \frac{\pi^4}{45} - \frac{\pi^2}{3}\right)\right] \nn\\
& + \left(\frac{2\pi^2}{3} -2\zeta_3  \right) \ln\left(2 + \frac{x_1}{x_2} + \frac{x_2}{x_1}\right) + \widetilde{\cF}_{\cal IT}\left(\frac{x_1}{x_2}\right)+ \widetilde{\cF}_{\cal IT}\left(\frac{x_2}{x_1}\right)\biggr\} \,,
\end{align}
where the last two terms are given by the integral
\begin{align}
\widetilde F_{IT} (b) &\equiv\int_0^1\frac{dr}{r} \ln\left(\frac{r+\frac{1}{r}+b+\frac{1}{b}}{2+b+\frac{1}{b}}\right) \left[\frac{4\ln(1+r)}{1+r}+ \frac{4\ln\left(1+\frac{1}{r}\right)}{1+\frac{1}{r}}  - 2\ln(1+r)\ln\left(1+\frac{1}{r}\right)  \right]  \nn \\
&= \widetilde{\cF}_{\cal IT} (b)+\widetilde{\cF}_{\cal IT} (1/b) 
\end{align}
where
\begin{align}
\label{eq:FITb}
\widetilde{\cF}_{\cal IT} (b) &\equiv -\frac{\pi^4}{36} + \frac{\ln^4 b}{24}+ \left(2\zeta_3 - \frac{2\pi^2}{3}\right)\ln(1+b)- \ln^2 b \ln(1-b) - \frac{\pi^2}{3}  \Li_2(1-b)\nn \\
&\qquad  - 4\ln b \Li_2(b)+ \big(\Li_2(1-b)\big)^2+6\Li_3(b) + 2\ln b \Li_3(1-b) 
\,. \end{align}
The sum of the two terms $\widetilde F_{IT}(b) = \widetilde{\cF}_{\cal IT} (b)+\widetilde{\cF}_{\cal IT} (1/b)$  is a finite function of $b$, in fact vanishing as $b\to 0$ or $b\to\infty$.  

In momentum space, the result for the $\cI_{C_F C_A},\cT$ diagrams' contribution to $\Sopp$ can be written 
\begin{align}
\label{eq:T+Imomentum}
\cI_{C_F C_A} + \cT &= A\,{C_F C_A} s^{-1-2\e} \bigg\{ \frac{2\pi^2}{3} + 4\zeta_3 \e + \frac{14\pi^2}{45}\e^2  + \hat{\cF}_{\cal IT}(r) +  \hat{\cF}_{\cal IT}(1/r) \bigg\} \\
&\equiv A\,{C_F C_A} s^{-1-2\e} \big[ F_0 + F_1(r) \big] \nn
\end{align}
where
\begin{align}
\label{eq:FITr}
  \hat{\cF}_{\cal IT}(r) &\equiv \ln(1+r) \ln\left(1+\frac{1}{r}\right) - \frac{4}{1+r}\ln(1+r) \nn\\
  & - \e \biggl[  \frac{\pi^2}{3}- 3\Li_3(-r) - \frac{4}{1+r}\ln r + \left(\frac{4r}{1+r} + 2\ln r\right)\Li_2(-r) \nn\\
&\qquad +\left(\frac{1}{2}\ln^2 r - 2\frac{1-r}{1+r}\ln r - 4 - \frac{\pi^2}{6}\right) \ln(1+r) \biggr] 
\,,\end{align}
The $r$-dependent functions appearing in \eq{eq:T+Imomentum} are the  $\cI_{C_F C_A} + \cT$ contribution to the function $F_1(r)$ defined in \eq{eq:Fdef}. This contribution is explicitly symmetric in $r\to \frac{1}{r}$ and   vanishes as $r\to 0$ or $r\to\infty$, as advertised.  This is also true of the other diagrams, which have the same form.

%=======================================================================
\subsection{$\cG$ and $\cH$ diagrams}

The matrix element for the gluon loop $\cG$ diagrams in \fig{fig:diagrams}(d) is
\begin{align}
\label{eq:Gdiagram}
\cA_{\cG} &= g^4 C_F C_A \,\mu^{4\e}  \frac{1}{(2k_1\cdot k_2)^2 (\kap+\kbp)^2(\kam+\kbm)^2 } \nn \\
&\quad   \Big\{ (\kap+\kbp)(\kam+\kbm)  \big[-16k_1\cdot k_2 - 2(1-\e)(\kap-\kbp)(\kam-\kbm) + 4(\kap + \kbp)(\kam+\kbm)   \nn \\
&\quad  + \kap\kbm + \kam\kbp \big] + (\kam+\kbm)^2 \left[ (1-\e) (\kap-\kbp)^2-2(\kap+\kbp)^2-\kap \kbp \right]  \nn \\
&\quad   + (\kap+\kbp)^2 \left[ (1-\e) (\kam-\kbm)^2-2(\kam+\kbm)^2- \kam \kbm \right] \Big\}  \,.
\end{align}
The matrix element for the ghost loop $\cH$ diagrams in \fig{fig:diagrams}(e) is
\begin{align}
\label{eq:Hdiagram}
\cA_{\cH} &= 
g^4 C_F C_A \,\mu^{4\e} \frac{1}{(2k_1\cdot k_2)^2 (\kap+\kbp)^2(\kam+\kbm)^2} \nn \\
&\quad \times \left\{ -(\kap+\kbp)(\kam+\kbm)  \left( \kap\kbm + \kam\kbp \right) + \kap \kbp (\kam + \kbm)^2+ \kam \kbm (\kap+\kbp)^2 \right\} \,.
\end{align}
The ghost loop diagram cancels a set of terms in the gluon loop diagram, and so
we evaluate them together.  Using the variables and transformations described in
Eqs.~(\ref{eq:transformstart}) -- (\ref{eq:transformend}), the intermediate step  with the remaining $z$ integral 
in the calculation of the diagrams' contribution to $\Sopp$ is
\begin{align}
\cG + \cH &= A\,{C_F C_A} s^{-1-2\e}
\int_1^{\infty} dz \,z^{-\e} \bigg\{ 4\frac{1}{z^2-1}\,g(z,r) f_1(z) 
\nn \\
&\qquad\quad\qquad\qquad\qquad\!\! 
+2 ( 1-\e) \frac{1}{z^2-1}\,g(z,r) [f_1(z) - 2f_2(z)] \nn \\
&\qquad\quad\qquad\qquad\qquad\!\! + 2(1-\e)\frac{z-1}{z+1} \frac{r}{(r+z)(1+rz)}\left[ f_1(z) - 2f_2(z)\right] \bigg\} \,.
\end{align}
In position space, the result for the diagrams' contribution to $\widetilde S^{\rm opp}$ is
\begin{align}
\label{eq:G+Hposition}
\widetilde{\cG} + \widetilde{\cH} &= A\,{C_F C_A} \, (ix_1 ix_2)^{2\e} \bigg\{  \Gamma(-4\e)  \left(\frac23 - \frac{10\pi^2}{9} \right) + \left( \frac{10}{9} - \frac{31\pi^2}{54} + \frac{5}{3}\zeta_3 \right)  \nn \\
& \qquad + \left(-\frac13 + \frac{5\pi^2}{9} \right) \ln\left(2 + \frac{x_1}{x_2} + \frac{x_2}{x_1}\right) + \widetilde{\cF}_{\cG\cH} \left(\frac{x_1}{x_2}\right) + \widetilde{\cF}_{\cG\cH} \left(\frac{x_2}{x_1}\right) \bigg\} \,,
\end{align}
where the last two terms are given by the integral
\begin{align}
\widetilde F_{GH}(b) &\equiv \int_0^1 \frac{dr}{r} \ln\left(\frac{r + \frac{1}{r} + b + \frac{1}{b}}{2 + b + \frac{1}{b}}\right)\frac{2}{3(1+r)^3}  \\
&\quad  \times \bigg[ -2r(1+r) + (5+9r+6r^2) \ln(1+r) + (6r + 9r^2 + 5r^3) \ln\left(1+\frac{1}{r}\right) \bigg] \nn \\
&= \widetilde{\cF}_{\cG\cH} (b)+\widetilde{\cF}_{\cG\cH} (1/b) \,, \nn
\end{align}
with
\begin{align}
\label{eq:FGHb}
\widetilde{\cF}_{\cG\cH} (b)
&\equiv \frac{1}{3(1-b)}\ln b + \frac{b}{6(b-1)^2} \ln^2 b +\left(\frac13 - \frac{5\pi^2}{9}\right)  \ln(1+b)  -\frac56 \ln^2 b \ln(1-b) \nn \\
&\qquad - \frac{10}{3}\ln b  \Li_2(b) + 5 \Li_3(b) \,. 
\end{align}
In momentum space, the diagrams' contribution to $\Sopp$ is 
\begin{align}
\label{eq:G+Hmomentum}
\cG + \cH &= A\,{C_F C_A} \, s^{-1-2\e} \Big\{\hat{\cF}_{\cG\cH}(r) +\hat{\cF}_{\cG\cH}(1/r)\Big\}
\end{align}
where
\begin{align}
\label{eq:FGHr}
 \hat{\cF}_{\cG\cH}(r) &= -\frac{2}{3(1+r)^3} \bigg[ -2r + (5+9r+6r^2) \ln(1+r)  \bigg]  \\
& \qquad - \frac{2\e}{3(1+r)^3} \Bigg[ \frac{16}{3} r + \frac{5\pi^2}{12} (1+r)^3  - \left(\frac{31}{3} +23r +8r^2 \right)\ln(1+r) \nn \\
& \qquad\qquad\qquad\qquad - \frac12 (5 + 3r - 3r^2 - 5r^3)  \ln^2 (1+r)   + (6r + 9r^2 + 5r^3) \Li_2(-r) \Bigg] \nn \,.
\end{align}
In this case \eq{eq:G+Hmomentum} takes the form given by \eqs{eq:generalmomspace}{eq:Fdef} with $F_0=0$ and $F_1(r) = \hat{\cF}_{\cG\cH}(r) + \hat{\cF}_{\cG\cH}(1/r)$.

%=======================================================================
\subsection{$\cQ$ diagrams}

The matrix element for the $\cQ$ diagrams in \fig{fig:diagrams}(f) is
\begin{align}
\label{eq:Qdiagram}
\cA_{\cQ} &=  8g^4 C_F T_R n_f \,\mu^{4\e}\frac{1}{(2k_1\cdot k_2)^2 (\kap+\kbp)^2(\kam+\kbm)^2} \nn \\
&\qquad \times \left[  \kap \kbp (\kam+\kbm)^2+\kam \kbm (\kap+\kbp)^2  \right. \nn \\
&\qquad \qquad\quad \left. + \left(2 k_1\cdot k_2 -  \kap\kbm - \kam\kbp\right)( \kap+\kbp) (\kam+\kbm)  \right] \,.
\end{align}
Using the variables and transformations described in Eqs.~(\ref{eq:transformstart}) -- (\ref{eq:transformend}), an intermediate step in the calculation for the $\cQ$ diagrams' contribution to $\Sopp$ is
\begin{align}
\cQ = A\,{C_F T_R n_f} s^{-1-\e} \int_1^{\infty} dz \,z^{-\e} &\bigg\{ -8\, \frac{1}{z^2-1}\,g(z,r) [f_1(z) - f_2(z)]   \nn \\
& \!\quad -4\,\frac{z-1}{z+1} \frac{r}{(r+z)(1+rz)}\left[ f_1(z) - 2f_2(z)\right] \bigg\}.
\end{align}
In position space, the result for the diagrams' contribution to $\widetilde S^{\text{opp}}$ is
\begin{align}
\label{eq:Qposition}
\widetilde{\cQ} = A\,{C_F T_R n_f}& \, (ix_1 ix_2)^{2\e} \bigg\{ \Gamma(-4\e) \left(-\frac43 + \frac{8\pi^2}{9}\right) 
- \left( \frac{17}{9}-\frac{16 \pi^2}{27} + \frac{4}{3} \zeta(3) \right) \nn \\
&  - \left( \frac{4 \pi^2}{9}- \frac23 \right) \ln \left( 2 + \frac{x_1}{x_2} + \frac{x_2}{x_1}\right) +\widetilde{\cF}_{\cQ} \left(\frac{x_1}{x_2}\right)+ \widetilde{\cF}_{\cQ} \left(\frac{x_2}{x_1}\right) \bigg\} ,
\end{align}
where the last two terms are given by the integral
\begin{align}
\widetilde F_Q(b) &\equiv\int_0^1 \frac{dr}{r} \ln\left(\frac{r + \frac{1}{r} + b + \frac{1}{b}}{2 + b + \frac{1}{b}}\right) \frac{-4}{3(1+r)^3}  \\
&\qquad\times \left[-2r(1+r) + (1+r)^3 \ln \left(2+r+\frac{1}{r}\right) + (1 - r^3 ) \ln r \right]  \nn \\
&= \widetilde{\cF}_{\cQ} \left(b\right)+\widetilde{\cF}_{\cQ} \left(1/b\right)  \nn
\end{align}
with
\begin{align}
\label{eq:FQb}
\widetilde{\cF}_{\cQ} \left(b\right) 
&\equiv  \frac{2 }{3 (b-1)} \ln b - \frac{b}{3(b-1)^2}\ln^2 b- \left(\frac23 - \frac{4\pi^2}{9}\right) \ln(1+b) \\
&\qquad + \frac23 \ln^2 b  \ln(1-b)+ \frac83\ln b \Li_2 (b)  -4 \Li_3(b)\nn \,.
\end{align}
In momentum space, the diagrams' contribution to $\Sopp$ is 
\begin{align}
\label{eq:Qmomentum}
\cQ =\, & A\,{C_F T_R n_f} \, s^{-1-2\e} \Big\{ \hat{\cF}_{\cQ}(r)+\hat{\cF}_{\cQ}(1/r)\Big\}
\,,\end{align}
where
\begin{align}
\label{eq:FQr}
\hat{\cF}_{\cQ}(r) &=
\frac{4}{3(1+r)^3}\left[-2r + (1+r)^3 \ln \left(1+r\right) +  \ln r \right]  \\
& + \frac{8\e}{3(1+r)^3}\bigg[ \frac{5}{3}r + \frac{\pi^2}{24}(1+r)^3- \frac{1}{8} (1+r)^3 \ln^2 r\nn \\
&\qquad\qquad\qquad - \left(\frac83 + 7r + r^2\right)\ln(1+r)  + \frac12 (r^3 - 1) \Big(\ln^2(1+r) + \Li_2 \left(-r \right)\Bigr) \bigg] \nn
 \,.\end{align}
 Again the form of \eq{eq:Qmomentum} is that given by \eqs{eq:generalmomspace}{eq:Fdef} with $F_0=0$ and $F_1(r) = \hat{\cF}_{\cQ}(r) + \hat{\cF}_{\cQ}(1/r)$.

%=======================================================================
\subsection{Total Opposite Hemisphere Soft Functions}

Adding up the results in Eqs.~(\ref{eq:T+Iposition}), (\ref{eq:G+Hposition}),
and (\ref{eq:Qposition}) the total opposite hemisphere soft function in position
space is
\begin{align}
\label{finalSoppposition}
& \widetilde S^{\text{opp}} (x_1, x_2) = \nn \\
& \left(\frac{\alpha_s}{2\pi}\right)^2 \frac{\left(e^{\gamma_E}\mu^2\right)^{2\e}}{\Gamma(1-\e)^2} \left(ix_1 ix_2\right)^{2\e} \Bigg\{ \Gamma(-2\e)^2 \left[ C_F^2 \frac{4}{\e^2} +  C_F C_A\left( \frac{2\pi^2}{3}  + 4\zeta_3 \e + \frac{14\pi^2}{45} \e^2 \right) \right] \nn \\
& \quad + \left[ 2 \Gamma(-4\e) - \ln\left(2 + \frac{x_1}{x_2} + \frac{x_2}{x_1}\right) \right] \nn \\
&\qquad\qquad \times \Bigg[ C_F C_A \bigg( \frac13 - \frac{11\pi^2}{9} + 2\zeta_3 + \e\bigg(-\frac{20}{9} + \frac{67\pi^2}{27} - \frac{4\pi^4}{45} - \frac{22}{3} \zeta_3 \bigg) \bigg)  \nn \\
& \qquad\qquad\qquad + C_F T_R n_f \bigg( -\frac{2}{3} + \frac{4\pi^2}{9} + \e \bigg(  \frac{34}{9} - \frac{32\pi^2}{27} + \frac83 \zeta_3 \bigg)\bigg) \Bigg] \nn \\
& \quad + C_F C_A \bigg( \widetilde{\cF}_{C_F C_A} \left(\frac{x_2}{x_1}\right)+ \widetilde{\cF}_{C_F C_A} \left(\frac{x_1}{x_2}\right) \bigg)
\nn\\ & \quad 
+ C_F T_R n_f \bigg( \widetilde{\cF}_{C_F T_R n_f} \left(\frac{x_2}{x_1}\right)+ \widetilde{\cF}_{C_F T_R n_f} \left(\frac{x_1}{x_2}\right)  \bigg) \Bigg\} 
\,, \end{align}
where 
\begin{align}
\widetilde{\cF}_{C_F C_A} \left(\frac{x_2}{x_1}\right) &= \widetilde{\cF}_{\cI\cT} \left(\frac{x_2}{x_1}\right) + \widetilde{\cF}_{\cG\cH} \left(\frac{x_2}{x_1}\right) \,,
\end{align}
which are given in~\eqs{eq:FITb}{eq:FGHb} and
\begin{align}
\widetilde{\cF}_{C_F T_R n_f} \left(\frac{x_2}{x_1}\right) = \widetilde{\cF}_{\cQ} \left(\frac{x_2}{x_1}\right) \,,
\end{align}
which is given in~\eq{eq:FQb}.  In the final result for the soft function $\cS(x_1,x_2,\mu)$ in position space given by \eqs{finalposition}{finalt2pos}, we chose to replace the single log $\ln(2+x_1/x_2+x_2/x_1)$ in \eq{finalSoppposition} with $\ln(x_1/x_2+x_2/x_1)$ and shift the difference into the sum of $\tilde \cF$ functions, resulting in the $F_{Q,N}$ functions given by \eq{FQFN}. This is so that the remaining terms $F_{Q,N}(x_1/x_2)+F_{Q,N}(x_2/x_1)$  in \eq{finalt2pos} are bounded along the real $b = x_1/x_2$ axis, as illustrated in \fig{fig:FQN}. Otherwise they would have a pole at $b=-1$.

In momentum space adding up the contributions from Eqs.~(\ref{eq:T+Imomentum}),
(\ref{eq:G+Hmomentum}), and (\ref{eq:Qmomentum}) the total opposite hemisphere
soft function is
\begin{align}
\label{Soppfinalmom}
&\Sopp(\la,\lb) = \left(\frac{\alpha_s}{2\pi}\right)^2 \frac{\left(e^{\gamma_E}\mu^2\right)^{2\e}}{\Gamma(1-\e)^2} \, C_F^2 \left(\la\lb\right)^{-1-2\e} \frac{4}{\e^2} \nn \\
&+ \left(\frac{\alpha_s}{2\pi}\right)^2 \frac{\left(e^{\gamma_E}\mu^2\right)^{2\e}}{\Gamma(1-\e)^2} \, (\la\lb)^{-1-2\e} \Bigg\{ C_F C_A \left( \frac{2\pi^2}{3} + 4\zeta_3\e + \frac{14\pi^2}{45}\e^2 \right) \nn \\
&\quad + C_F C_A \bigg[  \hat{\cF}_{\cal IT}(r) + \hat{\cF}_{\cal IT}(1/r) + \hat{\cF}_{\cG\cH}(r) +\hat{\cF}_{\cG\cH}(1/r) \bigg] \nn \\
&\quad + C_F T_R n_f  \bigg[ \hat{\cF}_{\cQ}(r)  + \hat{\cF}_{\cQ}(1/r) \bigg] \Bigg\} 
\,\end{align}
where $ \hat{\cF}_{\cal IT}(r) $,  $\hat{\cF}_{\cG\cH}(r)$, and $\hat{\cF}_{\cQ}(r) $ are defined in \eqss{eq:FITr}{eq:FGHr}{eq:FQr}, respectively.

\bibliography{NGL}

\end{document}